\documentclass[11pt]{article}
\usepackage{url}
\usepackage{tikz}
\usepackage{subfig}
\usepackage{transparent}

\usepackage{amsmath}
\usepackage{amssymb}
\usepackage{graphicx}
\usepackage{float}
\usepackage{caption}
\usepackage{amsthm}
\usepackage{color}
\usepackage[normalem]{ulem}
\usepackage{enumerate}
\definecolor{green}{rgb}{0,0.5977,0}
\usepackage[linesnumbered,algoruled,boxed,lined]{algorithm2e}
\usepackage{placeins} 

\newcommand{\rr}{\mathbb R}

\newcommand{\zz}{\mathbb Z}

\newcommand{\ep}{\varepsilon}
\newcommand{\rhat}{R'}
\newcommand{\cmin}{c_\textnormal{min}}
\newcommand{\cmax}{c_\textnormal{max}}
\newcommand{\imbalance}{\textnormal{imb}}

\newcommand{\cE}{\mathcal{E}}
\newcommand{\cR}{\mathcal{R}}
\newcommand{\cI}{\mathcal{I}}
\newcommand{\cS}{\mathcal{S}}
\newcommand{\cL}{\mathcal{L}}

\newcommand{\poly}{\mathrm{poly}}

\newcommand{\im}{\mathrm{im}}

\newcommand{\st}{\txt{sp}}

\newcommand{\floor}[1]{\left\lfloor#1\right\rfloor}
\newcommand{\ceil}[1]{\left\lceil#1\right\rceil}
\newcommand{\abs}[1]{\left|{#1}\right|}

\newcommand{\suchthat}{\ | \ }

\newcommand{\twocases}[4]{\begin{cases} #2 & #1 \\ #4 & #3 \end{cases}}

\newcommand{\txt}[1]{\text{#1}}

\newcommand{\red}[1]{{\color{red} #1}}

\newcommand{\stext}[1]{\ \ \ \ \ \text{(#1)}}

\newcommand{\push}{\\ & \ \ \ \ \ \ \ \ \ \ }

\makeatletter 
\g@addto@macro{\@algocf@init}{\SetKwInOut{Parameter}{Parameters}} 
\makeatother

\newcommand{\real}{\text{Re}}

\theoremstyle{plain}
\newtheorem{theorem}{Theorem}
\newtheorem{lemma}[theorem]{Lemma}
\newtheorem{proposition}[theorem]{Proposition}
\newtheorem{claim}{Claim}
\newtheorem{corollary}[theorem]{Corollary}
\newtheorem{conjecture}[theorem]{Conjecture}

\newtheorem{observation}[theorem]{Observation}
\theoremstyle{definition}
\newtheorem{definition}[theorem]{Definition}
\newtheorem{example}[theorem]{Example}

\newcommand{\rev}[1]{{#1}}
\newcommand{\new}[1]{{#1}}
\newcommand{\neww}[1]{{{#1}}}
\newcommand{\newww}[1]{{{#1}}}

\makeatletter
\newcommand{\eqnum}{\refstepcounter{equation}\textup{\tagform@{\theequation}}}
\makeatother

\title{Sampling Balanced Forests of Grids in Polynomial Time}

\author{Sarah Cannon\footnote{Claremont McKenna College, \texttt{scannon@cmc.edu}}, Wesley Pegden\footnote{Carnegie Mellon University, \texttt{wes@math.cmu.edu}}, and Jamie Tucker-Foltz\footnote{Yale University, \texttt{jtuckerfoltz@gmail.com}}}

\begin{document}
	\maketitle
	\thispagestyle{empty}
	\begin{abstract}
		We prove that a polynomial fraction of the set of $k$-component forests in the $m \times n$ grid graph have equal numbers of vertices in each component, for any constant $k$. This resolves a conjecture of Charikar, Liu, Liu, and Vuong, and establishes the first provably polynomial-time algorithm for (exactly or approximately) sampling balanced grid graph partitions according to the spanning tree distribution, which weights each $k$-partition according to the product, across its $k$ pieces, of the number of spanning trees of each piece. Our result follows from a careful analysis of the probability a uniformly random spanning tree of the grid can be cut into balanced pieces.
		
		Beyond grids, we show that for a broad family of lattice-like graphs, we achieve balance up to any multiplicative $(1 \pm \varepsilon)$ constant with constant probability, and up to an additive constant with polynomial probability. More generally, we show that, with constant probability, components derived from uniform spanning trees can approximate any given partition of a planar region specified by Jordan curves. These results imply polynomial-time algorithms for sampling approximately balanced tree-weighted partitions for lattice-like graphs.
		
		Our results have applications to understanding political districtings, where there is an underlying graph of indivisible geographic units that must be partitioned into $k$ population-balanced connected subgraphs. In this setting, tree-weighted partitions have interesting geometric properties, and this has stimulated significant effort to develop methods to sample them.
	\end{abstract}
	
	\setcounter{page}{1}
	
	\section{Introduction}\label{secIntro}
	
	We consider the following question: \neww{G}iven a graph $G$ and an integer constant $k$, how can one randomly sample partitions of $G$ into $k$ connected pieces, each of equal size?  We address this question in the context of the \emph{spanning tree distribution} on partitions, under which the weight of a partition is proportional to the product of the numbers of spanning trees in each partition class.  This distribution has been the subject of intense research in the context of mathematical approaches to the analysis of political districtings~\cite{revrecom, charikar2022complexity, recom,  SMC, clelland2021compactness, procacciaTuckerFoltz2021compactness, tappspanning} \new{(on which we elaborate in the next section)}.  While efficient algorithms exist to sample from this distribution when there are no size constraints on the partition classes, there is no general recipe for converting such a sampler to an efficient sampler for the \emph{balanced} spanning tree distribution, where we condition the spanning tree distribution on the event that the partition classes are equal in size.   For the prototypical case of grid graphs, the following conjecture of Charikar, Liu, Liu, and Vuong asserted rejection sampling would suffice:
	
	\begin{conjecture}[Charikar, Liu, Liu, and Vuong (2022)]\label{c.balance}
		For the $m \times n$ grid graph, the proportion of balanced $k$-partitions under the spanning tree distribution is at least $1/\poly(m, n)$, when $k = O(1)$.
	\end{conjecture}
	
	\noindent We confirm this conjecture as follows. 
	\begin{theorem}\label{thm:k-dist}      Let $G$ be an $m \times n$ grid graph where $m \geq n$ and $k|m$.  The probability that a $k$-partition from the spanning tree distribution is balanced is at least 
		\begin{equation}
			\frac{1}{\beta^{k^2}n^{5k-5}m^{3k-3}}
		\end{equation}
		for a fixed constant $\beta$.
	\end{theorem}
	\noindent We note that the assumption that $k$ divides the longer dimension is mostly for ease of exposition. With some more effort (and worse constant factors) one could require just $k|nm$, with essentially the same proof techniques. 
	Theorem \ref{thm:k-dist} will follow from Theorem \ref{thm:k-split}, which will assert that, for a uniformly random spanning tree of the $m \times n$ grid graph ($m\geq n, k|m$), there is a $1/\poly(mn)$ chance that there are $k-1$ edges whose removal divides the tree into equal-size components. 
	
	Our result implies the existence of the first provably polynomial-time sampling algorithms for balanced $k$-partitions of grids under the spanning tree distribution. Approximate sampling in polynomial time can be achieved in a straightforward way by adding a rejection step to the approximate sampler of~\cite{anari2021forestsampling} (Theorem~\ref{thmAlgUpDown}). We also give an alternative sampling algorithm (based on splitting uniformly random spanning trees) that produces perfect samples in polynomial time (Theorem~\ref{thmAlgWilson}). 
	
	Motivated by the application to redistricting, grid graphs are a natural focus as a prototype of graphs capturing adjacency relations among small geographic units. A natural follow\neww{-}up question is, does a similar balance theorem hold when the underlying graph is not a grid, but is sufficiently ``grid-like''? For any definition of ``grid-like'' that is general enough to model the kinds of graphs encountered in \neww{redistricting}, we cannot hope to get exact balance, corresponding to exactly equal populations in each district. In practice, the implemented requirement may instead be that districts are close in population \neww{(examples are discussed in Section~\ref{sec:dist_background})}. With this motivation we generalize Theorem \ref{thm:k-dist} to results beyond exact balance in grid graphs to allow multiplicative and even additive error from exact balance in more general families of graphs. In Section \ref{sec:partitioncurves}, we define general classes of ``lattice-like'' graphs (see Definition \ref{defLatticeSequence}) and consider partitions under multiple looser notions of balance. \rev{For example, our strictest definitions include standard deterministic lattices, such as the square, triangular, and hexagonal lattices, while our looser notion also covers random geometric examples such as the Poisson--Voronoi/Delaunay pair after restriction to any fixed bounded region, with the relevant high-probability interpretation as the intensity tends to infinity.}

\begin{figure}[t]
		\begin{center}
			\includegraphics[width=.5\linewidth]{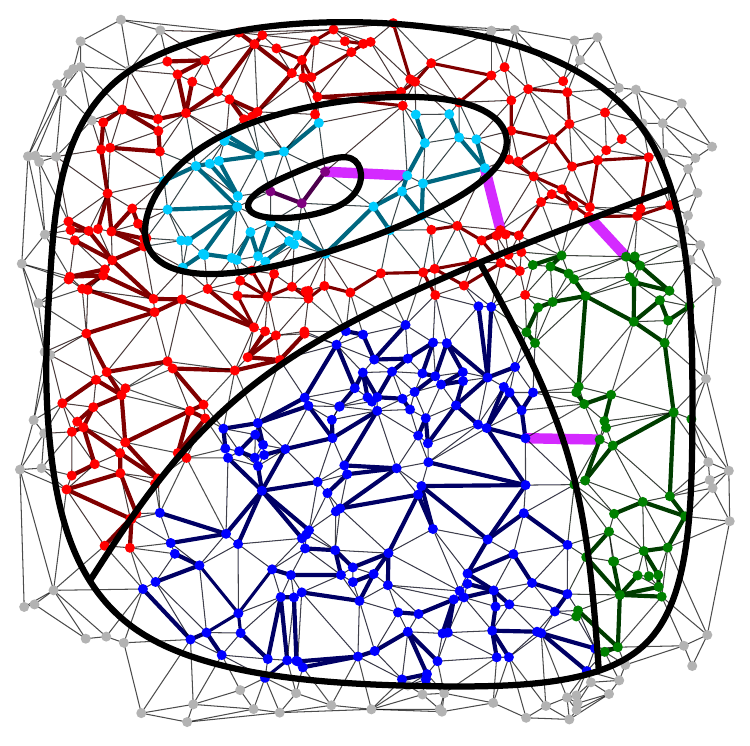}
		\end{center}
		\vspace{-1em}
		\caption{A partition of a region of a lattice-like graph approximating a division of the plane given by Jordan curves, and induced by the components remaining after deleting the four bright purple edges from a spanning tree of the region.   Theorem \ref{thm:planegraph} shows that given a division of the plane by curves, a random spanning tree of a sufficiently refined lattice-like graph can, with probability bounded below by a constant, be cut into components inducing a partition whose classes each has small Hausdorff distance from the corresponding face of the drawing.}
		\label{fig:curves} 
	\end{figure}

	If we are interested in dividing a random spanning tree into $k = O(1)$ components that are only approximately balanced (up to a $(1\pm \ep)$ multiplicative error), we show the surprisingly strong result on lattice-like graphs (including grids) that this is possible with \emph{constant} probability; Corollary \ref{cor:ep} gives the precise statement for grids. In fact, we prove a more general result, which is that 
	\neww{for any partition of a region of the plane given by a collection of Jordan curves,  
		a uniform spanning tree on a sufficiently refined lattice-like graph will, with probability bounded below by a constant, be splittable into components that approximately match this partition (see Figure \ref{fig:curves}).} 
	If $D$ is a fixed plane graph, and $\Omega_{D,\Lambda_n}$ denotes a region of $\Lambda_n$ whose boundary approximates the boundary of the outer face of $D$, we have that:
	
	\begin{theorem}[Informal version of Theorem \ref{thm:planegraph}]\label{thm:mult_intro}
		Given any plane graph $D$ with $k+1$ faces, let $\phi_1,\dots,\phi_k\subseteq \rr^2$ denote its inner faces.  For any $\ep>0$, as $n\to \infty$, there is a constant lower bound, depending only on the plane graph, $\ep$, \neww{and the lattice-sequence constant $\rho$}, on the probability that a random spanning tree $T$ of $\Omega_{D,\Lambda_n}$ contains $k-1$ edges whose removal disconnects $T$ into components $C_1,\dots, C_k$, where each $C_i$ is at Hausdorff distance $<\ep$ from a corresponding face $\phi_i$ of $D$.
	\end{theorem}
	
	If one is interested in the even stronger condition of \emph{additive} approximate balance, we show one can achieve this in lattice-like graphs as well, with just one additional assumption regarding the uniformity of the density of \neww{both primal and dual} vertices. This stronger assumption would fail, for example,  for the previously mentioned case of \neww{Delaunay} triangulations of random point sets, but would still \newww{hold} for any finite-degree, doubly-periodic connected plane graph. 
	
	\begin{theorem}[Informal version of Theorem \ref{thmAdditiveError}]\label{thm:add_intro}
		If $\Lambda_n$ is a sequence of infinite \newww{plane} graphs of decreasing scale embedded in $\rr^2$ which are \emph{uniformly} lattice-like (see Definition \ref{defStrongLatticeSequence}) and $\Omega_{D,\Lambda_n}$ denotes a region of $\Lambda_n$ whose boundary approximates the unit square, then there exists a constant $A > 0$ such that as $n \rightarrow \infty$, the probability a random spanning tree $T$ of $\Omega_{D,\Lambda_n}$ contains $k-1$ edges whose removal produces $k$ pieces whose sizes are equal up to an additive $A$ vertices in each part is at least $1/\poly(n)$.    
	\end{theorem}
	
	Balance up to an additive constant is the best one could hope for in the framework of Theorem~\ref{thm:add_intro}; exact balance may not be possible because of the structure of $\Lambda_n$ or the way it's trimmed to approximate the unit square. However, we expect that even exact balance would be possible with mild additional assumptions on the local behavior of random walks in the lattice (as well as the necessary divisibility conditions).
	
	Combining these approximate balance results for lattice-like graphs with known algorithms and rejection sampling gives the first provably polynomial-time sampling algorithms in all of the settings discussed.
	
	\subsection{Random Sampling of Political Districting Plans}\label{sec:dist_background}
	
	The question of efficient sampling of balanced graph partitions has been extensively studied over the last few years in the context of analysis of political districting plans. \neww{While regions are divided into political districts for the purpose of electing representatives in many countries around the world, most recent analysis using sampling methods has focused within the United States}. Sampling algorithms enable the generation of large {\it ensembles} of plans, which are useful for several purposes (detecting outliers, understanding the impacts of rules, evaluating the stated intentions of map-drawers, and more).  Ensemble analysis has been used in many academic studies, including~\cite{ForestReCom, ForestRecomMultiscale,  VRA-ELJ, benade2021rcv, Alaska, revrecom, chenstephanopoulos, cohen2021dp, competitiveness, recom,  MRL,ImaiFlip, NC,  ImaiReComDPpub, SMC, georgia}, as well as in mathematicians' expert reports in court cases~\cite{hirsch2022brief, chen2022brief, duchin2019brief, mattingly2021expert}.
	
	Randomly sampling political districting plans is equivalent to a sampling problem for suitable partitions of a graph, with vertices representing small geographic regions such as precincts or census blocks and edges representing adjacencies. A districting plan with $k$ districts is a partition of this graph into $k$ pieces, which are generally required to be connected. Throughout, we will call a partition of a graph into $k$ connected pieces a {\it $k$-partition}, and we will refer to the $k$ partition classes of a partition as {\it districts}.   Because they represent physical geography, these graphs are typically planar or nearly planar. While they are not usually perfect grids (except at times in cities), there is general consensus that grids are the logical simplified setting to first consider.  By going beyond grids to lattice-like graphs, we move to a much more expressive graph class that can describe significant additional real-world geography.

	In the context of redistricting, there are other constraints on partitions one must consider, including those related to population and shape. Our interest in balanced partitions stems from common requirements that districts have equal or near-equal populations. While our first main result resolves a conjecture about exactly balanced partitions, in practice most processes for sampling political districting plans do not aim for exact population balance but instead aim to keep the population to within a tolerance of $1$-$2\%$ \neww{~\cite{recom, revrecom, ForestReCom, SMC}}. This naturally corresponds to the setting of Theorem~\ref{thm:mult_intro}, where district sizes may vary by a multiplicative $1 \pm \varepsilon$ factor. 
	Related to district shape, the spanning-tree distribution we analyze is targeted by several sampling algorithms designed for redistricting analysis~\cite{revrecom, recom, SMC}, and has been shown to strongly correlate with geometric properties intended to capture legal requirements for `compactness' of districts~\cite{clelland2021compactness, procacciaTuckerFoltz2021compactness,tappspanning}. 
	
	The relative frequency of balanced partitions under the spanning tree distribution is particularly salient given the significant recent progress made in sampling algorithms for this distribution. 
	Note one can sample from the spanning tree distribution over $k$-partitions by sampling a $k$-component forest uniformly at random. 
	In 2020, leveraging recent breakthroughs in the polynomial-method approach to Markov chain mixing, Anari, Liu, Gharan, Vinzant, and Vuong gave an $O(N\log^2 N)$ approximate sampler for the latter problem based on the `down-up' walk on the complement of $k$-component forests of an $N$-vertex graph~\cite{anari2021forestsampling}. \neww{When viewed as a Markov chain on the forests themselves rather than their complements, this becomes an `up-down' walk, and we will refer throughout to this method as the up-down walk on forests.}
	In Section \ref{secAlgs}, we discuss the use of our results in the context of an additional rejection step for approximate samplers based on Markov chains, and also give a new perfect sampler for this problem that is not based on a Markov chain. 

	
	
	\neww{Our new perfect sampler and Markov chains such as the up-down walk operate in a context without balance constraints and only account for balance in an extra rejection step. This is different from the approaches cited above such as {\it recombination Markov chains}, which maintain balance throughout and} can have exponential mixing time for some special families of graphs (including carefully chosen subgraphs of the grid) \cite{charikar2022complexity}. Even on rectangular grids, recombination chains with strict balance constraints can fail to be ergodic if there are many small districts \cite{LPT}. Positive mixing time results for any reasonable class of graphs are not available. 
	\neww{However, most sampling methods that do maintain balance throughout use as a subroutine the process of drawing a random spanning tree and, if possible, splitting it into pieces of prescribed sizes~\cite{recom, revrecom, ForestReCom, SMC}. 
		Because of this, our results are also relevant to these methods, bounding the probability this subroutine successfully finding edges that cut  random spanning trees in balanced ways.   }
	

	Other Markov chains employed in the redistricting context include Glauber dynamics for contiguous partitions, which exchange individual vertices between districts.  Here, without any additional constraints or weighting, stationary distributions are uniform on partitions with connected districts.  Mixing time can again be exponential for some classes of graphs~\cite{frieze2022subexponential}. In fact, even in the absence of balance, it is not known whether the Glauber dynamics has polynomial mixing time for partitions of grid graphs into $k$ connected pieces, or indeed whether any polynomial-time algorithm to uniformly sample partitions of grid graphs uniformly randomly into $k$ connected pieces exists, even for $k=2$.
	
	\subsection{Approach}
	
	Rather than working with the tree distribution on partitions, we work with the uniform distribution on spanning trees. As we prove in Lemma~\ref{lem:balancedprob}, if there is a polynomial lower bound on the probability a random spanning tree can be split into $k$ equal-sized components, there is a polynomial lower bound on the probability a random tree-weighted forest with $k$ components is balanced.  The majority of our work therefore focuses on uniformly random spanning trees and the probability they can be split into components with desired properties.
	
	Wilson's algorithm for generating uniformly random spanning trees shows that a uniformly random spanning tree of a graph can be generated with a sequence of loop-erased random walks~\neww{\cite{wilson}}.  This allows us to couch questions about the structure of spanning trees in grids and grid-like graphs as questions about the behavior of random walks in such environments.  For our approximate results, we will leverage basic properties of random walks in grid-like environments to know that they have a good chance of behaving roughly any way we want; for exact cases like nice regions of the square and triangular lattices, these properties can be viewed as consequences of the conformal invariance of Brownian motion, as the scaling limit of the lattice walk.  But we will state and prove these properties in more direct ways, as we need less than this and thus can establish our approximate results in greater generality.
	
	This perspective offered by Wilson's algorithm would be most valuable if a small number of loop-erased random walks used by the algorithm \neww{were} enough to tell whether, e.g., the resulting spanning tree will be splittable into $k$ balanced components\neww{. This is because it would then suffice} to analyze the first phases of Wilson's algorithm, when the environment in which the loop-erased-random walks are taking place is mostly unperturbed.  But whether a spanning tree will be splittable into $k$ balanced parts is a property which may not be decided until the very last step of Wilson's algorithm.
	
	For this reason, we instead will analyze how Wilson's algorithm generates a spanning tree of the \emph{dual} of the graph that we are interested in, which suffices since spanning trees of \newww{plane} graphs are in bijective correspondence with the spanning trees of their dual graphs: If $T$ is a spanning tree of $G$, its dual spanning tree $T^*$ contains all edges in $G^*$ whose corresponding edges are missing from $T$.   If $T$ is a spanning tree of $G$ with dual tree $T^*$, then the $k$ connected components of $T\setminus e_1,\dots,e_{k-1}$ are bounded by $k$ cycles in $T^*\cup e_1\cup\dots\cup e_{k-1}$.  In particular, to show that components with certain sizes or structure can be created by removing edges in $T$, it suffices to show that suitable boundary cycles almost already exist in $T^*$.  Crucially, when $k$ is a constant, this is a property that can \neww{at times} be observed after a constant number of loop-erased-random walks have \neww{been} taken by Wilson's algorithm, in the generation of $T^*.$
	
	For some of our results on general lattice sequences, we will use a particular implementation of Wilson's algorithm (described in Section \ref{sec:modified_wilson}) in which, having completed one loop erased random walk, we (sometimes) choose the next starting point for a new loop-erased random walk as the exit vertex of simple random walk within the induced subgraph of the already-built tree itself. This allows us to analyze the progress of the algorithm in long phases that may include many separate loop-erased random walks, but for which these separate loop-erased random walks can all be seen as being generated using a single random walk on the graph.
	
	\subsection{Organization of paper}\label{subOutline}
	
	In Section \ref{secPrelim} we give the necessary background to understand our approach, including the relationships between various algorithms for sampling random spanning trees and forests. Section~\ref{secGrids} is devoted to proving our main result about $k$-partitions of grid graphs, Theorem \ref{thm:k-split}, along with stronger bounds for the special case of $k = 2$. Section~\ref{sec:partitioncurves} extends these results to approximate balance (multiplicative and additive) on lattice-like graphs. Finally, in Section \ref{sec:experiments}, we empirically evaluate the probabilities that random spanning trees of $10\times10$, $50\times50$, and $100\times100$ grids can be split at various locations into pieces of various sizes. These experiments visually confirm the analytical results in this paper and show some surprising patterns.
	
	\section{Preliminaries}\label{secPrelim}
	
	\subsection{Notation}\label{subNotation}
	
	For a positive integer $n$, we denote $[n] := \{1, 2, \dots, n\}$. Unless otherwise specified, all graphs we consider are undirected with no self-loops, but multiple edges may be allowed between any pair of vertices.
	The $m \times n$ grid graph is the graph with vertex set $[m] \times [n]$, with an edge between $(i, j)$ and $(i', j')$ whenever $\abs{i' - i} + \abs{j' - j} = 1$. We always draw grid graphs in a Cartesian coordinate system, with $m$ being the horizontal dimension and $n$ being the vertical dimension. We denote by $\zz^2$ the infinite grid graph, where the vertex set is $\zz \times \zz$ and the edge relation is the same as in finite grids.
	
	A forest is a graph with no cycles, and a tree is a connected forest. A \emph{$k$-forest} is a forest with $k$ connected components. A forest is \emph{balanced} if every connected component has exactly the same number of vertices. If $T$ is a tree and $S \subseteq E(T)$, we define $T \setminus S$ to be the forest $F$ with vertex set $V(F) := V(T)$ and edge set $E(F) := E(T) \setminus S$. Thus, a tree $T$ is \emph{$k$-splittable} if there is some set $S \subseteq E(T)$ of size $k - 1$ such that $T \setminus S$ is a balanced $k$-forest.
	
	For a graph $H$, we let $\st(H)$ denote the number of spanning trees of $H$. For a $k$-partition $P$ of $G$ with \neww{parts} $P_1$, $\ldots$, $P_k$, we denote by $\pi_\st$ the spanning tree distribution, given by 
	\[
	\pi_\st(P) = \frac{\prod_{i = 1}^k \st(P_i)}{Z},
	\]
	where $Z$ is the normalizing constant, also called the partition function, given by 
	\[
	Z = \sum_{k\text{-partitions } P} \ \ \prod_{i = 1}^k \st(P_i).
	\]
	Note that the uniform distribution over $k$-forests of $G$ is equivalent to the spanning tree distribution over $k$-partitions of $G$ when a forest is identified with its connected components.
	
	We write $d(x, y)$ to denote the Euclidean distance between two points $x, y \in \rr^2$. The \emph{Hausdorff distance} between two subsets $X, Y \subseteq \rr^2$, written $d(X, Y)$, is defined as
	\begin{equation} \label{eqn:haus} d(X, Y) := \max\{\sup_{x \in X} \inf_{y \in Y} d(x, y), \sup_{y \in Y} \inf_{x \in X} d(x, y)\}.
	\end{equation}
	
	\subsection{Duality}\label{subDuality}
	
	Let $G$ be a connected, planar graph, and fix an embedding of $G$ in the plane with no edges crossing. The \emph{dual graph} of $G$ (with respect to the embedding) is the graph $G^*$ whose vertices are faces of $G$, with an edge between two faces $a^* \in V(G^*)$ and $b^* \in V(G^*)$ whenever the two faces share a common boundary edge. Note that we count the outer face of $G$ as a vertex of $G^*$ as well.
	
	For any edge $e \in E(G)$, let $e^* \in E(G^*)$ be the edge between the faces it bounds. For any set of edges $S \subseteq E(G)$, we analogously define $S^* := \{e^* \suchthat e \in S\} \subseteq E(G^*)$. The following lemma is a standard result.
	
	\begin{lemma}\label{lemDuality}
		Assume that $G$ is connected and embedded in the plane such that no edge of $G$ has the same face on both sides.
		Then $e \mapsto e^*$ is a bijection between edges of $G$ and edges of $G^*$, and $T \mapsto T^* := (V(G^*), E(G^*) \setminus E(T)^*)$ is a bijection between spanning trees of $G$ and spanning trees of $G^*$.
	\end{lemma}
	\noindent Note that $T^*$ does not contain the edges $e^*$ for each $e \in T$, but rather those edges that are {\it not} in this set. 
	The hypotheses of Lemma \ref{lemDuality} hold for all $m \times n$ grid graphs with $m, n > 1$.

	\subsection{Wilson's Algorithm}\label{subWilson}
	
	Wilson's algorithm~\cite{wilson} is important for us not just because it samples uniformly random trees efficiently, thus serving as a key subroutine in our perfect sampling algorithm (See Section~\ref{secAlgsWilson}), but also because our proofs rely on running Wilson's algorithm in a specific way.
	
	For an input graph $G$, the steps of Wilson's algorithm are as follows: 
	\begin{enumerate}
		\item Set $T \leftarrow \{r\}$ for an arbitrary ``root'' vertex $r \in V(G)$
		\item While $T$ does not connect all vertices of $G$: 
		\begin{enumerate}
			\item Do a loop-erased random walk\footnote{That is, every time the random walk revisits a node $u$, erase the cycle and resume the random walk from $u$.} starting at an arbitrary vertex $v \notin T$ until it reaches a vertex of $T$
			\item Add all vertices and edges along this loop-erased random walk to $T$
		\end{enumerate}
		\item Return $T$
	\end{enumerate}
	
	Importantly, it does not matter which vertex is initially chosen as the root, and\neww{,} in each iteration of the while loop, it does not matter at which vertex not in $T$ the next loop-erased random walk begins. Regardless of what arbitrary choices are made at these steps, one can prove \neww{that} the end result is a uniformly random spanning tree of $G$. We use this crucial fact in our proofs, analyzing the process of Wilson's algorithm (in the dual graph $G^*$) from carefully-chosen starting vertices. 
	
	Recall that the hitting time $\tau_u(v)$ of \new{$v$} from \new{$u$} is the expected time before a simple random walk reaches $v$ from $u$, and the commute time between $u$ and $v$ is $\tau_u(v)+\tau_v(u)$. A \emph{$\pi$-random} vertex of $G$ is a vertex chosen according to the stationary distribution of the simple random walk on $G$, $\pi(v) = deg(v)/\neww{(2\abs{E(G)})}$. Wilson characterizes the expected \neww{runtime} of his algorithm (measured by the number of times we need to find a random neighbor of a vertex \neww{while taking a loop-erased random walk in step 2 (a)}) in terms of the commute time as follows:
	\begin{proposition}[Wilson]\label{p.WilsonRunning}
		The expected \neww{total} number of \neww{random walk steps taken} in the course of running Wilson's algorithm on a graph $G$ with root $r$ is precisely the expected commute time between $r$ and a $\pi$-random vertex $v$.\qed
	\end{proposition}
	
	For general graphs with $N$ vertices and $M$ edges, it is well-known that the hitting time and thus the commute time between any pair of vertices is at most $O(NM)$~\cite{markovmixing}; this implies that Wilson's algorithm runs in time $O(N^2)$ for any planar graph on $N$ vertices.  However, this can be improved for grid graphs by considering the dual graph and a carefully-chosen root: 
	\begin{proposition}\label{p.WilsonGrid}  Wilson's algorithm runs in expected time $O(N\log N)$ on the dual of any grid graph on $N$ vertices, when the root is chosen to be the dual vertex corresponding to the outer face of the grid graph.
	\end{proposition}
	\noindent This is easily proved using the characterization of the commute time in terms of effective resistance; we include a proof in the appendix.
	
	\subsection{Splittability and the Spanning Tree Distribution}\label{subTreeToForest}
	
	Here we explicitly connect the uniform distribution over spanning trees of a graph $G$ with the uniform distribution over $k$-forests of $G$. This enables us to analyze the likelihood of obtaining a balanced partition when sampling from the spanning tree distribution over forests, as the up-down walk of~\cite{charikar2022complexity} (approximately) does; see Section~\ref{secAlgsUpDown}.
	
	\begin{lemma}\label{lem:balancedprob}
		If the probability a uniformly random spanning tree of $G$ with $N$ vertices and $M$ edges is $k$-splittable is at least $\alpha$, then the probability a uniformly random $k$-forest of $P$ is balanced is at least 
		\[
		\frac{\alpha}{N^{k - 1}(M - N + \new{k})^{k-1}}.
		\] 
	\end{lemma}
	
	\begin{proof}[\new{Proof}]
		Note every $k$-forest of $G$ can be obtained by removing $k-1$ edges from some spanning tree of $G$.  For each spanning tree, the number of ways in which to do this is exactly $\binom{N-1}{k-1}$.  This means there are at most $\binom{N-1}{k-1} \cdot  \st(G)$ $k$-forests of $G$. 
		
		If the probability a uniformly random spanning tree of $G$ is $k$-splittable is at least $\alpha$, this means there are at least $\alpha \cdot \st(G)$ splittable spanning trees of $G$. 
		For a splittable spanning tree, exactly one of the ways of removing $k-1$ edge produces a balanced $k$-forest.  On the other hand, each forest can be obtained from a spanning tree in at most 
		\[
		\binom{M-(N-\new{k})}{k-1}\leq (M - N + \new{k})^{k-1}
		\]
		ways, since this is a bound for the number of choices for the additional $k-1$ edges which belong to a spanning tree which contains the forest.
		This means there are at least $\alpha \cdot \st(G)/(M - N + \new{k})^{k-1}$ balanced $k$-forests of $G$. 
		
		The probability a uniformly random $k$-forest is balanced is the number of balanced $k$-forests divided by the total number of $k$-forests.  This is at least 
		\[
		\frac{\alpha \cdot \st(G)/(M - N + \new{k})^{k-1}}{\binom{N-1}{k-1} \st(G)} \geq \frac{\alpha}{N^{k-1}(M - N + \new{k})^{k-1}}. 
		\]
	\end{proof}
	
	\begin{proof}[Proof that Theorem \ref{thm:k-split} implies Theorem \ref{thm:k-dist}] 
		Sampling a $k$-partition $P$ of $G$ according to the spanning tree distribution is the same as sampling a $k$-forest $F$ of $G$ uniformly at random and then considering its connected components. By Theorem~\ref{thm:k-split}, the probability that a uniformly random spanning tree $T$ of $G$ is $k$-splittable is a least $\frac{1}{\beta^{k^2}n^{3k-3}m^{k-1}}$ for some fixed constant $\beta$. By Lemma~\ref{lem:balancedprob}, as $G$ has $nm$ vertices and strictly less than $2nm$ edges, the probability a uniformly random $k$-forest is balanced is therefore at \new{least} $\frac{1}{\beta^{k^2}n^{5k-5}m^{3k-3} }$.
	\end{proof}
	
	\noindent There is hope this bound could be improved by studying random forests directly, rather than studying spanning trees and then considering cutting them to obtain forests. 
	
	\subsection{Algorithms for Sampling Balanced Tree-Weighted Partitions}
	
	\label{secAlgs}
	
	Our theorems imply that known approaches for sampling (not necessarily balanced) $k$-partitions according to the spanning tree distribution in polynomial time can be combined with a rejection sampling step to obtain an expected polynomial-time algorithm for sampling balanced $k$-partitions. Here we present two methods by which this could be done, for the case of sampling exactly balanced $k$-partitions.  
	
	\subsubsection{Perfectly Sampling Balanced $k$-Partitions with Wilson's Algorithm}
	
	\label{secAlgsWilson}
	
	Wilson's algorithm generates a uniform random spanning tree of a graph $G$.  We can use it to randomly sample a balanced $k$-partition as follows.

	\begin{enumerate}
		\item \neww{Sample a uniformly} random spanning tree $T$ of $G$ using Wilson's algorithm.
		\item Check if $T$ has $k-1$ edges whose removal disconnects $T$ into $k$ components of equal size.  If no\neww{t}, reject and return to step 1.  
		\item Create a $k$-partition $P$ of $G$ comprised of the connected components when these $k-1$ edges are removed from $T$. 
		\item Create a graph $G / P$ which contracts each \neww{part} of $P$ into a single point and retains all edges between components with the appropriate multiplicity.  
		\item Compute the number $s$ of spanning trees of $G / P$.
		\item Return $P$ with probability $1/s$.  With the remaining probability $(s-1)/s$ reject and return to step 1. 
	\end{enumerate}
	
	\begin{theorem}\label{thmAlgWilson}
		For $N$-vertex grid graphs, this algorithm produces a balanced $k$-partition drawn \neww{uniformly} from the spanning tree distribution in expected \neww{runtime} $O(N^{3k-2}\log N)$. 
	\end{theorem}
	\noindent See the appendix for a proof of this theorem. \neww{Briefly, it} takes expected time $O(N \log N)$ steps to sample a random spanning tree and check if it is $k$-splittable, $O(N^{2k-2})$ attempts in expectation to see a $k$-splittable tree \neww{(by Theorem \ref{thm:k-split})}, and $O(N^{k-1})$ attempts to be successful in the final rejection of Step \neww{6}.

	\subsubsection{Approximately Sampling Balanced $k$-Partitions with the Up-Down Walk}
	\label{secAlgsUpDown}
	
	An alternat\neww{iv}e method using \neww{matroid} Markov chain\neww{s} described in Charikar et al. can produce an approximately uniformly random $k$-forest~\cite{anari2021forestsampling, charikar2022complexity}. \new{While current upper bounds on the runtime of this method are worse than the exact sampling method presented above, we do not believe these bounds are tight and suspect this method may be more effective in practice. Thus,} we briefly motivate and describe this \new {alternative} approach here.   
	
	On any graph $G$, the spanning forests with at least $k$ components form a matroid whose bases are exactly the $k$-component spanning forests of $G$. \neww{There are two well-known markov chains on matroid bases: down-up and up-down. The down-up chain} mixes in time $O(r (\log r + \log \log n))$ when bases have $r$ elements and the matroid has $n$ total elements~\cite{cryan2021matroid}\neww{. W}hen run for longer than its mixing time, this chain produces an approximately uniformly random basis. For $k$-component forests, this down-up chain randomly removes an edge of the forest (to produce $k+1$ components), and then randomly adds back in an edge connecting two different components. Its mixing time is $O((N-k) ( \log (N-k) + \log \log M))$ for graphs with $N$ vertices and $M$ edges\neww{. F}or constant $k$, this becomes $O(N \log N)$. However, naively implementing one down-up step requires $O(M)$ time, making the overall time for this chain to produce an approximate sample $O(NM \log N)$.
	
	This was improved in~\cite{anari2021forestsampling} by considering the up-down walk instead. This walk randomly adds an edge to the forest. If adding this edge creates a cycle, a random edge of the cycle is removed. If adding this edge did not create a cycle (e.g. the edge connected two components of the forest) then a random edge of the forest is removed. 
	This chain has mixing time $O(M \log M)$ for graphs with $M$ edges, as the up-down walk can also be viewed as the down-up walk on the dual matroid whose bases are the complements of $k$-component forests. Using a link-cut tree data structure, each up-down step can be implemented in amortized quasi-constant time, resulting in an overall runtime of $O(M\log^2 M)$ to produce one random sample. For planar graphs where $M = O(N)$, this mixing time is at most $O(N \log^2 N)$\neww{. F}or the sake of specificity, suppose the mixing time \new{is} at most $C_0 N\log^2 N$ for a constant $C_0$. It is this up-down chain, rather than the usual down-up chain, that we will use. 
	
	The following is our algorithm for approximately\footnote{\neww{It will not be a perfectly uniform sample. There is a small convergence error term depending on the number of up-down steps taken. See the proof in the Appendix.}} randomly sampling a balanced $k$-forest of a $N$-vertex graph. 
	\begin{enumerate}
		\item Run the up-down Markov chain on $k$-forests for $\kappa C_0 N\log^2 N$ steps; here increasing $\kappa$ will increase the accuracy of the approximate sampling.
		\item If the current state of the chain is a balanced $k$-forest, return the partition $P$ consisting of the connected components of the forest. Else, return to step 1 and repeat.
	\end{enumerate}

	\begin{theorem}\label{thmAlgUpDown}
		For $N$-vertex grid graphs, this algorithm produces a balanced $k$-partition drawn approximately from the spanning tree distribution in expected \neww{runtime} $O(N^{4k-3} \log^4 N)$.
	\end{theorem}
	\noindent See the appendix for a proof of this theorem\neww{. Briefly, it} takes $\widetilde O(N)$ steps to approximately sample a random $k$-forest and $O(N^{4k-4})$ attempts in expectation to see a balanced one, by Theorem \ref{thm:k-dist}. Note that, with the relatively crude estimates we employ to deduce Theorem \ref{thm:k-dist} from Theorem \ref{thm:k-split} (in Section \ref{subTreeToForest}), the runtime we prove for this approximate sampling approach is actually worse than for the exact sampler above.  There is little reason to believe this to be the truth, however.
	
	\section{Exact Balance on Grid Graphs}\label{secGrids}
	
	\subsection{Exactly Balanced Bipartitions}\label{secBipartitions}
	
	In this section we prove the following:
	\begin{theorem}\label{thm:2-split}
		Let $G$ be a grid graph with $N$ vertices, where $N$ is even. The probability that a uniformly random spanning tree $T$ of $G$ is $2$-splittable is at least $1/N^2$. 
	\end{theorem}
	
	In fact, we prove a stronger result, namely that specific edges near the center of the grid have a decent probability of being the edge that splits the tree. Formally, If $G$ is an $m \times n$ grid graph, we define a \emph{horizontal central edge} of $G$ to be an edge of the form $\{(i, j), (i\new{ + 1}, j)\}$ that is as close to the center of $G$ as possible. Note that there may be 1, 2, or 4 horizontal central edges depending on the parities of $m$ and $n$.
	
	\begin{lemma}\label{lemCentralEdgeBound}
		Let $G$ be an $m \times n$ grid graph where $m \geq n$, and $mn$ is even, and let $e \in E(G)$ be any horizontal central edge of $G$. Then the probability that a uniformly random spanning tree $T$ of $G$ contains $e$, and $T \setminus \{e\}$ is a balanced 2-forest, is at least
		$$\twocases{\txt{if $m$ is even}}{\frac{1}{m n^3}}{\txt{if $m$ is odd}}{\frac{1}{4 m n^3}}.$$
	\end{lemma}
	
	To prove Lemma \ref{lemCentralEdgeBound}, we will require two further lemmas.
	
	\begin{lemma}\label{lemRandomWalkToBoundary}
		Let $G$ be the $m \times n$ grid graph induced by the subset $[m]\times [n]$ of the grid $\zz^2$, and $(i_0, j_0) \in V(G)$. The probability that a random walk from $(i_0, j_0)$ in $\zz^2$ exits $G$ for the first time to a vertex $(i', j')$ with $j' > 0$ is at least $\frac{j_0}{n + 1}$.
	\end{lemma}
	
	\begin{proof}
		Let $(X_0, Y_0) = (i_0, j_0), (X_1, Y_1), (X_2, Y_2), \dots$ be an infinite random walk in $\zz^2$. Let $\widetilde{Y}_0 = j_0, \widetilde{Y}_1, \widetilde{Y}_2, \dots$ be the subsequence of vertical coordinates $Y_0, Y_1, Y_2, \dots$ where we delete terms that repeat the previous value (because the horizontal coordinate changed instead). A sufficient condition for the original two-dimensional walk to first exit $G$ to a vertex $(i', j')$ with $j' > 0$ is for $\widetilde{Y}_0, \widetilde{Y}_1, \widetilde{Y}_2, \dots$ to reach $n + 1$ before it reaches zero, as reaching $n + 1$ necessitates exiting $G$. Since $\widetilde{Y}_0, \widetilde{Y}_1, \widetilde{Y}_2, \dots$ is just an unbiased one-dimensional random walk from $j_0$, it is a well-known fact\footnote{This is an instance of the classic ``Gambler's Ruin'' problem. This proof can be found in \cite[Section 7.2.1]{Mitz}, where $\ell_1 := j_0$ and $\ell_2 := (n + 1) - j_0$.} that this probability is precisely $\frac{j_0}{n + 1}$.
	\end{proof}
	
	\begin{lemma}\label{lemSoS}
		Let $X$ be a discrete probability distribution supported on a set of size $k$. Then
		$$\Pr_{x_1, x_2 \sim X \times X}(x_1 = x_2) \geq \frac{1}{k}.$$
	\end{lemma}
	\begin{proof}
		Let the probabilities of each element in the support of $X$ be $p_1, p_2, \dots, p_k$, where these values sum to one. Then the probability that two independent samples are the same is given by
		$\sum_{i = 1}^k p_i^2$. Let $\mathbf{p} = (p_1, p_2, \ldots, p_k)$ and let $\mathbf{v} = \left(\frac 1 k, \frac 1 k, \ldots, \frac 1 k\right)$ be length $k$ vectors.  We see \neww{$\|\mathbf{p}\|^2 = \sum_{i = 1}^k p_i^2$, $\|\mathbf{v}\|^2 = 1/k$}, and $|\langle \mathbf{p}, \mathbf{v} \rangle |^2 = 1/k^2$, so the lemma follows immediately from the \neww{Cauchy--Schwarz} inequality.
	\end{proof}
	
	\begin{proof}[Proof of Lemma \ref{lemCentralEdgeBound}]
		Assume $n > 1$ (otherwise there is nothing to show; the probability is one). Then note that Lemma \ref{lemDuality} applies to $G$. We first consider the case where $n$ is odd (so $m$ must be even). In this case there is a unique horizontal central edge $e$, connecting the vertices $(m/2, (n + 1)/2)$ and $(m/2 + 1, (n + 1)/2)$. Let $G^*$ be the dual graph of $G$ in the plane, and denote the outer face by $r^* \in V(G^*)$. Let $a^* \in V(G^*)$ be the face above $e$ and let $b^* \in V(G^*)$ be the face below $e$, as in Figure \ref{figCentralEdge}.
		
		Consider the following algorithm for generating a uniformly random spanning tree $T$ of $G$. Run Wilson's algorithm on $G^*$ with $r^*$ as the root, starting the first loop-erased random walk from $a^*$ and the second random walk from $b^*$ (if it is not already added to the tree in the first random walk). The remaining random walks in Wilson's algorithm can be executed from arbitrary starting points. This gives us spanning tree $T^*$ of $G^*$. We then output the primal tree $T$ whose dual is $T^*$. Since Wilson's algorithm gives a uniformly random sample from the set of spanning trees of $G^*$, and those dual trees are in bijection with the primal spanning trees of $G$ (Lemma \ref{lemDuality}), this algorithm gives us a uniformly random sample from the set of spanning trees of $G$.
		
		We apply Lemma \ref{lemRandomWalkToBoundary} to the $(m - 1) \times \frac{n - 1}{2}$ dual sub-grid outlined in the top (red) rectangle in Figure \ref{figCentralEdge}, with initial vertex $a^*$. Note that $j_0 = 1$ because the coordinate system is shifted so that $a^*$ is in the bottom row. Lemma \ref{lemRandomWalkToBoundary} says that a random walk from $a^*$ will first exit the sub-grid above, to the left, or to the right (just not below) with probability at least
		$$\frac{1}{\frac{n - 1}{2} + 1} > \frac{1}{(n - 1) + 1} = \frac1n.$$
		This clearly applies to our loop-erased random walk as well: The probability that the first walk in Wilson's algorithm, which starts from $a^*$, makes it to the outer face $r^*$ without ever entering the bottom half of the grid is at least $\frac1n$. Assuming this happens, we may then apply the same argument to the bottom (blue) rectangle, for the next random walk starting from $b^*$. By independence, with probability at least $\frac{1}{n^2}$, both paths will have made it to $r^*$ without crossing the horizontal midline.
		
		\begin{figure}[t]\centering
			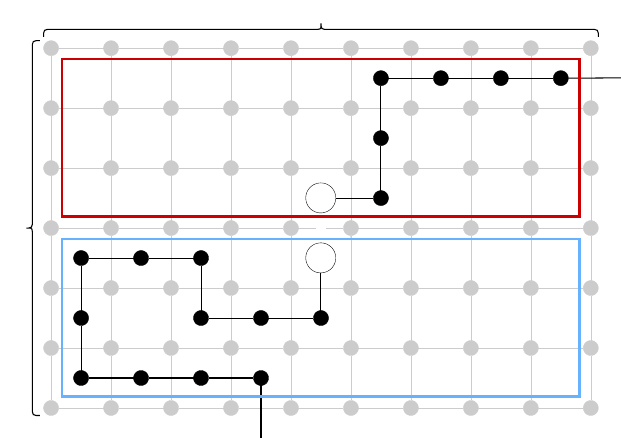
			\caption{A possible run of the dual graph spanning tree sampling algorithm in the proof of Lemma \ref{lemCentralEdgeBound} when $m$ is odd. In this example, $m = 10$ and $n = 7$. The primal graph $G$ is depicted in gray, and the first two random walks in the dual graph $G^*$ are depicted in black.}
			\label{figCentralEdge}
		\end{figure}
		
		Assume that this happens, as it does in Figure \ref{figCentralEdge}. Let $P^*_a$ be the path from $a^*$ to the boundary, and let $P^*_b$ be the path from $b^*$ to the boundary. Since both $P^*_a$ and $P^*_b$ will be included in $T^*$, we know that $T$ cannot cross these paths. This means $e$ must be included in $T$. Moreover, $P^*_a$ and $P^*_b$ completely determine the number of vertices on each side of $e$ in $T$, as follows. Suppose there are $X$ vertices in the top-half of the grid to the left of $P^*_a$, $Y$ vertices in the top-half of the grid to the right of $P^*_a$, $Z$ vertices in the bottom-half of the grid to the left of $P^*_b$, and $W$ vertices in the bottom-half of the grid to the right of $P^*_b$. Then the subtree of $T$ to the left of $e$ will have $X + Z + \frac{m}{2}$ vertices, and the subtree to the right of $e$ will have $Y + W + \frac{m}{2}$ vertices (the $\frac{m}{2}$ terms come from the vertices on the horizontal midline). Observe that the distribution, over the random path $P^*_a$, of the possible values of $X - Y$ is independent of and identical to the distribution, over the random path $P^*_b$, of the possible values of $W - Z$. Both distributions can take any integral value from $-\frac{m - 1}{2} n$ to $\frac{m - 1}{2} n$. Thus, applying Lemma \ref{lemSoS}, we know that, with probability at least
		$$\frac{1}{(m\frac{n - 1}{2}) - (-m\frac{n - 1}{2}) + 1} = \frac{1}{mn - m + 1} > \frac{1}{mn},$$
		we have $X - Y = W - Z$, which implies
		$$X + Z + \frac{m}{2} = Y + W + \frac{m}{2},$$
		i.e., the subtrees are balanced.
		
		\neww{W}e have shown that the probability $e$ is included in a uniformly random spanning tree $T$ of $G$ and splits it into a balanced 2-forest is at least
		\begin{equation*}
			\frac{1}{n^2} \cdot \frac{1}{mn} = \frac{1}{mn^3}.
		\end{equation*}
		
		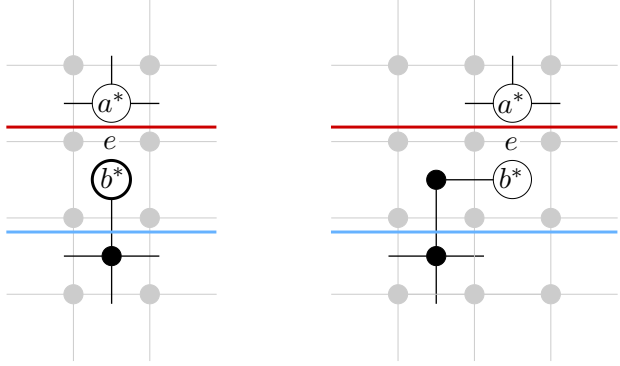
\begin{figure}\centering
\begingroup%
  \makeatletter%
  \providecommand\color[2][]{%
    \errmessage{(Inkscape) Color is used for the text in Inkscape, but the package 'color.sty' is not loaded}%
    \renewcommand\color[2][]{}%
  }%
  \providecommand\transparent[1]{%
    \errmessage{(Inkscape) Transparency is used (non-zero) for the text in Inkscape, but the package 'transparent.sty' is not loaded}%
    \renewcommand\transparent[1]{}%
  }%
  \providecommand\rotatebox[2]{#2}%
  \newcommand*\fsize{\dimexpr\f@size pt\relax}%
  \newcommand*\lineheight[1]{\fontsize{\fsize}{#1\fsize}\selectfont}%
  \ifx\svgwidth\undefined%
    \setlength{\unitlength}{230.36103058bp}%
    \ifx\svgscale\undefined%
      \relax%
    \else%
      \setlength{\unitlength}{\unitlength * \real{\svgscale}}%
    \fi%
  \else%
    \setlength{\unitlength}{\svgwidth}%
  \fi%
  \global\let\svgwidth\undefined%
  \global\let\svgscale\undefined%
  \makeatother%
  \begin{picture}(1,0.59351812)%
    \lineheight{1}%
    \setlength\tabcolsep{0pt}%
    \put(0,0){\includegraphics[width=\unitlength,page=1]{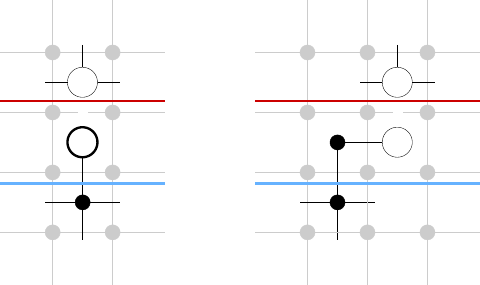}}%
    \put(0.14928545,0.40738019){\color[rgb]{0,0,0}\makebox(0,0)[lt]{\lineheight{1.25}\smash{\begin{tabular}[t]{l}$a^*$\end{tabular}}}}%
    \put(0.15355712,0.28344353){\color[rgb]{0,0,0}\makebox(0,0)[lt]{\lineheight{1.25}\smash{\begin{tabular}[t]{l}$b^*$\end{tabular}}}}%
    \put(0.15852649,0.34649295){\color[rgb]{0,0,0}\makebox(0,0)[lt]{\lineheight{1.25}\smash{\begin{tabular}[t]{l}$e$\end{tabular}}}}%
    \put(0.80341414,0.40610799){\color[rgb]{0,0,0}\makebox(0,0)[lt]{\lineheight{1.25}\smash{\begin{tabular}[t]{l}$a^*$\end{tabular}}}}%
    \put(0.80549274,0.28272682){\color[rgb]{0,0,0}\makebox(0,0)[lt]{\lineheight{1.25}\smash{\begin{tabular}[t]{l}$b^*$\end{tabular}}}}%
    \put(0.81469197,0.34473694){\color[rgb]{0,0,0}\makebox(0,0)[lt]{\lineheight{1.25}\smash{\begin{tabular}[t]{l}$e$\end{tabular}}}}%
  \end{picture}%
\endgroup%

			\caption{The cases in the proof of Lemma \ref{lemCentralEdgeBound} when $m$ is even, in which we must assume that the initial steps of the random walk from $b^*$ takes a specific path into the blue rectangle, from which it never leaves until hitting the outer face.}
			\label{figMEvenCase}
		\end{figure}
		
		The remaining cases, where $n$ is even, are almost the same. There are just a few minor additional assumptions we must impose about what happens to the random walks at the very beginning, as illustrated in Figure \ref{figMEvenCase}.
		
		If $m$ is even as well, there are two horizontal central edges bordering the unique central face of the grid. Without loss of generality, take $e$ to be the top one, then define $a^*$ and $b^*$ as before. We suppose that the random walk from $b^*$ first steps directly downward, as in Figure \ref{figMEvenCase} (left). This happens with probability $\frac14$. From there, by the same arguments as before, noting that the two subgrids are now each $m \times (n/2 - 1)$, the probabilities that the paths leave their respective red and blue rectangles at the boundary of the grid are both at least 
		\[
		\frac{1}{\left(\frac{n}{2}-1\right)+1} = \frac2n. 
		\]
		The probabilities that the number of vertices on each side are the same is at least
		$$\frac{1}{(m\frac{n - 2}{2}) - (-m\frac{n - 2}{2}) + 1} = \frac{1}{mn - 2m + 1} > \frac{1}{mn}.$$
		Thus, the probability that $e$ splits a uniformly random spanning tree $T$ into a balanced 2-forest is at least
		\begin{equation*}
			\frac14 \cdot \left(\frac{2}{n}\right)^2 \cdot \frac{1}{mn} = \frac{1}{mn^3}. 
		\end{equation*}
		
		Finally, consider the case where $m$ is odd and $n$ is even. Now there are four horizontal central edges, of which we pick the top-right one without loss of generality. With probability $\frac{1}{16}$, the random walk from $b^*$ first steps to the left and then down into the blue rectangle, as in Figure \ref{figMEvenCase} (middle). Now we can again apply the same arguments as above to the subgrids of dimensions $m \times (n/2 -1)$ showing the probability the remaining paths leave their subgrids at the boundary of $G$ are both at least $\frac2m$. While the random walk in the top grid no longer begins exactly in the center of the top grid (it cannot, because this grid is now of even width), the top and bottom grids are rotationally symmetric, with the top walk beginning just one unit left of center and the bottom walk beginning one unit right of center. As before, the distributions of difference of the number of vertices on each side of the path are identical and are supported on sets of size at most $mn$, so by Lemma~\ref{lemSoS}, the probability these differences are identical is at least $\frac{1}{mn}$. Thus, the probability that $e$ splits a uniformly random spanning tree $T$ into a balanced 2-forest is at least
		\begin{equation*}
			\frac{1}{16} \cdot \left(\frac{2}{n}\right)^2 \cdot \frac{1}{mn} = \frac{1}{4mn^3}. 
		\end{equation*}
	\end{proof}
	
	\noindent We now use this lemma to prove Theorem~\ref{thm:2-split}. 
	
	\begin{proof}[Proof of Th\neww{eo}rem \ref{thm:2-split}]
		Recall that $N = nm$ is the total number of vertices. Assuming $m \geq n$, we know that $mn^3 \leq m^2n^2 = N^2$. Thus, in the case where $m$ is even, we simply choose one of the horizontal central edges, which, by Lemma \ref{lemCentralEdgeBound}, splits a random tree into a balanced 2-forest with probability at least $\frac{1}{mn^3} \geq \frac{1}{N^2}$. In the case where $m$ is odd (and so $n$ must be even) there are 4 horizontal central edges, each of which will split a random tree into a balanced 2-forest with probability at least $\frac{1}{4mn^3}$. Since these 4 events are mutually exclusive, one of these four will give a balanced split with probability at least $\frac{1}{mn^3} \geq \frac{1}{N^2}$.
	\end{proof}
	
	\begin{figure}\centering
		\includegraphics[scale=.13]{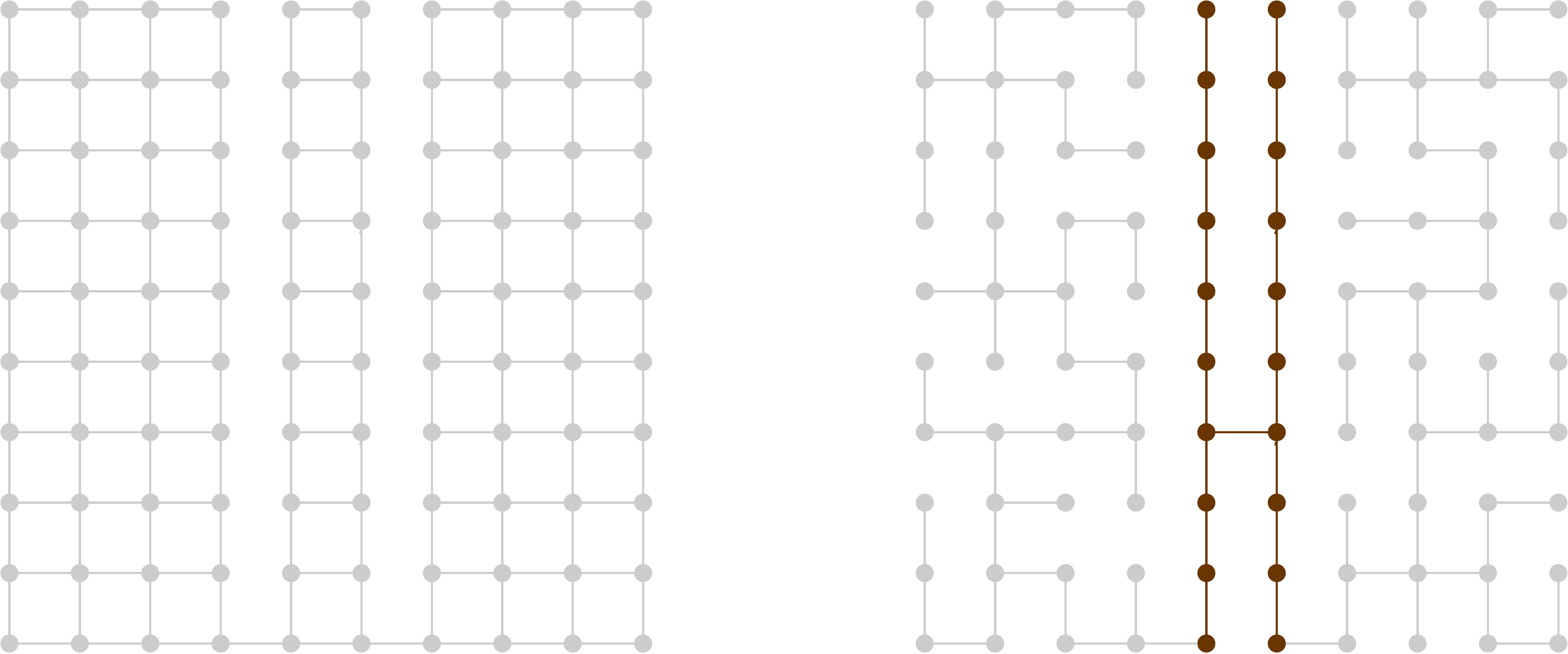}
		\caption{\label{figCounterexample}An $n \times n$ grid subgraph and one of its few 2-splittable spanning trees. There are only $n$ possible configurations for the part highlighted in brown, each with a single horizontal edge and all vertical edges.}
	\end{figure}
	
	We remark that this result does not extend from grid graphs to arbitrary grid subgraphs. In fact, even after removing only $O(\sqrt{N})$ edges from a square grid graph, it may be exponentially unlikely that a uniformly random tree is 2-splittable. To see this, consider the family of graphs $G$ depicted on the left in Figure~\ref{figCounterexample}, formed by removing $2(n - 1)$ edges from the $n \times n$ grid graph, for any even $n \geq 4$. There are only $n$ possible choices for what a 2-splittable spanning tree could look like in the middle two columns, as shown on the right. On the other hand, the total number of trees is clearly exponential\neww{.}\footnote{Asymptotically, this grows as $\Theta((2 + \sqrt{3})^n)$. See: \url{http://oeis.org/A001353}} Similar obstructions arise for $k > 2$ as well.
	
	\subsection{Exactly Balanced $k$-Partitions}\label{secKParititions}
	
	\label{sec:k-split}
	
	In this section we prove the following:
	\begin{theorem}\label{thm:k-split}
		For $m \geq n$, let $G$ be an $m \times n$ grid graph, and let $k$ be a positive integer dividing $m$. There exists a set $S \subseteq E(G)$ of size $k - 1$ such that the probability a uniformly random spanning tree $T$ of $G$ contains each edge in $S$, and $T \setminus S$ is a balanced $k$-forest, at least
		\begin{equation}
			\frac{1}{\beta^{k^2}n^{3k-3}m^{k-1}}
		\end{equation}
		for a fixed constant $\beta$.  
	\end{theorem}
	The proof proceeds along similar lines as the proof of Lemma~\ref{lemCentralEdgeBound}. We require the following stronger lemmas about random walks on grids. This is similar to Lemma \ref{lemRandomWalkToBoundary}, except that now we are also not allowed to hit the left or right sides, which makes the proof significantly more involved. We begin by proving a result about square grids, then extend this to tall thin rectangles.\footnote{\new{While these facts can be derived from standard results on harmonic functions on subsets of $\zz^2$~\cite[Chapter 8]{MondernIntro}, we give a self-contained and elementary proof here. We also note that the argument applied in Lemma~\ref{lemRandomWalkToFarSide} to extend from the square to rectangular case will be used later in a more general context, in Lemma~\ref{lemCurveWalkSimple}.}}  
	\begin{lemma}\label{lemSquareFarSide}
		For any \neww{positive} integer $\ell$, let $G$ be the $(2\ell + 1) \times (2\ell+1)$ grid graph whose coordinates are given by $\{-\ell, \ldots, \ell\} \times \{0, \ldots 2\ell\}$.  For a random walk in $\zz^2$ beginning at vertex $(0,0)$, the probability this random walk first exits $G$ to a vertex with coordinate $(i, 2\ell + 1)$ for some integer $-\ell \leq i \leq \ell$, is at least $1/B \ell$ for some constant $B$.  
	\end{lemma}
	\begin{proof}
		Let $\cL$ denote the event we are interested in, that a random walk in $\zz^2$ beginning at vertex $(0,0)$ first exits $G$ to a vertex with coordinate $(i, 2\ell + 1)$. Our proof giving a lower bound on $\Pr(\cL)$ will use the following three claims:
		\begin{claim} \label{c.reachesn} A simple random walk on $\zz$ from 0 reaches vertex $n$ within $2n^2$ steps with probability at least $\frac 1 4$.
		\end{claim}
		\begin{claim} \label{c.avoids0} For any $j$, conditioned on the event that a simple random walk on $\zz$ from 0 of fixed length $j\geq x$ ends at the vertex $x > 0$, the probability that the walk never revisits $0$ is $\frac{x}{j}$.
		\end{claim}
		\begin{claim}\label{c.epinterval} There is a constant $\lambda$ such that for any $D$, a simple random walk on $\zz$ from 0 of length $\leq D \ell^2$ has probability $\geq e^{-\lambda D}$ of never leaving the interval $[-\ell, \ell]$.
		\end{claim}
		\noindent We give elementary proofs of these claims below, but first let us use them to prove that for an absolute constant $B$, 
		\begin{equation}
			\Pr(\cL)\geq \frac{1}{B\ell}.\label{leavebox}
		\end{equation}
		In particular, we will consider a walk of fixed length $40\ell^2$, and show that with probability at least $\frac 1 {B\ell}$, such a walk from $(0,0)$ will exit $G$ from the top side and do so before exiting any other side.
		
		The random walk of length $40\ell^2$ is associated to an i.i.d.\txt{} sequence of length $40\ell^2$ over the four directions in the lattice, which we denote (in cyclic order) by N, E, S, W.  We generate the random walk in two steps:
		\begin{enumerate}
			\item We generate a uniform binary sequence of length $40\ell^2$ over the symbols $\{H,V\}$.
			\item We independently replace each symbol $H$ with a symbol $E$ or $W$ with probability $\tfrac 1 2$, and each symbol $V$ with a symbol $N$ or $S$ with probability $\tfrac 1 2$.
		\end{enumerate}
		Observe first that, with high probability, there are between $18\ell^2$ and $22 \ell^2$ symbols of each type $H$ and $V$ after the first step. \neww{This event has probability at least an absolute positive constant, by any standard binomial concentration bound; we will condition on it and absorb this constant into the final value of $B$.} In particular, we will simply condition on the event that the number of symbols $H$ and $V$ fall in this particular range. By associating steps $S$ and $W$ to $-1$ and $N$ and $E$ to $+1$, we can associate to the sequence of replacements for the $H$'s a random walk $W_H$ on $\zz$, and to the sequence of replacements for the $V$'s a 1D random walk $W_V$ on $\zz$.  We need to put a lower bound on the probability that $W_H$ never leaves the interval $[-\ell, \ell]$ AND $W_V$ reaches $2\ell+1$ without visiting $-1$.  Conditioned on the lengths of the walks $W_V,W_H$, these are independent events; thus it \neww{suffices} to prove lower bounds on the probabilities of these two events conditioned on any fixed lengths for $W_V$ and $W_H$ between $18\ell^2$ and $22\ell^2.$
		
		We first consider the case of $W_V$. We define the events $\cE',\cE_j$ by
		\begin{align*}
			\cE'&=\{W_V\text{ reaches $2\ell+1$ within \neww{$18 \ell^2$} steps}\}\\
			\cE_j&=\{j\text{ is the index of last visit of $W_V$ to $2\ell+1$ among first $18\ell^2$ steps}\}.\\ 
		\end{align*}
		Observe that
		\begin{equation}\label{conditionalsum}
			\bigcup_{j\leq 18\ell^2}\cE_j=\cE'.
		\end{equation}
		Since $\ell \geq 1$ implies $2(2\ell+1)^2 \leq 18\ell^2$, we have from Claim \ref{c.reachesn} that
		\begin{equation}\label{prob14}
			\Pr(\cE')\geq \tfrac 1 4.
		\end{equation}
		Now fixing some $j\leq 18\ell^2$ steps and conditioning on the event that $W_V$ is at $2\ell+1$ on the $j$th step, we have from Claim \ref{c.avoids0} that
		\begin{align}
			&\Pr\left(W_V\text{ never  visits $-1$ before step $j$}\mid W_V\text{ at }2\ell+1\text{ on step }j\right) \nonumber
			\\&\hspace*{5mm}\geq \Pr\left(W_V\text{ never revisits $0$ before step $j$}\mid W_V\text{ at }\neww{2\ell+1}\text{ on step }j\right) 
			\label{fromballot}\\&\hspace*{5mm}\geq \frac{2\ell+1}{18\ell^2}\geq \frac{1}{9\ell}.\nonumber 
		\end{align}
		
		Now \eqref{conditionalsum}, \eqref{prob14}, and \eqref{fromballot} imply that
		\begin{align*}
			\Pr&\left(W_V\text{ visits $2\ell+1$ before $-1$}\right) \\
			&= \Pr(\cE') \Pr(W_V \txt{ visits $2\ell+1$ before $-1$} \suchthat \cE')\\
			&= \Pr(\cE')\sum_{j\leq \neww{18}\ell^2}\Pr(\cE_j\mid \cE') \Pr\left(W_V\text{ visits $2\ell+1$ before $-1$}\mid \cE',\cE_j\right)\\
			&\geq \Pr(\cE')\sum_{j\leq \neww{18}\ell^2}\Pr(\cE_j\mid \cE') \Pr\left(W_V\text{ avoids $-1$ through step $j$}\mid \cE',\cE_j\right)\\
			&\geq \frac14 \cdot \frac{1}{9\ell} = \frac{1}{36\ell}.
		\end{align*}
		In the final inequality, we have used the fact that for $j\leq 18\ell^2$, when conditioning on both events $\cE'$ and $\cE_j$, the initial segment of $W_V$ of length $j$ is still a uniformly random walk among all walks of length $j$ from $0$ to $2\ell+1$, as replacing such an initial segment with any other does not change the outcome of the events $\cE',\cE_j$.
		
		Finally, since $W_H$ is of length $\leq 22\ell^2$, we have from Claim \ref{c.epinterval} (with $D = 22$) that with probability at least $\exp(-22\lambda)$, $W_H$ never leaves the interval $[-\ell, \ell]$. As we have conditioned ahead of time on the lengths of these two walks, the events corresponding to the two probabilities are independent, and so we have
		\begin{multline*}
			\Pr(\text{$W_V$ visits $2\ell+1$ before $-1$ AND $W_H$ stays in $[-\ell, \ell]$,}\\\neww{\text{ and the numbers of $H$ and $V$ steps lie in the above range}})\geq\frac{1}{B\ell},
		\end{multline*}
		for an absolute constant $B$.
		This completes the proof of \eqref{leavebox}; it only remains to prove the three claims.
		
		\noindent\textbf{Proof of Claim \ref{c.reachesn}}.   Recall that simple random walk begun from $s$ on $\{0,1,\dots,M\}$ reaches an endpoint of this interval in expected time $s(M-s)$ (see e.g., \cite[Proposition 2.1]{markovmixing}).  Thus a simple random walk on $\zz$ from 0 reaches either $-n$ or $n$ in expected time $n^2$.  This means that a simple random walk from 0 of length $\geq 2n^2$ reaches $-n$ or $n$ with probability at least $\frac 1 2$, as Markov's inequality implies that the probability that the hitting time exceeds $2n^2$ is  $\leq \frac{n^2}{2n^2}$. By symmetry the probability it reaches $n$ is thus at least $\frac 1 4$.  This proves Claim \ref{c.reachesn}.
		
		\smallskip
		
		\noindent\textbf{Proof of Claim \ref{c.avoids0}}.  This is a consequence of Bertrand's ballot theorem. 
		If the walk of length $j\geq x$ ends at the vertex $x>0$, it takes $R=\frac{j+x}{2}$ steps to the right and $L=\frac{j-x}{2}$ steps to the left.  Bertrand's ballot theorem then implies that the probability that among any initial nonempty segment, there \neww{are} never as many left steps as right steps, is precisely
		\[
		\frac{R-L}{R+L}=\frac{x}{j}.
		\]
		\smallskip
		
		\noindent\textbf{Proof of Claim \ref{c.epinterval}}.  Choose $K=\frac{1}{20}\ell^2$.  Consider $K$-step simple random walk on $\zz$ starting from $0$. Observe that by \neww{Hoeffding's inequality}, if $S_K$ is the location after $K$ steps,  
		\[
		\Pr(S_K\geq \ell/2)\leq e^{-(\ell/2)^2/2K} = e^{-5/2} <  \frac 1 {10},
		\]
		It follows that with probability $>\frac{9}{10}$, a simple random walk of length $K = \frac{1}{20}\ell^2$ from $0$ in $\zz$ will not end at a vertex $S_K\geq \ell$. Recall the standard reflection principle for simple random walk:
		\begin{equation}
			\Pr(\max_{j\leq n} S_j\geq T)=\Pr(S_n\geq T)+\Pr(S_n\geq T+1).
		\end{equation}
		It follows that with probability $\geq \frac{8}{10}$, the walk $S_0,\dots,S_K$ will never exceed $\ell/2$.  By symmetry, with probability $\geq \frac 6 {10}\geq \frac 1 2$, it will never leave the interval $\cI=[-\ell/2, \ell/2]$.  Thus, with probability at least $\frac 1 4$, the random walk will both not leave the interval $\cI$ \neww{AND} end in the nonnegative (respectively, nonpositive) side of the interval.  Repeated applications of this simple fact then give Claim 3 as follows\neww{.}
		
		We break the random walk $S_0,S_1,\dots,S_{D\ell^2}$ into $D\ell^2/K=20D$ walks, each of length $K$.  From the observation of the previous paragraph, each such walk has probability at least $\frac 1 4$ of never deviating more than $\ell/2$ from its starting point \emph{and} ending to the right or left of its starting point, according to whichever direction the origin 0 is from the starting point of the walk.  In this way, with probability $\geq \left(\frac{1}{4}\right)^{20D}$, every walk begins and ends within $\ell/2$ of the origin, and never ventures more than $\ell/2$ from its starting point. \neww{When this happens,} the whole concatenated walk remains within $\ell$ of the origin.
		This proves Claim \ref{c.epinterval} with $\lambda = 20 \ln 4$, and thus completes the proof of the Lemma.
	\end{proof}
	
	The following lemma extends these results beyond a square region to a narrow rectangle, provided its width is at least a constant fraction of its height.  
	
	\begin{lemma}\label{lemRandomWalkToFarSide}
		Suppose there is a constant $\ep>0$ such that  $m > \ep n$.  Let $G$ be the $(m+1) \times (n+1)$ grid graph induced by the subset $\{0,\dots,m\}\times \{0,\dots,n\}$ of the grid $\zz^2$, where $m+1$ is odd, and let $(i_0,j_0)=(\tfrac{m}{2},0)$.  The probability that a random walk from $(i_0,j_0)$ in $\zz^2$ exits $G$ for the first time to a vertex $(i',j')$ with $j'=n+1$ is at least $\frac{1}{A n e^{A/\ep}}$, for a fixed constant $A$.
	\end{lemma}
	
	\begin{proof}
		First, we consider the case where $m \geq n$.  In this case, set $\ell = \lfloor (n+1) /2 \rfloor$, so that $2 \ell + 1$ is equal to $n+1$ or $n+2$, whichever is odd. By Lemma~\ref{lemSquareFarSide}, there is a constant $B$ such that the probability the a random walk starting at $(m/2, 0)$ first exits the rectangle at some $(i', n+1)$ for some $i' \in [m/2 - \ell, m/2 + \ell]$ without ever leaving this interval of $x$-coordinates is at least $1/Bn$ for some constant $B$.  
		When $m \geq n$, this interval is a subset of $[0,m]$, and so the probability a random walk from $(m/2)$ first exits $G$ to some $(i', n+1)$ is at least $1/Bn$ as well; \neww{an} appropriate choice of $A$ such that $A e^A > B$ proves the lemma. 
		
		Now, suppose $m < n$.  
		For $\ell=\floor{\tfrac{m - 2}{4}}$, define $\cS_0$ to be the $(2\ell + 1) \times (2\ell + 1)$ square subset of $G$ induced by the vertices $(x,y)$ with $\floor{\frac{m}{2}} - \ell \leq x \leq \floor{\frac{m}{2}} + \ell$ and $0\leq y < (2\ell + 1)$.  Given a random walk in $\zz^2$ 
		beginning at $(\frac m 2,0)$, we define $\cL_0$ to be the event that the random walk exits the square $\cS_0$  for the first time along its top side. By Lemma~\ref{lemSquareFarSide}, $\Pr(\cL_0) \geq \frac{1}{B\ell} \geq \frac{1}{B m}$ as $\ell \leq m$. We now show how one can extend this path from the top of $\cS_0$ to the top boundary of $G$ in a reasonably likely way.  
		
		For any vertex $v\in \zz^2$ and positive integer $\ell$, for a random walk in $\zz^2$ starting from $v$, define events $\cL_{v,\ell}^{\mathrm{NNW}}$ and $\cL_{v,\ell}^{\mathrm{NNE}}$, as the events that random walk from $v$ first exits the $(2\ell + 1) \times (2\ell + 1)$ square centered at $v$ along the left half or right half, respectively, of the top side of the square. In the case where the random walk exits at the middle of the top side of the square, we say both events occur.  In this way, by the symmetries of the square, we have that
		\begin{equation}
			\Pr(\cL_{v,\ell}^{\mathrm{NNE}})=\Pr(\cL_{v,\ell}^{\mathrm{NNW}})\geq \frac 1 8.
		\end{equation}
		Now, given $v$ and $\ell$, we can define $\cL_{v,\ell}^{N0}$ to be the event  $\cL_{v,\ell}^{NNE}$ whenever the horizontal coordinate of $v$ is $<\frac m 2$, and to be $\cL_{v,\ell}^{NNW}$ whenever the horizontal coordinate of $v$ is $\geq \frac m 2$. 
		
		\begin{figure}[t]\centering
\begingroup%
  \makeatletter%
  \providecommand\color[2][]{%
    \errmessage{(Inkscape) Color is used for the text in Inkscape, but the package 'color.sty' is not loaded}%
    \renewcommand\color[2][]{}%
  }%
  \providecommand\transparent[1]{%
    \errmessage{(Inkscape) Transparency is used (non-zero) for the text in Inkscape, but the package 'transparent.sty' is not loaded}%
    \renewcommand\transparent[1]{}%
  }%
  \providecommand\rotatebox[2]{#2}%
  \newcommand*\fsize{\dimexpr\f@size pt\relax}%
  \newcommand*\lineheight[1]{\fontsize{\fsize}{#1\fsize}\selectfont}%
  \ifx\svgwidth\undefined%
    \setlength{\unitlength}{141.3886528bp}%
    \ifx\svgscale\undefined%
      \relax%
    \else%
      \setlength{\unitlength}{\unitlength * \real{\svgscale}}%
    \fi%
  \else%
    \setlength{\unitlength}{\svgwidth}%
  \fi%
  \global\let\svgwidth\undefined%
  \global\let\svgscale\undefined%
  \makeatother%
  \begin{picture}(1,1.80422314)%
    \lineheight{1}%
    \setlength\tabcolsep{0pt}%
    \put(0,0){\includegraphics[width=\unitlength,page=1]{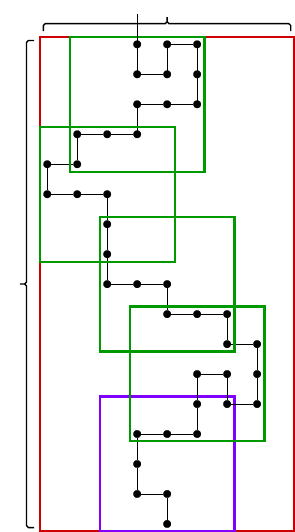}}%
    \put(0.50162928,0.19911617){\color[rgb]{0.49803922,0,1}\makebox(0,0)[lt]{\lineheight{1.25}\smash{\begin{tabular}[t]{l}$S_0$\end{tabular}}}}%
    \put(0.55405293,0.5093412){\makebox(0,0)[lt]{\lineheight{1.25}\smash{\begin{tabular}[t]{l}$v_1$\end{tabular}}}}%
    \put(0.58755952,0.81820081){\makebox(0,0)[lt]{\lineheight{1.25}\smash{\begin{tabular}[t]{l}$v_2$\end{tabular}}}}%
    \put(0.3839468,1.12769091){\makebox(0,0)[lt]{\lineheight{1.25}\smash{\begin{tabular}[t]{l}$v_3$\end{tabular}}}}%
    \put(0.35374132,1.4242718){\makebox(0,0)[lt]{\lineheight{1.25}\smash{\begin{tabular}[t]{l}$v_4$\end{tabular}}}}%
    \put(-0.00989211,0.82205986){\makebox(0,0)[lt]{\lineheight{1.25}\smash{\begin{tabular}[t]{l}$n$\end{tabular}}}}%
    \put(0.51977989,1.75504655){\makebox(0,0)[lt]{\lineheight{1.25}\smash{\begin{tabular}[t]{l}$m$\end{tabular}}}}%
  \end{picture}%
\endgroup%

			\caption{A random walk from the bottom of the red rectangle that first exits at the very top because each of the events $\cL_0, \cL_{v_1,\ell}^{N0}, \cL_{v_2,\ell}^{N0}, \dots, \cL_{v_L,\ell}^{N0}$ occur. Here $m = 10$, $n = 16$, $\ell = 2$, and $L = 4$. The event $\cL_0$ says that the walk from the bottom first exits the bottom purple outlined square to some vertex $v_1$ above the top. From there, each subsequent event $\cL_{v_i, \ell}$ says that the walk exits the next green square along the top, on the side of the top boundary that is closer to the center.}
			\label{figSquareLadder}
		\end{figure}
		
		Let $v_0 = (\frac m 2 , 0)$, and let $v_1$ be the first vertex outside $\cS_0$ encountered in a random walk in $\zz^2$ starting at $v_0$.  For each $i \geq 1$, define $\cS_i$ to be the $(2\ell + 1) \times (2\ell + 1)$ square centered at $v_i$, and let $v_{i+1}$ be the first vertex outside $\cS_i$ in a random walk beginning at $v_i$; see Figure~\ref{figSquareLadder}. To prove the lemma, note that if $\cL_0$ occurs and, for each subsequent $v_i$ for $i = 1, \ldots, L$ where $L = \ceil{\frac{2(n - 2\ell)}{2\ell + 2}}$,  event $\cL_{v,\ell}^{N0}$ occurs, then the random walk from $(\frac m 2 , 0)$ first exits $G$ to a vertex $(i', n+1)$: the initial event $\cL_0$ ensures it never exits the bottom of the rectangle; the choice of $N0 = NNW$ or $N0 = NNE$ ensures it never exits the side of the rectangle; and $L$ was chosen large enough to ensure exiting $\cS_L$ along its top implies exiting $G$ along its top.
		Therefore, to prove the lemma, it suffices to prove a lower bound, for any fixed sequence $v_1,\dots,v_L$\neww{,} on the product
		\begin{align}\label{equProdL}
			\Pr(\cL_0)\prod_{i=1}^{L}\Pr(\cL_{v_i,\ell}^{N0}\mid v_i)=\Pr(\cL_0)\prod_{i=1}^{L}\Pr(\cL_{v_i,\ell}^{N0}) \geq \frac{1}{B m 8^{L}} \geq \frac{1}{B n 8^{L}}.
		\end{align}
		
		Note that the side length of the square is $2\ell + 1 \geq \frac m 2 -1$, and so 
		\begin{equation}\label{equLbound}
			L \leq  \frac{2n}{2\ell+2} \leq \frac{2n}{m/2} = \frac{4n}{m} \leq \frac{4n}{\varepsilon n} = \frac{4}{\varepsilon}.
		\end{equation}
		Combining (\ref{equProdL}) and (\ref{equLbound}), we conclude that the probability the random walk $S_0, S_1,$ $S_2 \dots, $ exits at the very top is at least
		\begin{equation}
			\frac{1}{B n 8^{L}} \geq  \frac{1}{B n 8^{4/\varepsilon}} 
		\end{equation}
		For an appropriate choice of constant $A \geq  \max\{B, 4 \ln 8\}$, this is of the form $\frac{1}{A n e^{A/\varepsilon}}$ claimed in the statement of the Lemma; note using the same constant $A$ twice is just a simplification for convenient bookkeeping. 
	\end{proof}
	
	We now use these lemmas to prove our main theorem. 
	
	\begin{proof}[Proof of Theorem \ref{thm:k-split}]
		As in the proof of Lemma \ref{lemCentralEdgeBound}, we assume $n > 1$ so that Lemma \ref{lemDuality} applies, and we first consider the case where $n$ is odd. For each $1 \leq i \leq k - 1$, we define $e_i := \{(\frac{m}{k}i, \frac{n + 1}{2}), (\frac{m}{k}i + 1, \frac{n + 1}{2})\}$. As shown in Figure~\ref{figKPartitions}, these edges lie on the horizontal midline of the grid and divide it vertically into $k$ equal pieces. For each index $i$, let $a^*_i$ and $b^*_i$ be the faces respectively above and below $e_i$.
		\begin{figure}[ht]\centering
			\input{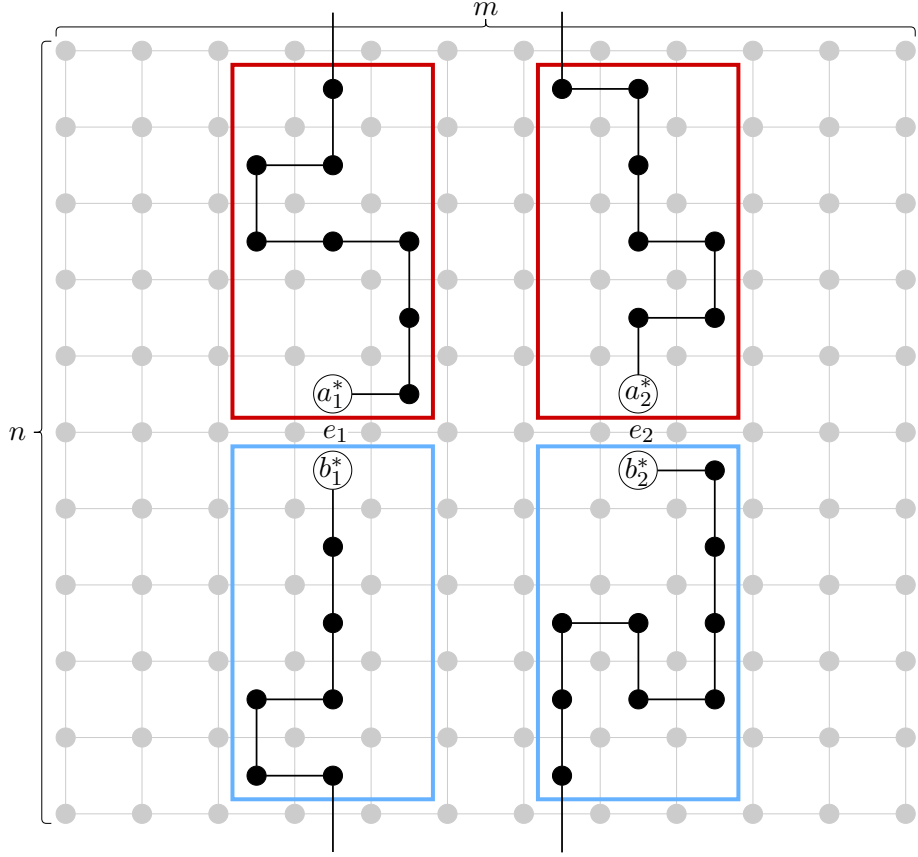}
			\caption{A possible run of the dual graph spanning tree sampling algorithm in the proof of Theorem \ref{thm:k-split} when $m$ is odd. In this example, $m = 12$, $n = 11$, and $k = 3$. The primal graph $G$ is depicted in gray, and the first four random walks in the dual graph $G^*$ are depicted in black.}
			\label{figKPartitions}
		\end{figure}
		
		We generate a random spanning tree $T$ of $G$ by running Wilson's algorithm on the dual graph $G^*$ with the outer face $r^*$ as the root. We start the first $2(k - 1)$ random walks from $a^*_1, b^*_1, a^*_2, b^*_2, \dots, a^*_{k - 1}, b^*_{k - 1}$, assuming each of these vertices of $G^*$ has not yet been added to the tree. The remaining starting points can be determined arbitrarily, and we output the tree $T$ that is dual to the uniformly random spanning tree $T^*$ of $G^*$ sampled by Wilson's algorithm. We will show that $S := \{e_1, e_2, \dots, \neww{e_{k-1}}\}$ splits $G$ into a balanced $k$-forest with sufficiently large probability.
		
		We apply Lemma \ref{lemRandomWalkToFarSide} to each of the red and blue rectangles depicted in Figure \ref{figKPartitions}, which all have grid dimensions $(\frac{m}{k} - 1) \times \frac{n - 1}{2}$ if $\frac{m}{k}$ is even and $\frac{m}{k} \times \frac{n - 1}{2}$ if $\frac{m}{k}$ is odd. Independently, by applying Lemma~\ref{lemRandomWalkToFarSide} to each of these grids with $\varepsilon = 2/k$ and recalling $m \geq n$, each random walk makes it to the far side of its rectangle with probability at least
		\begin{equation}
			\frac{1}{Ane^{Ak/2}},
		\end{equation}
		for a fixed constant $A$.
		
		When this happens, we then apply the same symmetry argument from the proof of Lemma \ref{lemCentralEdgeBound} to each pair of rectangles to conclude that, within each red-blue pair of rectangles, the number of vertices of $G$ to the left of the paths is equal to the number of vertices to the right of the paths with probability at least $\frac{1}{2p + 1}$, where 
		$$p := \left(\frac{m}{k} - 1\right) \cdot \left(\frac{n - 1}{2} - 1\right)$$
		is (an upper bound on) the number of vertices of $G$ in each rectangle. Simplifying this expression, a lower bound on this probability is $\frac{k}{mn}$.    By independence, all pairs of rectangles will be balanced (as they are in Figure \ref{figKPartitions}) with probability at least
		$$\left(\frac{k}{mn}\right)^{k - 1},$$
		in which case $T \setminus S$ will be a balanced $k$-forest. Altogether, this will happen with probability at least
		$$\left(\frac{1}{Ane^{Ak/2}}\right)^{2(k - 1)} \cdot \left(\frac{k}{mn}\right)^{k - 1} \geq \frac{1}{\beta^{k^2}m^{k-1}n^{3k-3}}$$
		for $\beta$ a fixed constant.
		
		In the case where $n$ is even, we may apply the same trick as in the proof of Lemma \ref{lemCentralEdgeBound}, assuming that each path goes downward one step first. This simply tacks on another factor of $\frac{1}{4^{k - 1}}$ to the final probability bound, which can be incorporated into the $\beta^{k^2}$ term.
	\end{proof}

	\section{Approximating Partitions of Lattice-Like Graphs}
	\label{sec:partitioncurves}
	
	In this section we move beyond grid graphs to a more general class of lattice-like graphs, to include examples like that sketched in Figure 	\ref{fig:curves}. \new{This is particularly relevant for the application to political districting, where the graphs considered are based on real-world geography, which is rarely a perfect grid.  This newer, more general setting also makes obtaining {\it perfectly} balanced partitions (where the number of vertices in each component is exactly the same) more challenging and even at times impossible. Therefore we consider both multiplicative approximate balance and additive approximate balance in this new setting. These are reasonable generalizations, as in many political districting settings exact balance of populations is not required.  While U.S. Congressional districts must be balanced to within an additive error of one person (if possible), many state and local offices require only require multiplicative approximate balance of population between districts.}
	
	In Section \ref{subLatticeSequencesDef}, we begin by defining what we mean by lattice-like graphs, including both the more general definition which we use in our multiplicative approximate balance results and the stronger version ({\it uniform} lattice sequences) we use for our additive approximate balance results. \new{Intuitively, these definitions abstract the properties of grids necessary to make it possible for similar arguments to those employed in Section~\ref{secGrids} to succeed (when coupled with a few new key ideas).  At a high level, we consider plane graphs whose dual plane graphs satisfy: 
		\begin{itemize}
			\item adjacent vertices are close to each other in the plane; 
			\item every small ball has at least one \neww{dual} vertex in it; and 
			\item a random walk \newww{on the dual graph} is roughly equally likely to move in any direction.
		\end{itemize}
		Furthermore, the additive approximation case also requires a stronger condition guaranteeing that there is a uniform density of \neww{both primal and dual} vertices throughout. As small finite-sized graphs pose problems for our analysis, we consider sequences of graphs satisfying these properties and require $n$, the number of vertices, to be sufficiently large. 
		
		Surprisingly, these are the only conditions we need in order for an analysis inspired by the same ideas presented in Section~\ref{secGrids} to be successful: we analyze the process by which Wilson's algorithm builds random spanning trees in the dual, with a carefully chosen starting point for each loop-erased random walk.  We are therefore able to prove results about both additive and multiplicative balance partitions in sufficiently refined grids; hexagonal lattices; triangular lattices; finite-degree, doubly-periodic, connected plane graphs; and any sequence of lattices for which the scaling limit of a random walk on the dual is \neww{Brownian} motion.  \rev{After restriction to any fixed bounded region,} our multiplicative approximate balance results also apply to} \neww{Delaunay} triangulations of random point clouds in $\mathbb{R}^2$ of rate $n$; this is an example where the uniform density condition is not satisfied, and so our additive approximation results do not hold.
	
	Despite relaxing the strict balance condition, our results about multiplicative approximate balance are stronger than the results of Section~\ref{secGrids} in other ways. Notably, we are able to show that for any fixed \newww{plane} graph $D$ with $k+1$ faces, there is a {\it constant} lower bound on the probability a random spanning tree $T$ of a lattice-like graph contains \neww{$k-1$ edges whose removal} disconnects $T$ into $k$ components that each approximately resemble one face of $D$. This surprising result uses the same ideas as in previous sections, with a new twist to how we implement Wilson's algorithm: Rather than picking the next starting point for a loop erased walk (in the dual) to be adjacent to a particular central edge, we pick it using another random walk. Not picking the edges whose removal disconnects the (primal) spanning tree ahead of time, but rather letting the random walk process pick them along the way, is the key that enables these much strong\neww{er} probability bounds.

	\new{After formally defining lattice-like graphs in Section~\ref{subLatticeSequencesDef},} in Section \ref{subLatticeStatementResults} we state our results for both lattice classes, including an interesting corollary for grids. In Section \ref{subStayNearCurve}, we give lower bounds on the probability that a random walk in a lattice\new{-like graph} follows a given curve in the plane. These bounds are used in the proofs of both our multiplicative and additive approximate balance results. In Section \ref{sec:modified_wilson} we prove our multiplicative approximate balance bound, and in Section \ref{subAdditive} we prove additive approximate balance.

	\subsection{Lattice Sequences and Uniform Lattice Sequences}\label{subLatticeSequencesDef}
	
	\new{In this section we formally define the lattice-like graphs we consider for our multiplicative and additive approximate balance results. }
	Recall that $d(x,y)$ denotes the Euclidean distance between points and $B_\varepsilon(p)$ is the ball of radius $\ep$ centered at $p$, that is, all points $x$ such that $d(p,x) \leq \ep$.

	We state our results in terms of sequences of infinite lattice-like plane graphs that get finer and finer. The following is the formal definition of the sequences of lattice-like graphs we consider for our multiplicative approximate balance results. 
	
	\begin{definition}\label{defLatticeSequence}
		A \emph{lattice sequence} is a pair $(\{\Lambda_n\}, \rho)$, where $\{\Lambda_n\}$ is sequence of plane graphs with vertex sets $V(\Lambda_n)\subseteq\rr^2$ for which there are corresponding dual plane graphs $\Lambda_n^*$, with each vertex $v\in V(\Lambda_n^*)$ a point in the corresponding face of $\Lambda_n$, \neww{such that the following local conditions hold: for every bounded set $K\subseteq\rr^2$ and every $\ep>0$, there exists $N=N(K,\ep)$ such that for all $n>N$,}
		\begin{enumerate}[(a)]
			\item\label{itmLatticeSequenceClose} \neww{\emph{[\textbf{Edges are short}]}} \neww{For any adjacent pair $x,y\in V(\Lambda^*_n)$ with $x\in K$ or $y\in K$,} $d(x,y)<\ep$,
			\item\label{itmLatticeSequenceDense} \neww{\emph{[\textbf{There is a vertex near every point in the plane]}}} \neww{For any $p\in K$,} the ball $B_\ep(p)$ contains a vertex of $\Lambda^*_n$,
			\item \label{circleproperty} \neww{\emph{[\textbf{Random walks are sufficiently diffuse}]}} \neww{For any dual vertex $v\in V(\Lambda^*_n)\cap K$,} there is a division of the circle $C_{\ep,v}$ of radius $\ep$ into arcs $A_1,\dots,A_s$ each of length at most $\frac 1 8 2\pi \ep$, such that the following holds for each $i \in [s]$: In a simple random walk $v = v_0, v_1, v_2, \dots$ \new{in $\Lambda^*_n$}, with probability at least $\rho > 0$, letting $j$ be the first index where $d(v_0, v_{j}) > \varepsilon$, the straight line segment joining $v_{j - 1}$ with $v_j$ passes through $C_{\ep, v}$ in $A_i$.
		\end{enumerate}
	\end{definition}
	
	\new{Informally, property (\ref{itmLatticeSequenceClose}) says that as $n$ goes to $\infty$ then adjacent vertices in the dual graph $\Lambda_n^*$ get arbitrarily close together \neww{inside each fixed bounded region}.  Property (\ref{itmLatticeSequenceDense}) says that there is no empty space \neww{\textemdash} for sufficiently large $n$ every ball of radius $\varepsilon$ \neww{with center in the fixed bounded region} contains at least one vertex of $\Lambda^*$, meaning the dual vertices also become arbitrarily dense \neww{locally}.}
	
	\new{Property \eqref{circleproperty} is significantly more involved. Its purpose is largely to ensure that random walks in the dual behave well, as these will be a main focus of our analysis. For each arc of a circle surrounding dual vertex $v$, there is a lower bound on the probability that a simple random walks exits the circle through that arc. This lower bound probability, $\rho$, is part of our definition of the lattice sequence, and our bounds on balance probabilities will depend on $\rho$.  The choice of $\frac 1 8$ is made to enable a geometric argument (in Lemma \ref{lemCurveWalkSimple}); replacing it with any smaller constant would give exactly the same family of lattice sequences. 
		
		The following example shows that finer and finer grid graphs are one example of a lattice sequence satisfying this definition.}
	
	\begin{example}
		The \neww{family} of plane graphs $\{\zz^2, \frac12\zz^2, \frac13\zz^2, \dots\}$ is a lattice sequence with $\rho = \frac18$. Properties (\ref{itmLatticeSequenceClose}) and (\ref{itmLatticeSequenceDense}) say that, as we increase $n$, neighboring vertices in $\frac1n \zz^2$ become arbitrarily close, and the vertex set becomes arbitrarily dense. Property (\ref{circleproperty}) holds using the partition of $C_{\ep, v}$ obtained by drawing horizontal, vertical, and both 45 degree diagonal lines through $v$, that is, splitting $C_{\ep, v}$ into 8 arcs of equal length. By symmetry, a random walk from $v$ is equally likely to exit through each of these 8 arcs.
	\end{example} 
	
	The definition also applies to scalings of lattices like the triangular or hexagonal lattice. \new{This is because property \eqref{circleproperty} holds for any sequence of lattices for which the scaling limit of random walk on the dual is Brownian motion, that is, where the random walk is equally likely to move in any direction.} \rev{For Poisson clouds in bounded regions, the Poisson--Delaunay random-walk input is the localized consequence of the quenched invariance-principle theory, see \cite{pointbrownian}.}

	\new{Perhaps surprisingly, these three conditions are all that is required for our multiplicative approximate balance results.  For our additive approximate balance results (Theorem \ref{thmAdditiveError}), we need a stronger condition on the density of points of \neww{both $\Lambda_n$ and $\Lambda_n^*$}, namely that density fluctuations are bounded at some constant scale. This motivates the following definition  of a  \textit{uniform lattice sequence}. Note that here $R/n$ for some positive constant $R$ plays the same role as $\ep$ in our previous definition. }
	
	
	\begin{definition}\label{defStrongLatticeSequence}
		A \emph{uniform lattice sequence} is a tuple $(\{\Lambda_n\}, \rho, R, C_1, C_2)$, where $\{\Lambda_n\}$ is sequence of plane graphs with vertex sets $V(\Lambda_n)\subseteq\rr^2$ for which there are corresponding dual plane graphs $\Lambda_n^*$, with each vertex $v\in V(\Lambda_n^*)$ a point in the corresponding face of $\Lambda_n$, and positive constants $\rho, R,C_1,C_2$ so that, for all sufficiently large $n$:
		\begin{enumerate}[(a)]
			\item\label{itmStrongLatticeSequenceClose} \neww{\emph{[\textbf{Edges are short}]}} Any adjacent pair $x,y\in V(\Lambda^*_n)$ satisfies $d(x,y)<\frac R n$,
			\item\label{itmStrongLatticeSequenceDense} \neww{\emph{[\textbf{There is a consistent density of primal and dual vertices near every point in the plane}]}} For any $p\in \rr^2$, the ball $B_{R/n}(p)$ contains between $C_1$ and $C_2$ vertices of \neww{each of $\Lambda_n$ and $\Lambda^*_n$},
			\item \label{Strongcircleproperty}  \neww{\emph{[\textbf{Random walks are sufficiently diffuse}]}} For any dual vertex $v$ and any  $r\geq \frac R n$, there is a division of a circle $C_{r,v}$ of radius $r$ into arcs $A_1,\dots,A_s$ each of length at most $\frac 1 8 2\pi {\red r}$, such that the following holds for each $i \in [s]$: In a simple random walk $v = v_0, v_1, v_2, \dots$, with probability at least $\rho > 0$, letting $j$ be the first index where $d(v_0, v_{j}) > {\red r}$, the straight line segment joining $v_{j - 1}$ with $v_j$ passes through $C_{r, v}$ in $A_i$.
		\end{enumerate}
	\end{definition}
	
	\new{\noindent In this definition, $R/n$ for positive constant $R$ is playing a similar role as $\ep$ in our previous definition of lattice sequences; this more tightly controls the rate at which vertices in \neww{both $\Lambda_n$ and $\Lambda_n^*$} get more and more dense. Note that, with this substitution, property \eqref{itmStrongLatticeSequenceClose} is equivalent to that for lattice sequences. Meanwhile, property (c) simply substitutes $r \geq R/n$ for $\ep$ in property \eqref{Strongcircleproperty} for lattice sequences.} \neww{As before, it is a way of requiring that random walks will never exit a ball in a predictable direction.}  \new{The strengthening in this definition mainly comes from property~\eqref{itmStrongLatticeSequenceDense}; rather than simply requiring that all sufficiently small balls contain at least one dual vertex, instead we now require both a constant lower bound and a constant upper bound on the number of \neww{primal and dual} vertices falling within every small ball. }

	This definition is also satisfied for grids, the triangular lattice, the hexagonal lattice, or indeed any finite-degree, doubly-periodic connected plane graph. However, it is not satisfied for triangulations of random point clouds, since as $n \to \infty$, the \neww{primal and dual} fluctuations in density \neww{are} too large \new{and Property~\eqref{itmStrongLatticeSequenceDense} will not be satisfied. }

	\subsection{Statement of Results}\label{subLatticeStatementResults}
	
	\new{In this section we state our two main results: our multiplicative approximate balance results for lattice sequences in Theorem~\ref{thm:planegraph} and our \newww{additive} approximate balance results for uniform lattice sequences in Theorem~\ref{thmAdditiveError}.  We will prove these theorems in subsequent subsections. We begin with relevant definitions related to planar geometry and plane graphs. }

	\new{For a point $p$ and set $X$, we let $d(p, X) = \min_{x \in X} d(p,x)$.} For two sets $X$ and $Y$, we let $d(X,Y)$ denote the Hausdorff distance between these two sets (recall the Hausdorff distance was defined in equation~\eqref{eqn:haus}).  
	Recall that a \emph{curve} $\gamma$ is a continuous function $\gamma:[a,b]\to \rr^2$ for $a<b$.  Except where specified otherwise, we take $a=0,b=1$. A plane graph $D=(V,\Gamma)$ is a drawing of a (planar) graph in the plane without intersections.   In particular, $V=V(D)$ is a finite set of points in the plane $\rr^2$.  $\Gamma$ is a finite collection of curves given by continuous functions $\gamma_i:[0,1]\to \rr^2$ such that no two such curves intersect except possibly at their endpoints, and such that if $E$ is the set of pairs of endpoints:
	\[
	E=\{ \{\gamma_i(0),\gamma_i(1)\}\mid i=1,\dots,|\Gamma|\},
	\]
	then $(V,E)$ is a graph.  The faces of $D$ are the connected components of $\rr^2\setminus \bigcup_{i=1}^{|\Gamma|} \im(\gamma_i)$ (here $\im(\gamma)$ is the image of the function $\gamma$ \new{in $\rr^2$}), and we refer to the unique unbounded face as the \emph{outer} face. 
	
	Theorem~\ref{thm:planegraph} considers a fixed bounded plane graph $D$ that gives the partition structure we are looking to approximate. In doing this, we will need to restrict the infinite graphs in the lattice sequence to a reasonable bounded subgraph that falls inside $D$, and we do this as follows. Let $D$ be a bounded plane graph, and fix $\delta>0$ that will be chosen later in terms of $D$ (in Lemma \ref{lem:epdelta}).  Given plane graph $\Lambda_n$ with dual $\Lambda_n^*$ from a lattice sequence, and a cycle $C^*$ in $\Lambda_n^*$ at Hausdorff distance $<\delta\new{/10}$ from the outer face boundary of $D$ (see Figure \ref{figGammaPartition} for an example of $C^*$),  we let $\Omega_{D,\Lambda_n}$ be the subgraph \neww{of} $\Lambda_n$ lying inside $C^*$.  We let $\Omega^*_{D,\Lambda_n}$ be the subgraph of $\Lambda_n^*$ induced by all vertices of $C^*$ along with the vertices of $\Lambda_n^*$ lying inside $C^*$.  In this way, we can consider the planar dual of $\Omega_{D,\Lambda_n}$ to be $\Omega^*_{D,\Lambda_n}$  with \emph{wired boundary condition}, where the entire cycle $C^*$ (rather than a single dual vertex) corresponds to the outer face of $\Omega_{D,\Lambda_n}$. (In particular, for our proofs, we will run Wilson's algorithm on $\Omega^*_{D,\Lambda_n}$ with the cycle $C^*$ identified as a single root.) 
	
	Given this plane graph $D$ describing the partition structure we are looking to approximate, let $k+1$ be the number of faces and denote the $k$ bounded faces by $\phi_1,\dots,\phi_k$. 
	We say a partition of the graph $\Omega_{D,\Lambda}$ into connected components $C_1,\dots,C_k$ is $\ep$-\emph{compatible} with $D$ if for all $i$ and vertices $v\in \Omega_{D,\Lambda}$, the implication
	\begin{equation}\label{eq:compatible}
		v\in C_i\implies d(v,\phi_i)\leq \ep
	\end{equation}
	holds.  By subdividing edges if necessary, we will assume that $D$ has no loops, so that $\gamma(0)\neq \gamma(1)$ for all $\gamma\in \Gamma(D)$.
	
	For a lattice sequence $(\{\Lambda_n\}, \rho)$, a probability space on the set of spanning trees of $\Omega_{D,\Lambda_n}$, and $\ep>0$, we define the event  $\cE_{D,\Lambda_n,\ep}$, which \neww{occurs} whenever \new{a random spanning tree $T$ of $\Omega_{D,\Lambda_n}$ has $k-1$ edges} whose removal results in a forest with components $C_1,\dots,C_k$ that is $\ep$-compatible with $D$.  The following is our main result for multiplicative balance (the formal version of Theorem~\ref{thm:mult_intro}).
	
	\begin{theorem}\label{thm:planegraph}
		Let $(\{\Lambda_n\}, \rho)$ be a lattice sequence, let $D$ be a \neww{bounded} plane graph with $k+1$ faces, and let $\Omega_{D,\Lambda_n}$ be as above.  For the uniform probability space on the set of spanning trees of a graph $\Omega_{D,\Lambda_n}$, we have that as $n\to \infty$, $\Pr(\cE_{D,\Lambda_n,\ep})$ is bounded below by a \new{positive} constant depending only on $D$, $\ep$, \neww{and $\rho$}.
	\end{theorem}
	\noindent \new{We prove Theorem \ref{thm:planegraph} in Section \ref{sec:modified_wilson}, after establishing preliminary results in Section~\ref{subStayNearCurve}.} 
	\new{For the interested reader, the precise lower bound and its dependence on $D$, $\ep$, \neww{and $\rho$} can be found within the proof as Equation~\ref{eq:mult_prob} (where $\delta < \ep$ is chosen as in Lemma~\ref{lem:epdelta}); it is omitted here as it includes several terms we have yet to define. }
	
	As a consequence of Theorem \ref{thm:planegraph}, if we draw the partition so that the parts contain approximately equal numbers of vertices, we can conclude that random trees are splittable into approximately balanced pieces with constant probability. 
	This is possible so long as $\{\Lambda_n\}$ has the property that for any $\delta>0$ and $R$, there is an $\ep>0$ so that \neww{ for every point $p$ in the bounded region under consideration,} the $\ep$-ball $B_\ep(p)$ satisfies $|B_\ep(p)\cap V(\Lambda_n)|\leq \delta |B_R(0)\cap V(\Lambda_n)|$. In the case of grid graphs, for instance, we obtain the following corollary.  \new{We say a tree is $(k,\newww{\xi})$-\emph{approximately splittable} if there are $k-1$ edges whose removal gives a forest} \newww{with $k$ components that each have between $(1-\xi) |V(T)|/k$ and $(1+\xi) |V(T)|/k$ vertices.}
	
	\begin{corollary}\label{cor:ep}
		Fix $\newww{\xi} \geq 0$ and a positive integer \neww{$k$}. Let $m,n$ be positive integers such that $n \leq m$, $k | m$, and $\newww{400k}/n \leq \newww{\xi} \leq \newww{2/3}$. 
		Let $G$ be an $m \times n$ grid graph. There is a constant $C(k, \newww{\xi}) > 0$ such that the probability a uniformly random spanning tree of $G$ is $(k, \newww{\xi})$-approximately splittable is at least $C(k, \newww{\xi})$. 
	\end{corollary}
	\newww{We will prove this corollary by applying Theorem~\ref{thm:planegraph} with $\ep = \xi/(\rev{3}k)$ to the plane graph $D$ that is the rectangle $[0,m/n]\times[0,1]$ divided vertically into $k$ strips, using the usual square-lattice sequence under the isotropic scaling by $1/n$. The upper bound $\xi \leq 2/3$ implies $\ep \leq 1/3k$, which ensures the boundaries of each of the $k$ regions are sufficiently separated from one another, that is, distance at least $3 \ep$ apart.}
	\neww{The lower bound on \newww{$\xi$} can instead be seen as imposing a lower bound on $n$: }\newww{while Theorem~\ref{thm:planegraph} makes a statement about lattice sequences as their index tends to infinity, the condition $n \geq 400k/\xi$ makes precise exactly how far along this lattice sequence we must proceed until the desired conclusions hold. It makes sense that as one considers splits into more parts and requires additional (smaller) accuracy, the size of the graphs considered must grow as well.}  
	
	
	As our final main result, we give an \new{inverse polynomial} \new{(in $n$)} lower bound on the probability that a random spanning tree in a square region of a uniform lattice sequence can be cut into pieces that differ in size by an \emph{additive constant} (the formal version of Theorem~\ref{thm:add_intro}). \new{Rather than considering an arbitrary plane graph $D$, we consider a specific plane graph whose outer boundary is the unit square $[0,1]^2$.} 
	
	\begin{theorem}\label{thmAdditiveError}
	\new{Let $k \in \mathbb{Z}_{\geq 2}$ be a constant}, and suppose $(\{\Lambda_n\}, \rho, R, C_1, C_2)$ is a uniform lattice sequence, \new{and there exists an $N$ such that} for all $n>N$, $C^*_n$ in $\Lambda^*_n$ is a cycle in the dual that is contained within the unit square $[0, 1]^2$ and at Hausdorff distance $<R/n$ from its boundary. Let $\Omega_{\Lambda_n}$ denote the subgraph of $\Lambda_n$ enclosed by the cycle \neww{$C^*_n$}. \new{There is a constant $\eta$} such that with probability at least $1/\poly(n)$, a uniformly random spanning tree of $\Omega_{\Lambda_n}$ can be disconnected by the removal of $k-1$ edges into $k$ components that differ in size by at most $\eta$.
	\end{theorem}
	
	
	\new{As we remark later, this inverse polynomial probability lower bound is of the form 
	$1/\beta^{k^3} n^{k^2}$
	for some constant $\beta$, although we do not believe this bound is tight. }
	
	Note that balance up to an additive constant is the best we can hope for in the generality we are working in here\neww{. F}or example, an $n \times n$ grid graph with $n$ odd and an additional leaf appended to every vertex is a case to which Theorem \ref{thmAdditiveError} applies, but \neww{which} does not admit an exactly balanced partition despite the fact that the total number of vertices is even.
	
	\new{We develop some necessary lemmas in Section~\ref{subStayNearCurve}, prove Theorem~\ref{thm:planegraph} in Section~\ref{sec:modified_wilson}, and prove Theorem~\ref{thmAdditiveError} in Section~\ref{subAdditive}. }
	
	\subsection{Probability of Staying Near a Curve}\label{subStayNearCurve}
	
	\noindent \new{In this section, we prove bounds on the probability that a random walk in a graph in a lattice sequence \neww{stays} near a given planar curve. These bounds are needed to prove both Theorem~\ref{thm:planegraph} and Theorem~\ref{thmAdditiveError}. Note that because a uniform lattice sequence is also a lattice sequence, the following lemma applies to both.}
	
	\new{
	We begin by noting that for our purposes, the following lemma allows us to assume when proving Theorem \ref{thm:planegraph} that we are only working with rectifiable curves (thus with well defined lengths):
	\begin{lemma}
		For any $\ep>0$ and any  curve $\gamma:[0,1]\to \mathbb{R}^2$, there is a piecewise linear curve $\gamma'$ with the same endpoints such that the Hausdorff distance between the image of $\gamma$ and $\gamma'$ is at most $\ep$.
	\end{lemma}
	\begin{proof}
		By uniform continuity of $\gamma$, there is a $\delta$ such that \neww{for any} $x_0,x_1\in [0,1]$ with $|x_1-x_0|<\delta$ we have $|\gamma(x_0)-\gamma(x_1)|<\ep$.  Choose now a sequence of points $0=x_0,x_1,x_2,\dots,x_\ell=1$ such that $|x_{i+1}-x_i|<\delta$ for each $i$. \neww{We} define the curve $\gamma'$ by the piecewise linear approximation of $\gamma$ through the points $x_0,x_1,\dots,x_\ell$.
	\end{proof}
	}
	\noindent \new{Thus in what follows we will assume all curves are rectifiable.}  The following lemma is a first step to proving Theorem \ref{thm:planegraph}.

	\begin{lemma}\label{lemCurveWalkSimple}
	Let $(\{\Lambda_n\}, \rho)$ be a lattice sequence, let $\ep>0$, and let $\gamma$ be a curve in the plane of length $\new{|\gamma|}$. \new{Let $N$ be large enough such that for all $n \geq N$, adjacent vertices in $\Lambda^*_n$ are at distance at most $\ep/20$ and property~\eqref{circleproperty} holds, in a fixed bounded neighborhood of the image of $\gamma$, for radius $\ep/2$. For all $n \geq N$,} for all $v_0\in V(\Lambda^*_n)$ with $d(v_0,\gamma(0))\leq \ep/2$, and for a random walk in $\Lambda^*_n$ started from $v_0$, let $\cE = \cE_{n, v_0, \ep, \gamma}$ be the event that the walk reaches a point within $\frac{\varepsilon}{2}$ of $\gamma(1)$ before ever reaching a vertex at distance $>\ep$ from the curve $\gamma$.  For all sufficiently large $n$,
	$$\Pr(\cE_{n, v_0, \ep, \gamma}) \geq \rho^{\new{1 +} 20\new{|\gamma|}/\ep }\neww{.}$$
	\end{lemma}
	\begin{proof}
	The proof of this lemma is reminiscent of the proof of Lemma \ref{lemRandomWalkToFarSide}. The differences are that now we consider a sequence of circles that the random walk must escape from rather than squares, and the direction to which we hope it escapes will be determined by the path $\gamma$, not just always going upward (see Figure~\ref{figCircleLadder}).
	
	Choose $n$ large enough so that neighboring vertices in $\Lambda^*_n$ are at distance at most $\frac{\ep}{20}$ \neww{and property~\eqref{circleproperty} holds in the bounded neighborhood of the image of $\gamma$ under consideration}. \newww{Let $v \in \Lambda^*_n$ be such that $d(v, \gamma(1)) \geq \ep/2$.} We may partition $C_{\ep/2, v}$ (the circle of radius $\ep/2$ around $v$) into arcs $A_1, A_2, \dots, A_s$ as in Definition \ref{defLatticeSequence} (\ref{circleproperty}) applied to $\frac{\varepsilon}{2}$. For any $t \in [0, 1]$ such that $v$ is within Euclidean distance $\varepsilon/2$ of $\gamma(t)$ \newww{(}but not within Euclidean distance $\varepsilon/2$ of $\gamma(1)$\newww{)}, that is, 
	\begin{equation}\label{equInfWellDefined}
		d(v, \gamma(t)) \leq \frac{\varepsilon}{2} \leq d(v, \gamma(1)),
	\end{equation}
	we define $A^*(v, t)$ to be \new{the arc} $A_i$ in $C_{\ep/2, v}$ such that the next time $\gamma$ reaches $C_{\ep/2, v}$ \neww{after time $t$} it is in the arc $A_i$, that is, \new{the} arc $A_i$ such that     
	$$\gamma(\inf\{t' \in (t, 1] \suchthat d(v, \gamma(t')) = \varepsilon/2\}) \in A_i.$$
	Note the infimum is well-defined by (\ref{equInfWellDefined}) \neww{in the applications below, where $v$ is strictly inside $C_{\ep/2,v}$ at time $t$ unless the conclusion has already been reached}. 
	We then define $\cL_{v, t}$ to be the event that a random walk in \neww{$\Lambda^*_n$} from $v$ first crosses $C_{\ep/2, v}$ through $A^*(v, t)$\new{, that is, the event that the random walk from $v$ is proceeding in the direction of $\gamma$ as desired.}
	
	\begin{figure}[t]\centering
		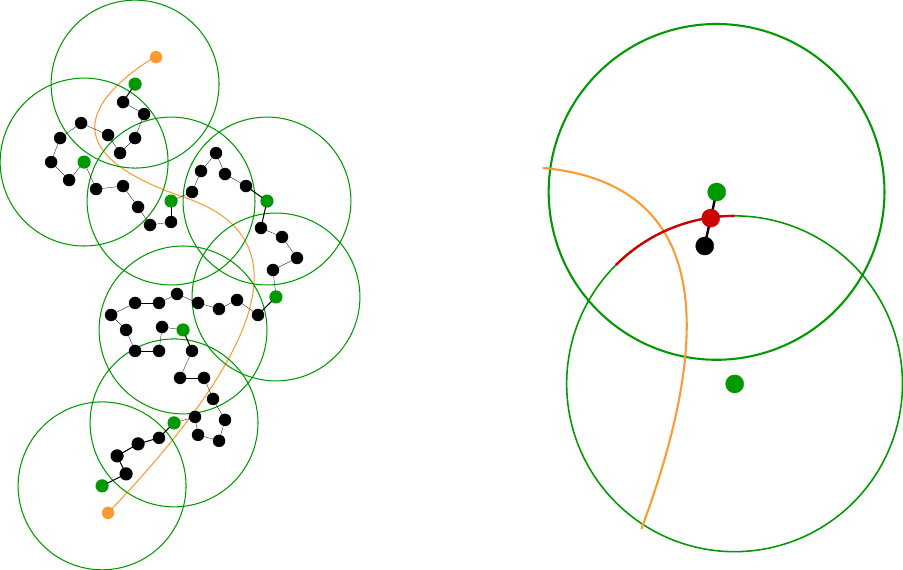
		\caption{Illustration accompanying the proof of Lemma \ref{lemCurveWalkSimple}. A random walk from $v_0$ in the dual lattice eventually reaches a vertex $v_L$ within $\frac{\ep}{2}$ of $\gamma(1)$ while staying within $\ep$ of the curve $\gamma$ because each of the events  $\cL_{v_0, 0}, \cL_{v_1, t_1}, \cL_{v_2, t_2}, \dots, \cL_{v_{L - 1}, t_{L - 1}}$ occur\neww{s}. A key step of the proof is lower-bounding the distance between $\gamma(t_{j + 1})$ and $\gamma(t_j)$, which we accomplish by observing that $v_j$ is closer to $\gamma(t_j)$ than $\gamma(t_{j + 1})$, since it is close to $a_j$ and the arc $A^*(v_{j - 1}, t_{j - 1})$ is \neww{short}.}
		\label{figCircleLadder}
	\end{figure}
	
	Assume that $v_0$ is not already within $\frac{\varepsilon}{2}$ of $\gamma(1)$, otherwise there is nothing to show. Then let $t_1$ be the first time such that $d(v_0, \gamma(t_1)) = \frac{\ep}{2}$. By property (\ref{circleproperty}), we know that $\cL_{v_0, 0}$ will occur with probability at least $\rho$. Assuming this happens, suppose the random walk exits to a point $v_1 \in \neww{\Lambda^*_n} \setminus B_{\frac{\ep}{2}}(v_0)$ from a previous point $u_1 \in \neww{\Lambda^*_n} \cap B_{\frac{\ep}{2}}(v_0)$, where the line segment between $u_1$ and $v_1$ passes  
	through a point $a_1 \in A^*(v_0, 0)$ as in Figure \ref{figCircleLadder}. \new{As each arc $A_i$ has length at most $\frac{1}{8} 2\pi \frac{\ep}{2} = \frac{\pi}{8} \ep$ by Definition~\ref{defLatticeSequence}\eqref{circleproperty}, and \newww{$\gamma(t_1)$} and $a_1$ are both in this arc}, the arc length along \new{$C_{\ep/2, v_0}$} from $a_1$ to $\gamma(t_1)$ is at most $\frac{\pi}{8} \ep$. \new{It follows that}
	\begin{equation}\label{equDistanceThroughCirclePlusArc}
		d(v_1, \gamma(t_1)) \leq d(v_1, a_1) + d(a_1, \gamma(t_1)) \leq d(v_1, u_1) + \frac{\pi}{8} \ep \leq \frac{\ep}{20} + \frac{\pi}{8} \ep \leq \frac{9}{20}\ep.
	\end{equation}
	If $d(v_1, \gamma(1)) \leq \frac{\varepsilon}{2}$, we  have shown $\Pr(\cE) \geq \rho$, \new{and the lemma follows}.  

	If, instead, $d(v_1, \gamma(1)) > \frac{\varepsilon}{2}$, then both inequalities in (\ref{equInfWellDefined}) hold \new{for $v_1$}. We then let $t_2 > t_1$ be the first time (later than $t_1$) such that $d(v_1, \gamma(t_2)) = \frac{\ep}{2}$ and observe that, with probability at least $\rho$, the event $\cL_{v_1, t_1}$ will occur, at which point the random walk will exit $B_{\ep/2}(v_1)$ to some vertex $v_2 \in B_{\frac{\ep}{2}}(\gamma(t_2))$ \rev{by the same estimate as in \eqref{equDistanceThroughCirclePlusArc}}. Continuing inductively until $d(v_L, \gamma(1)) \leq \frac{\varepsilon}{2}$, we see that we will continue to follow the curve $\gamma$ for a sequence of $L$ steps, through events $\cL_{v_0, 0}, \cL_{v_1, t_1}, \cL_{v_2, t_2}, \dots, \cL_{v_{L - 1}, t_{L - 1}}$, with probability at least $\rho^L$. Observe that the bound in (\ref{equDistanceThroughCirclePlusArc}) applies to each round, choosing $a_j$ to be the point on the circle between $v_j$ and the previous step $u_j$. Thus, \new{as $\gamma(t_j)$ is necessarily closer to $v_j$ than $\gamma(t_{j+1})$}, we    
	may bound the distance between consecutive points on the path as
	$$d(\gamma(t_{j + 1}), \gamma(t_{j})) \geq d(\gamma(t_{j + 1}), v_j) - d(v_j, \gamma(t_{j})) \geq \frac{\ep}{2} - \frac{9\ep}{20} = \frac{\ep}{20}.$$
	Since $\new{|\gamma|}$ is \new{the} total length of the curve $\gamma$, \new{and the above equation is true for all $i$ from $1$ to $L-1$ (but, notably, not true for $d(\gamma(t_1), \gamma(0))$), it follows that $|\gamma| \geq (L-1) \ep/20$.  This means
		that $L \leq 1 + 20 |\gamma|/\ep$, and so  }      
	it follows that we only need at most $1 + \frac{20|\gamma|}{\ep}$ steps of the correct events $\cL_{v_j, t_j}$ occurring until we reach a vertex within $\frac{\ep}{2}$ of $\gamma(1)$. All the while, we know that the path is always within a ball of radius $\frac{\ep}{2}$ that contains parts of $\gamma$, so it never is more than $\ep$ away from $\gamma$. Thus, we have
	\begin{equation*}
		\Pr(\cE) \geq \rho^{\new{1 +} 20\new{|\gamma|}/\ep }. 
	\end{equation*}
	\end{proof}
	
	We will also require the following slightly stronger version of this lemma with some extra conditions about how the path ends. \new{Again, note the following lemma applies to both lattice sequences and uniform lattice sequences, as the second is a special case of the first.} 
	
	
	\begin{lemma}\label{lem:curveswalk}
	Let $(\{\Lambda_n\}, \rho)$ be a lattice sequence, and let $\gamma$ be a curve in the plane of positive length $|\gamma|$, \new{and let $\varepsilon > 0$} \neww{with $d(\gamma(0),\gamma(1))\geq \varepsilon$}. \new{Let $N$ be large enough such that for all $n \geq N$, adjacent vertices in $\Lambda^*_n$ are at distance at most $\ep/100$ and property (\ref{circleproperty}) in the definition of a lattice sequence holds in a fixed bounded neighborhood of the image of $\gamma$. For all \neww{$n \geq N$},} for any $v_0\in \Lambda^*_n$ with $d(v_0,\gamma(0))\leq \ep/2$, and for a random walk in $\Lambda^*_n$ started from $v_0$, let $\cE' = \cE'_{n, v_0, \gamma}$ be the event that the walk never reaches a vertex at distance $>\ep$ from the curve $\gamma$, \neww{and that its trace contains a subpath contained in $B_{\ep}(\gamma(1))\setminus B_{\ep/5}(\gamma(1))$ that encircles $\gamma(1)$}.
	We have $$\Pr(\cE'_{n, v_0, \gamma}) \geq \rho^{100\new{|\gamma|}/\ep + \rev{501}}.$$
	\end{lemma}
	
	\begin{figure}\centering
	\includegraphics[scale=.2]{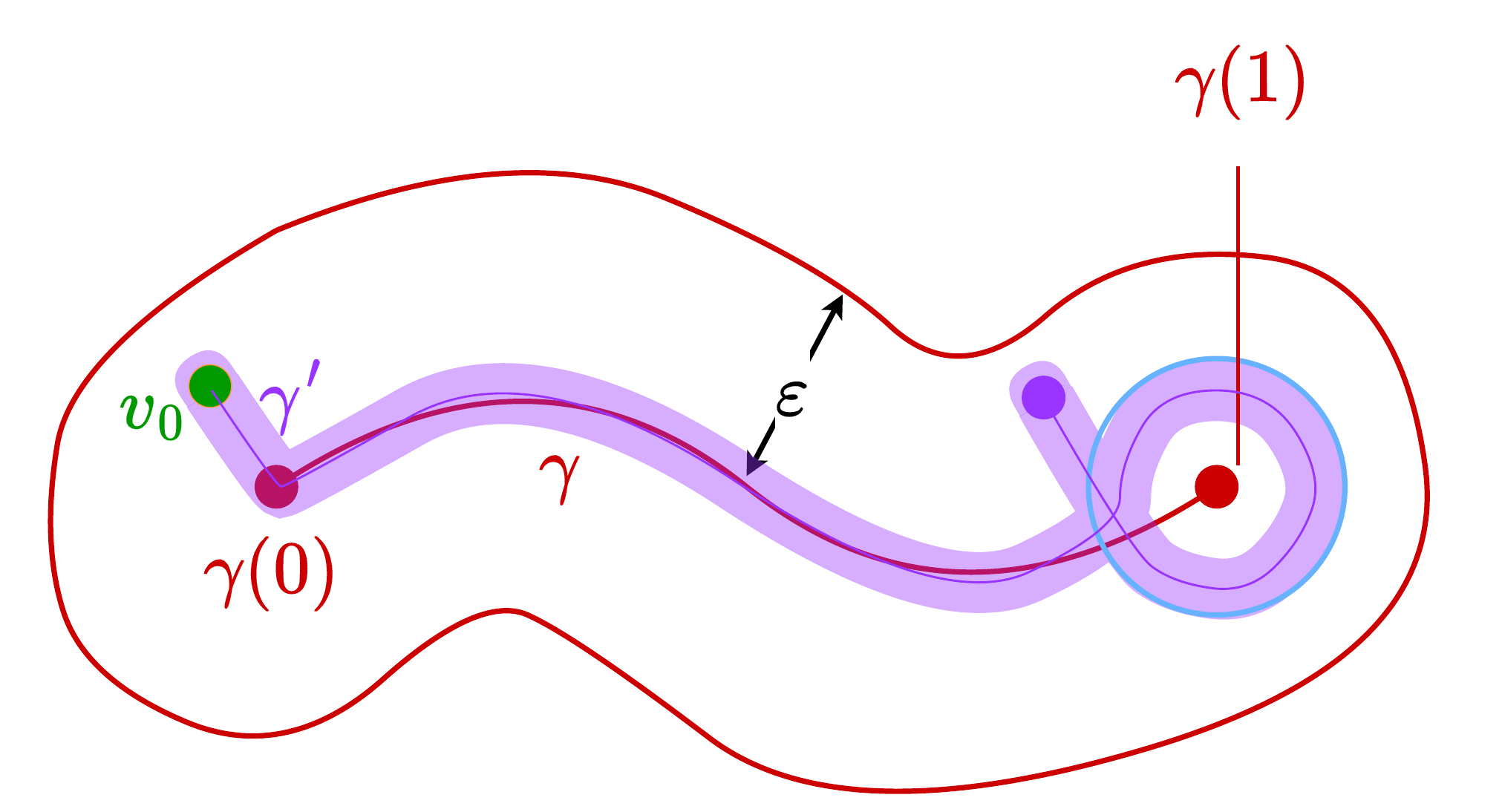}
	\caption{Illustration accompanying the proof of Lemma \ref{lem:curveswalk}. We apply Lemma \ref{lemCurveWalkSimple} to the orange path $\gamma'$ and $\frac{\varepsilon}{5}$, which is the radius of the orange tube around $\gamma'$. The quantities $\frac{\ep}{5}$ and $\frac{3\varepsilon}{2}$ from the proof are crude bounds on the width and length of the last part of the tube that must encircle $\gamma(1)$ while staying within the blue circle of radius $\frac{3\ep}{5}$.}
	\label{figLoopDeLoop}
	\end{figure}
	
	\begin{proof}
	We apply Lemma \ref{lemCurveWalkSimple} with $\frac{\varepsilon}{5}$ to the path $\gamma'$ drawn in Figure \ref{figLoopDeLoop}. The path consists of 3 parts: a straight line from $v_0$ \new{to} $\gamma(0)$, then $\gamma$ until it enters $B_{\ep/2}(\gamma(1))$, then a loop that crosses itself and remains at distance greater than $2\ep/5$ and less than $3\ep/5$ from $\gamma(1)$ while encircling this point. Then observe that $\cE \subseteq \cE'$, where $\cE$ is the event from Lemma \ref{lemCurveWalkSimple} applied to this new path using $\varepsilon/5$. This is because any random walk in $\neww{\Lambda^*_n}$ satisfying $\cE$ stays within $\varepsilon$ of the original curve $\gamma$; \rev{the endpoint-separation assumption ensures that the initial straight segment and the portion of $\gamma$ used before entering $B_{\ep/2}(\gamma(1))$ stay at distance at least $\ep/2>2\ep/5$ from $\gamma(1)$, and the endpoint-encircling loop itself stays between distances $2\ep/5$ and $3\ep/5$ from $\gamma(1)$}. Since the random walk is within $\ep/5$ of this path, the walk satisfie\newww{s} the conditions of the lemma. 
	We may upper bound the length of this new combined curve $\gamma'$ as
	$$\new{|\gamma'|} \leq \frac{\ep}{2} + \new{|\gamma|} + \rev{4}\ep \leq \new{|\gamma|} + \rev{5}\ep.$$
	\new{By substituting $ \ep \mapsto \ep/5$ and $|\gamma| \mapsto |\gamma| + \rev{5}\ep$ in the bound of Lemma~\ref{lemCurveWalkSimple}, the result follows.} \new{Note making this substitution now also requires assuming that neighboring vertices of $\Lambda_n^*$ are at distance at most $\ep/100$. }
	\end{proof}
	
	Note that when a disconnected plane graph $D$ has bounded faces $\phi_1,\dots,\phi_k$ and satisfies the hypotheses of Theorem \ref{thm:planegraph}, we can add curves to $D$ to create a connected plane graph $D'$ whose bounded faces satisfy $\phi_i'\subseteq \phi_i$ for all $i$.  We have then that $\cE_{D',\Lambda_n,\ep}\subseteq \cE_{D,\Lambda_n,\ep}$, and thus it suffices to prove Theorem \ref{thm:planegraph} in the case where $D$ is connected.  Therefore, we assume that $D$ is connected for the rest of \new{this section}.  
	
	\new{Our results will also require that we consider small enough scales so that different curves in the plane graph $D = (V, \Gamma)$ are well-separated. This is necessary as we will need that our random walks that stay near curves $\gamma \in \Gamma$, as described in Lemmas~\ref{lemCurveWalkSimple} and~\ref{lem:curveswalk}, do not interact except as desired. The following lemma makes this precise. }
	
	\begin{lemma}\label{lem:epdelta} Let $D = (V, \Gamma)$ be a plane graph. For $0<\delta<\ep$ sufficiently small, we have that:
	\begin{enumerate}[({A}1)]
		\item \label{epchoicevxs} The distance between any two \new{distinct} vertices $u,v$ is at least $3\ep$.
		\item \label{epchoicecurves} If $a_\gamma$ denotes the last time $t$ at which $\gamma(t)$ is in the closed ball of radius $\ep$ about $\gamma(0)$, and $b_\gamma$ denotes the first time $t$ at which $\gamma(t)$ is in the closed ball of radius $\ep$ about $\gamma(1)$, the restricted curves $\bar \gamma=\gamma_{|[a_\gamma,b_\gamma]}$ are all at distance at least $3\newww{\delta}$ from each other.
		\item \label{epdeltachoice} For any two points $p,q$ in a common face of $D$ and both at distance at least $\ep$ from any curve of $D$, there is a curve $\gamma$ (not a curve of $D$) joining $p$ to $q$ whose distance to every curve of $D$ is at least $3\delta$.
	\end{enumerate}
	\end{lemma}
	\noindent This is a straightforward consequence of compactness of the curves \new{$\gamma \in \Gamma$}; we include a proof in the appendix for completeness.
	
	\subsection{Multiplicative Approximate Balance}
	\label{sec:modified_wilson}
	~\new{In this section we prove Theorem \ref{thm:planegraph}, our multiplicative approximate balance result for lattice sequences and arbitrary plane graphs $D$. } From here on, we let $\ep,\delta$ be as promised by Lemma \ref{lem:epdelta}. We call a curve of $D$ an \emph{outer curve} if every point of the curve lies on the boundary of the outer face, and an \emph{inner curve} if no point does, other than possibly its endpoints.  Note that every curve must be one of these two types.    We let $\Gamma^I$ and $\Gamma^O$ denote set of inner curves and outer curves, respectively.  
	
\begin{figure}[t]\centering
\begingroup%
  \makeatletter%
  \providecommand\color[2][]{%
    \errmessage{(Inkscape) Color is used for the text in Inkscape, but the package 'color.sty' is not loaded}%
    \renewcommand\color[2][]{}%
  }%
  \providecommand\transparent[1]{%
    \errmessage{(Inkscape) Transparency is used (non-zero) for the text in Inkscape, but the package 'transparent.sty' is not loaded}%
    \renewcommand\transparent[1]{}%
  }%
  \providecommand\rotatebox[2]{#2}%
  \newcommand*\fsize{\dimexpr\f@size pt\relax}%
  \newcommand*\lineheight[1]{\fontsize{\fsize}{#1\fsize}\selectfont}%
  \ifx\svgwidth\undefined%
    \setlength{\unitlength}{302.32704163bp}%
    \ifx\svgscale\undefined%
      \relax%
    \else%
      \setlength{\unitlength}{\unitlength * \real{\svgscale}}%
    \fi%
  \else%
    \setlength{\unitlength}{\svgwidth}%
  \fi%
  \global\let\svgwidth\undefined%
  \global\let\svgscale\undefined%
  \makeatother%
  \begin{picture}(1,0.82247944)%
    \lineheight{1}%
    \setlength\tabcolsep{0pt}%
    \put(0,0){\includegraphics[width=\unitlength,page=1]{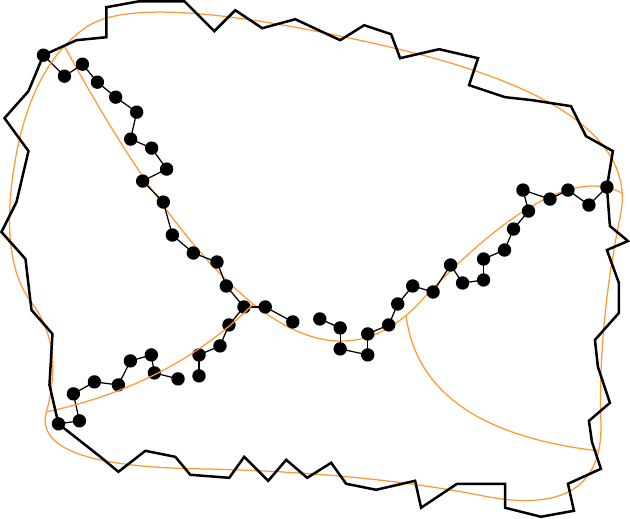}}%
    \put(0.14081938,0.57123606){\color[rgb]{1,0.6,0.2}\makebox(0,0)[lt]{\lineheight{1.25}\smash{\begin{tabular}[t]{l}$\gamma_1^I$\end{tabular}}}}%
    \put(0.20654833,0.17717535){\color[rgb]{1,0.6,0.2}\makebox(0,0)[lt]{\lineheight{1.25}\smash{\begin{tabular}[t]{l}$\gamma_2^I$\end{tabular}}}}%
    \put(0.44912814,0.2502235){\color[rgb]{1,0.6,0.2}\makebox(0,0)[lt]{\lineheight{1.25}\smash{\begin{tabular}[t]{l}$\gamma_4^I$\end{tabular}}}}%
    \put(0.71833815,0.46226981){\color[rgb]{1,0.6,0.2}\makebox(0,0)[lt]{\lineheight{1.25}\smash{\begin{tabular}[t]{l}$\gamma_3^I$\end{tabular}}}}%
    \put(0.75556998,0.17291822){\color[rgb]{1,0.6,0.2}\makebox(0,0)[lt]{\lineheight{1.25}\smash{\begin{tabular}[t]{l}$\gamma_5^I$\end{tabular}}}}%
    \put(0.76629708,0.75013215){\color[rgb]{0,0,0}\makebox(0,0)[lt]{\lineheight{1.25}\smash{\begin{tabular}[t]{l}$C^*$\end{tabular}}}}%
  \end{picture}%
\endgroup%

	\caption{A run of the first four phases of Wilson's algorithm approximating the first four inner curves $\gamma_1^I, \gamma_2^I, \gamma_3^I, \gamma_4^I$. Note that paths corresponding to $\gamma_2^I$ and $\gamma_4^I$ each are missing an edge, and in particular, the edges present do not disconnect the interior of $C^*$.  These missing edges are dual to the edges in the corresponding primal spanning tree whose removal would disconnect the tree into components approximating the faces of this drawing.}
	\label{figGammaPartition}
	\end{figure}

	\new{We now order the inner curves of $D$ such that the endpoint $\gamma(1)$ of each curve is contained in the previous curves or an outer curve\newww{; this is} possible since $D$ is connected. This will be the order in which we consider random walks resembling these curves. Formally,} we first order the inner curves of $D$ as $\gamma^{I}_1,\dots,\gamma^{I}_{\ell}$ such that for all $i$, the plane graph $D_i$ of $D$ with edges
	\[
	\Gamma_i=\Gamma^O\cup \{\gamma^{I}_1,\dots,\gamma^{I}_i\}
	\]
	and vertex set $V_i=\{\gamma(x)\mid \gamma\in \Gamma_i, x\in \{0,1\}\}$
	is connected (see Figure \ref{figGammaPartition}). \new{Additionally,} without loss of generality, we assume that the orientation of each curve is such that $\gamma^I_{i}(1)$ is a vertex of $\Gamma_{i-1}$ for all $i=1,\dots,\ell$ (the curves are oriented ``towards the outer face''). \neww{We also choose the order so that whenever a vertex not incident to an outer curve first appears, it appears as the initial endpoint of the curve that introduces it; equivalently, each curve introducing a new vertex is oriented from that new vertex toward the already constructed graph.}
	
	\new{At a high level, } we prove the theorem by analyzing how Wilson's algorithm constructs spanning trees of $\Omega^*_{D,\Lambda_n}$ with wired boundary conditions, where the whole boundary cycle $C^*$ is used as the root of Wilson's algorithm\newww{.} In particular, it is equivalent to view Wilson's algorithm as building a unicylic graph, initialized with the boundary cycle of $\Omega_{D,\Lambda_n}^*$.

	The freedom to choose the starts of loop-erased random walks in Wilson's algorithm allows us to use the following implementation, which determines \new{the start of each random walk} by using additional random walks. We proceed in phases, creating a sequence of trees $T_i^j$ where the subscript $i$ denotes the phase and the superscript $j$ denotes the step within phase $i$. By $T_i$ we will denote the tree at the end of phase $i$, and note $T_{i+1}^0 = T_i$. For simplicity, we use $T_0$ to denote just the root $C^*$.   \new{Each phase $i$ corresponds to approximately tracing out one inner curve $\gamma^I_i \in \Gamma^I$ between two vertices of $D$. Each phase begins at a source vertex in \newww{$\Lambda^*_n$} near $\gamma^I_i(0)$, and ends when a vertex that is already in $T_{i-1}$ and is near $\gamma^I_i(1)$ is reached.} 
	
	Within each phase $i$, we alternate a loop-erased random walk from a vertex outside $T_i^j$ to a vertex in $T_i^j$ that gets added to the tree to obtain  $T_i^{j+1}$ (one \neww{loop-erased random walk} of Wilson's algorithm) with a random walk among the vertices in $T_i^{j+1}$ until a vertex outside $T_i^{j+1}$ is reached (choosing the starting point for the next step of Wilson's algorithm). Here we describe the procedure for one phase of this process.
	
	\new{{\bf In Phase $i$:} At the start of phase $i$, we have existing tree $T_{i-1}$ that has already been built (For $i=1$, $T_0$ consists of just the root vertex). In this phase, we:}
	\begin{enumerate}
	\item \label{case.source}\new{Choose an initial source vertex $v$: 
		\begin{enumerate}[(A)]
			\item \label{case.source.empty} If $T_{i-1} \cap \newww{B_{\ep}} ( \gamma_i^I(0)\neww{)} = \emptyset$, let $v$ be a vertex at minimum distance from $\gamma_i^I(0)$.
			\item  \label{case.source.not.empty} If $T_{i-1} \cap \newww{B_{\ep}} ( \gamma_i^I(0)\neww{)} \neq \emptyset$, \rev{let $v$ be any vertex of $T_{i-1}\cap B_{\delta/5}(\gamma_i^I(0))$ if this set is nonempty; if it is empty, choose an arbitrary source vertex, since this fallback case will not be used in the successful event analyzed below.} 
		\end{enumerate}
		\item Select target set $U_i = T_{i-1} \cap \newww{B_{\ep}}( \gamma_i^I(1))$. }
	\item Initialize $T_{i}^0=T_{i-1}$.
	\item We begin each step of this phase with $T_{i}^{j}$ (at the beginning of the phase, for $j=0$) and a source vertex.  We do one of two things according to whether the source belongs to the tree $T_{i}^{j}$: \begin{enumerate}
		\item \label{case.notin} If the source vertex is not in $T_{i}^{j}$, we conduct a loop-erased random walk from the source until it hits $T_{i}^{j}$ at a vertex $u$.  This loop-erased random walk is added to $T_{i}^{j}$ to create $T_{i}^{j+1}$.  If $u \in U_i$\neww{, then} this phase ends, and we set $T_{i+1}=T_{i}^{j+1}$.  Otherwise, we increment $j$, and continue this phase with $u$ as the new source vertex (we will be in case (b) next).
		\item \label{case.in} If the source vertex is in $T_{i}^{j}$, we take a random walk from the source until we reach a vertex $u$ outside of $T_{i}^{j}$, and then increment $j$ and restart this step from the vertex $u$ as the new source vertex, and $T_{i}^{j+1}=T_{i}^{j}$ (we will be in case (a) next).
	\end{enumerate} 
	\item The previous loop continues until either the target is eventually hit by an instance of the loop-erased random walk, or the entire spanning tree is completed.
	\end{enumerate}
	
	Note that we can use a single random walk $W_i$ \rev{from the chosen source vertex $v$ (which may already lie in $T_{i-1}$ in Step~\ref{case.source.not.empty})} to implement each phase of the algorithm (with loop erasure while in case \eqref{case.notin}, and without loop erasure while in case \eqref{case.in}).  \new{This means we will be able to use Lemma~\ref{lemCurveWalkSimple} to bound the probability that the additions to $T_i$ to form $T_{i+1}$ approximately trace out $\gamma_{i}$.} 
	
	\new{The following observation about this process for general graphs formalizes this intuition.  While we prove a general claim, the reader should think of the set $S$ in this observation as the set of vertices close to curve $\gamma_i$; set $S$ will be made precise later. While it may not be immediately obvious to the reader, we will later see that case \eqref{case.genwalkwilson.empty} of this observation corresponds to a source chosen as in Step \ref{case.source.empty} of the above process, and case \eqref{case.genwalkwilson.not.empty}  of this observation corresponds to a source chosen as in  Step \ref{case.source.not.empty} of the above process.}

	\begin{observation}\label{obs:genwalkwilson}
	Suppose that we run the implementation of Wilson's algorithm above on a graph $G$, and have built the tree $T_{i-1}$ after the first $i-1$ phases. 
	
	\begin{enumerate}[(A)]
		\item \label{case.genwalkwilson.empty} For a connected set of vertices $S$, if the walk $W_{i}$ begins from a vertex $v\in S$ and ends phase $i$ by hitting $T_{i-1}$ for the first time at the target $U_{i}\subseteq S$ and without leaving $S$, then after this phase, there is a path $P$ in \newww{$T_{i} \cap S$} joining $v$ to $U_{i}$.
		
		\item \label{case.genwalkwilson.not.empty} For a connected set of vertices $S$, suppose the walk $W_{i}$ begins \rev{either from a vertex $w \in (T_{i-1}\cap S) \setminus U_i$ or from a vertex $v\in S$ adjacent to such a vertex $w$}, and ends phase $i$ by hitting the target $U_{i} \subseteq S$ without leaving $S$. Then after this phase, there is a path $P$ from a vertex $v'\in (S\cap T_{i-1})\setminus U_{i}$ to $U_{i}$, all of whose vertices belong to $S$, and all but one of whose edges belongs to $T_{i}$. 
	\end{enumerate}
	\end{observation}
	\begin{proof}
	In (A), the path $P$ consists of loop-erased $W_{i}$ from $v$ until the first time it reaches $T_{i-1}$\new{; this entire path is added to $T_{i-1}$ to form $T_{i}$. }
	
	In (B), we construct the desired path $P$ backwards from $U_{i}$.  Let $Q$ be the last loop-erased part of $W_{i}$ before it reaches $U_{i}$, corresponding to an entire step as described in Case \eqref{case.notin} of our implementation of Wilson's algorithm. We first add $Q$ to $P$. 
	Note $Q \subseteq T_{i}$, and it is entirely contained in $S$ because it is a subset of $W_{i}$ and all vertices of $W_{i}$ are in $S$. 
	
	Let $y$ be the first vertex of $Q$. \rev{If $y$ is the first vertex of $W_i$, then we are in the adjacent-start alternative above: $y$ is adjacent to $w$, and adding $w$ to the start of $Q$ produces a path $P$ with the desired properties where $v'=w$.}
	
	Otherwise, let $x$ be the vertex preceding $y$ in $W_{i}$. \new{As $x \in W_i$, we know $x \in S$; as phase $i$ continues after reaching $x$, we know $x \notin U_i$.  When walk $W_i$ reaches $x$, the current tree is $T_i^j = T_i \setminus Q$, as $Q$ has not yet been added.  It must be that $x \in T_{i} \setminus Q$ and there is no edge in $T_{i}$ between $x$ and $y$; this is because $x$ is the penultimate vertex in a step (as described in case \eqref{case.in} of our implementation of Wilson's algorithm) of $W_{i}$ that is entirely contained in $T_{i} \setminus Q$ until its last vertex, $y$, is outside $T_{i} \setminus Q$.} 
	
	\new{If $x \in T_{i-1}$, adding $x$ to the start of $Q$ produces a path $P$ with the desired properties where $v'=x$. Therefore, we assume $x \notin T_{i-1}$.     
		It must be that $x$ was added to the tree as part of a loop-erased random walk in some previous step of phase $i$. Loop-erased walks are only added to the tree when they reach a vertex already in the tree, and we know no such loop erased random walks reached $U_i \subseteq T_{i-1}$ as the phase had not concluded before reaching $x$, so there must exist a path of newly-added edges (possibly from multiple loop-erased walks) in $T_{i} \setminus T_{i-1}$ from $x$ to some vertex $v' \in T_{i-1} \setminus U_i$.  As all vertices in $T_i \setminus T_{i-1}$ are vertices of $W_i$ and $W_i \subseteq S$, it follows that there exists a path in $T_i \cap S$ between $x$ and $v'$.} 
	We then let path $P$  consists of the path from $v'$ to $x$ in $T_{i} \cap S$, the edge $\{x,y\}$, and path $Q$ from $y$ to $U_{i}$. All vertices of this path are in $S$, and all edges except $\{x,y\}$ are in $T_{i}$, as desired.  
	\end{proof}
	
	\new{The following result specializes this observation to the case where $S$ is all vertices within distance $\delta$ of curve $\gamma_i$, with additional hypothesis and results about the start of $W_i$ being in $\newww{B_{\ep}}(\gamma_{i}(0))$. }

	\begin{observation}\label{obs:walkwilson}    Suppose that we run the implementation of Wilson's algorithm above on the dual graph $\Omega^*_{D,\Lambda_n}$, and have built the tree $T_{i-1}$ after the first $i-1$ phases. Then:  
	\begin{enumerate}[(A)]
		\item \label{firstcase} If the walk $W_{i}$ begins from a vertex $v$ and ends phase $i$ by hitting $T_{i-1}$ for the first time at the target $U_{i}$ while also staying within distance $\delta$ of the curve $\gamma_{i}$, then after this phase, there is a path $P$ in $T_{i}$ joining $v$ to $U_{i}$ whose vertices are all within distance $\delta$ from the curve $\gamma_{i}$.
		
		\item \label{latercase} Suppose the walk $W_{i}$ begins \rev{from a vertex of $T_{i-1}\cap B_{\delta/5}(\gamma_i(0))$} and ends phase $i$ by hitting the target $U_{i}$ while also staying within distance $\delta$ of the curve $\gamma_{i}$ throughout the phase. Suppose all vertices of $T_{i-1}$ that are within $\delta$ of the curve $\gamma_{i}$ are in either $\new{\newww{B_{\ep}}(\gamma_{i}(0))}$ or \new{$\newww{B_{\ep}}(\gamma_{i}(1))$}. Then after this phase, there is a path $P$ in the dual graph joining some vertex in $T_{i-1} \cap \newww{B_{\ep}}(\gamma_{i}(0))$ to $U_{i}$ whose vertices are all within distance $\delta$ from the curve $\gamma_{i}$, and such that \neww{all but one edge} of $P$ belongs to the tree $T_{i}$.
	\end{enumerate}
	\end{observation}
	\begin{proof}
	This follows from Observation \ref{obs:genwalkwilson}, with $G=\Omega^*_{D,\Lambda_n}$ and $S$ \neww{taken} to be all the vertices of $\Omega^*_{D,\Lambda_n}$ within distance $\delta$ of the curve $\gamma_{i}$. \rev{In Case~\ref{latercase}, the starting vertex lies in $(T_{i-1}\cap S)\setminus U_i$, since Lemma~\ref{lem:epdelta}\eqref{epchoicevxs} separates the two endpoint balls.} \new{In Case \ref{latercase}, the fact that the path stretches from $T_{i-1} \cap \newww{B_{\ep}}(\gamma_{i}(0))$ to $U_i$ follows from the fact that \newww{$v' \in (T_{i-1}\cap S) \setminus U_i$}, and by hypothesis all such vertices are in $T_{i-1} \cap \newww{B_{\ep}}(\gamma_{i}(0))$.}
	\end{proof}
	
	\new{For a given plane graph $D$, we now consider how partial tree structures in the dual graph $\Omega^*_{D,\Lambda_n}$ lead to spanning trees in the primal graph $\Omega_{D, \Lambda_n}$ that correspond to $D$. The following definition formalizes the dual graph structures we will be interested in.  }  
	\begin{definition} \label{def.corr}\new{Let $D$ be a plane graph with $k +1$ faces.} 
	Let $T$ be a (not-necessarily-spanning) tree in the dual graph $\Omega^*_{D,\Lambda_n}$. We say $T$ \newww{$(\delta,\ep)$-\emph{corresponds}} to a plane graph $D$ with inner curves $\gamma^I_1,\dots,\gamma^I_{\ell}$ if there are paths $P_1,\dots,P_{\ell}$ in $\Omega^*_{D,\Lambda_n}$, such that \neww{the following conditions are satisfied} for each $i$:
	\begin{enumerate}[(a)]
		\item \label{corr.allbutone} For \neww{$k-1$} of the paths, all but one edge of $P_i$ belongs to the tree $T$, while for the rest of the paths, the whole path belongs to $T$. \rev{Moreover, after deleting these $k-1$ missing edges from $\bigcup_i P_i$, the remaining path-union is a tree contained in $T$.} (Intuitively, $T$ is the dual tree that partitions the lattice, and the $P_i$ are obtained by adding the ``missing'' edges whose duals are the split edges of the tree in the primal graph, such as the gap in the center of Figure \ref{figGammaPartition}.)
		\item \label{corr.intersectiff} $P_i$ and $P_j$ intersect if and only if $\gamma^I_i$ and $\gamma^I_j$ share a common endpoint.
		\item \label{corr.intersectorder} For any point $p$, if $\gamma^I_{i_1},\dots,\gamma^I_{i_s}$ are all the curves of $D$ that have $p$ as an endpoint, then the union of the paths $P_{i_1},\dots,P_{i_s}$ is a tree.
		\item \label{corr.close} Every vertex of $P_i$ is within distance $\delta$ of the curve $\gamma^I_i$, \newww{or within $\ep$ of one of the endpoints of $\gamma^I_i$.}
	\end{enumerate}  
	\end{definition}
	
	We then have the following lemma, which explicitly relates \newww{$(\delta,\ep)$-correspondence} in $\Omega_{D,\Lambda_n}^*$ to $\ep$-compatibility in  $\Omega_{D,\Lambda_n}$. \new{Recall that in a plane graph with $k+1$ faces, there are $k$ bounded faces, and removal of $k-1$ cutedges from a spanning subgraph suffices to separate it into $k$ components. }
	
	\begin{lemma}\label{lem:treetopartition}
	\new{Let $D$ be a plane graph with $k+1$ faces, and} let $T$ be a tree in the dual graph $\Omega_{D,\Lambda_n}^*$, and let $H_T$ be the spanning subgraph of the primal graph \neww{$\Omega_{D,\Lambda_n}$} obtained by removing from $\Omega_{D,\Lambda_n}$ all the edges $e$ for which $e^*\in T$.  If $T$ \newww{$(\delta,\ep)$-corresponds} to $D$, then there are $k-1$ edges of $H_T$ whose removal results in connected components $C_1,\dots,C_k$ \neww{which induce a partition of $\Omega_{D,\Lambda_n}$ that} is $\ep$-compatible with $D$.
	\end{lemma}
	
	\new{We defer the proof of this lemma to the end of this subsection, and proceed first to using it to prove our main result, which has been reduced to bounding the probability that our algorithm produces a tree $T$ that $\delta$-corresponds to $D$.}
	
	\begin{proof}[Proof of Theorem~\ref{thm:planegraph}]
	\noindent We are given a plane graph $D$ and desired multiplicative approximation factor, and have chosen $\delta,\ep$ so that the conditions of Lemma~\ref{lem:epdelta} are satisfied. \new{Next, fix a bounded set $K$ containing the $\delta$-neighborhood of the drawing $D$ and \rev{of the bounded region enclosed by the outer boundary of $D$; since $C^*$ is at Hausdorff distance $<\delta/10$ from that boundary, this same fixed $K$ contains every admissible $C^*$ for sufficiently small $\delta$}. By the localized definition of a lattice sequence, let $n$ be large enough such that for any adjacent pair $x,y \in V(\Lambda_n^*)$ with $x\in K$ or $y\in K$, $d(x,y) < \delta/100$, and such that for any $p \in K$, the ball $B_{\delta/5}(p)$ contains a vertex of $\Lambda_n^*$.  Our results will hold for all such $\Lambda_n$\newww{. Since they hold for all sufficiently large $n$, they hold as $n \rightarrow \infty$. }}

	\new{If $D$ has $\ell$ inner curves labeled as $\gamma_1^I, \gamma_2^I, \ldots, \gamma_\ell^I$, recall that $D_i = \Gamma^O \cup \gamma_1^I \cup \gamma_2^I \cup \ldots \cup \gamma_i^I$, with $D_\ell = D$. Let $f(D_i)$ be the number of interior faces of plane graph $D_i$, where $f(D_\ell) = f(D) = k$.}
	
	\neww{We} use Lemma \ref{lem:treetopartition} and apply induction on the sequence of plane graphs $D_i$. Initially, tree $T_0$ consisting of just the root vertex of $\Omega_{D, \Lambda_n}^*$ (equivalently, of just the wired boundary cycle of $\Omega_{D, \Lambda_n}^*$) trivially $\newww{(\delta,\ep)}$-corresponds to the plane graph $D_0$ with no interior curves.

	We use the following induction invariant in addition to $(\delta,\ep)$-correspondence.  For every vertex $p$ of $D_i$ that is not on the outer face, $T_i$ contains a vertex in $B_{\delta/5}(p)$; for every vertex $p$ of $D_i$ that is on the outer face, the boundary cycle $C^*$ contains a vertex in $B_{\delta/5}(p)$.  The outer-face part holds by the Hausdorff assumption on $C^*$ and the short-edge assumption, and the non-outer part is initially vacuous.
	
	Suppose that $T_{i-1}$ $(\delta,\ep)$-corresponds to $D_{i-1}$ and satisfies this endpoint-cluster invariant, and begin phase $i$ as in Steps~\ref{case.source.empty}--\ref{case.source.not.empty}.  Let $q=\gamma_i^I(0)$, let $p=\gamma_i^I(1)$, and let $U_i=T_{i-1}\cap B_\ep(p)$.  We first check that the selected source vertex is within distance $\delta/2$ of $q$, so that Lemma~\ref{lem:curveswalk} can be applied with $\delta$ in place of $\ep$.  If Step~\ref{case.source.empty} is used, then $q$ is a new vertex of the partial drawing and the closest dual vertex to $q$ is within $\delta/5$ by the local density condition.  If Step~\ref{case.source.not.empty} is used, then $q$ is already represented either by the boundary cycle $C^*$ or by the endpoint-cluster invariant, and the step chooses $v\in T_{i-1}\cap B_{\delta/5}(q)$.  In both cases $d(v,q)\le \delta/5<\delta/2$.
	
	By Lemma~\ref{lem:curveswalk}, with probability at least
	\[
	\rho^{100|\gamma_i^I|/\delta+\rev{501}},
	\]
	the walk $W_i$ stays within distance $\delta$ of $\gamma_i^I$ and makes an encircling subpath contained in $B_{\delta}(p)\setminus B_{\delta/5}(p)$.  We claim that on this event phase $i$ hits $U_i$ and that the resulting tree $T_i$ $(\delta,\ep)$-corresponds to $D_i$ while preserving the endpoint-cluster invariant.
	
	First suppose that $p$ lies on the outer face of $D$.  The boundary cycle $C^*$ is at Hausdorff distance at most $\delta/10$ from the boundary of the outer face, and consecutive vertices of $C^*$ are at distance at most $\delta/100$.  Hence some vertex of $C^*$ lies in $B_{\delta/5}(p)$.  A walk inside $\Omega^*_{D,\Lambda_n}$ cannot encircle this boundary-cycle vertex without first hitting $C^*$.  Since the encircling portion of $W_i$ is contained in $B_{\delta}(p)\subseteq B_\ep(p)$, this hit of $C^*$ is a hit of $U_i$.
	
	Now suppose that $p$ is not on the outer face.  Since $p=\gamma_i^I(1)$ belongs to the previous drawing $D_{i-1}$, the endpoint-cluster invariant gives a vertex of $T_{i-1}\cap B_{\delta/5}(p)$.  The tree $T_{i-1}$ also contains the root $C^*$, which is outside $B_{\delta}(p)$ by the separation assumptions.  The unique path in $T_{i-1}$ between these two vertices must cross any curve in $B_{\delta}(p)\setminus B_{\delta/5}(p)$ that encircles $p$.  Therefore the encircling portion of $W_i$ intersects $T_{i-1}$, and because this portion is contained in $B_{\delta}(p)\subseteq B_\ep(p)$, the intersection is a hit of $U_i$.
	
	It remains to check that the correspondence is updated correctly.  By Lemma~\ref{lem:epdelta} and the induction hypothesis, every vertex of $T_{i-1}$ within distance $\delta$ of the restricted part of $\gamma_i^I$ is contained in one of the two endpoint balls $B_\ep(\gamma_i^I(0))$ or $B_\ep(\gamma_i^I(1))$.  Thus Observation~\ref{obs:walkwilson} applies.  If Step~\ref{case.source.empty} is used, Observation~\ref{firstcase} gives a new path entirely contained in $T_i$, and adding $\gamma_i^I$ to $D_{i-1}$ attaches a new vertex to the existing drawing, so $f(D_i)=f(D_{i-1})$.  If Step~\ref{case.source.not.empty} is used, Observation~\ref{latercase} gives a path with exactly one edge missing from $T_i$, and adding $\gamma_i^I$ connects two vertices already present in the connected drawing, so Euler's formula gives $f(D_i)=f(D_{i-1})+1$.  The separation conclusions of Lemma~\ref{lem:epdelta} ensure that the new path intersects the previous paths exactly at the endpoint clusters prescribed in Definition~\ref{def.corr}.  Finally, if $q$ was a new non-outer vertex, the source chosen in Step~\ref{case.source.empty} lies in $B_{\delta/5}(q)$ and is added to the tree in this phase; all other endpoint clusters are preserved because vertices are never removed from the Wilson tree.  Hence $T_i$ $(\delta,\ep)$-corresponds to $D_i$ and the endpoint-cluster invariant holds for the next phase.

	In particular, after phase $i = \ell$ (recall $\ell$ is the total number of interior curves in $D$) we have that with \new{probability at least 
		\begin{equation} \label{eq:mult_prob}
			\rho^{\sum_{i = 1}^\ell \left(100|\gamma^I_i|/\delta +\rev{501} \right) }
	\end{equation}} our tree $T_{\ell}$ consists only of the root and vertices within distance $\delta$ of the internal curves of $D = D_\ell$, and there is a collection of paths $P_1,\dots,P_{\ell}$ satisfying properties (a), (b), (c), and (d) above, and additionally all but precisely $f(D)-1 = k-1$ of the paths belong entirely to tree $T_{\ell}$ \rev{while deleting the missing edges leaves a path-union tree contained in $T_\ell$}. 
	
	From here, we complete Wilson's algorithm with arbitrary choices for starting vertices to produce a final tree $T$, which still contains all but at most one edge of each path $P_i$, and the whole path in all but precisely $k-1$ cases.  In particular, in the primal graph, the tree $T$ corresponds to a tree from which $k-1$ edges can be deleted, to produce a partition that is $\ep$-compatible with the drawing $D$. \new{This concludes our proof of Theorem~\ref{thm:planegraph}. }
	\end{proof}
	
	\noindent It only remains to prove Lemma \ref{lem:treetopartition}\new{, which we now do}. 
	
	\begin{proof}[Proof of Lemma \ref{lem:treetopartition}]
	
	Let $k=f(D)$, and let $P_1,\dots,P_\ell$ be the paths from Definition~\ref{def.corr}.  The case $k=1$ is immediate, so assume $k>1$.  Let $Q=\bigcup_i P_i$, and let $P$ denote $Q$ together with the wired boundary cycle $C^*$.  Let $M$ be the set of edges of the paths $P_i$ that do not belong to $T$.  By Definition~\ref{def.corr}\eqref{corr.allbutone}, $|M|=k-1$.  The primal edges dual to the edges of $M$ belong to $H_T$, and these will be the $k-1$ edges removed from $H_T$.
	
	First remove from $\Omega_{D,\Lambda_n}$ every primal edge whose dual lies in one of the paths $P_i$.  The resulting components are the lattice regions corresponding to the bounded faces of the plane graph $P$.  We claim that $P$ has exactly the same bounded-face structure as $D$.  Around each vertex $p$ of $D$, Definition~\ref{def.corr}\eqref{corr.intersectiff}--\eqref{corr.intersectorder} says that the paths corresponding to curves incident to $p$ form a tree; by Lemma~\ref{lem:epdelta} and Definition~\ref{def.corr}\eqref{corr.close}, this tree is contained in the $\ep$-neighborhood of $p$.  After contracting each such endpoint tree to a point, the remaining middle portions of the paths are pairwise disjoint and are in one-to-one correspondence with the inner curves of $D$.  Adding these middle portions one at a time either joins a new vertex to the existing drawing or creates one new bounded face, exactly as adding the corresponding curves to $D$ does.  Hence $P$ has $k$ bounded faces.
	
	The same separation gives $\ep$-compatibility.  Points of a bounded face of $D$ that are at distance greater than $\ep$ from the curves of $D$ can be joined inside that face by curves staying at distance at least $3\delta$ from the curves of $D$, by Lemma~\ref{lem:epdelta}\eqref{epdeltachoice}; such curves do not cross any $P_i$, since each $P_i$ lies within distance $\delta$ of the corresponding curve away from endpoint balls.  Thus each component obtained after deleting the path-dual edges is associated to one bounded face of $D$, and any vertex assigned to that component but not lying in the face itself is within the allowed $\ep$-neighborhood of that face.
	
	It remains only to check that deleting the additional primal edges whose duals lie in $T\setminus Q$ does not split these components further.  By Definition~\ref{def.corr}\eqref{corr.allbutone}, $Q\setminus M$ is a tree contained in the wired dual tree $T$.  Since $T$ is a tree after the boundary cycle is wired to the root, every component of $T\setminus (Q\setminus M)$ attaches to $(Q\setminus M)\cup C^*$ in at most one wired vertex; otherwise the wired tree would contain a cycle.  Thus the extra dual edges of $T\setminus Q$ are only dangling trees attached to the already present dual drawing.  Adding such dangling trees to a plane drawing does not change its bounded faces.  Equivalently, their dual primal deletions do not further split any component already cut out by the paths $P_i$ and the $k-1$ primal edges dual to $M$.  Consequently $H_T$ with the $k-1$ primal edges dual to $M$ removed has exactly the $k$ $\ep$-compatible components described above.
	
	\end{proof}
	
	Corollary \ref{cor:ep} now follows with some additional observations about how the parameters of Theorem \ref{thm:planegraph} applies to grids. 
	

	\begin{proof}[Proof of Corollary \ref{cor:ep}]
	
	Let $G$ be an $m\times n$ grid graph with $n\le m$, and scale it isotropically by $1/n$.  Thus the embedded grid lies in the rectangle $[0,m/n]\times[0,1]$ and is a subgraph of the lattice sequence $\{(1/\ell)\mathbb Z^2\}$, whose dual random-walk constant is $\rho=1/8$, independently of the aspect ratio $m/n$.
	
	Let $D$ be the rectangle $[0,m/n]\times[0,1]$ divided vertically into $k$ strips by placing the $q$th separating curve between columns $qm/k$ and $qm/k+1$ of the scaled grid, for $q=1,\ldots,k-1$. Thus each strip contains exactly $mn/k$ grid vertices. \rev{Set $\ep=\xi/(3k)$ and $\delta=\xi/(4k)$.}  Since $m/n\ge1$, the vertical separating curves of $D$ are pairwise separated, and separated from the vertical sides of the rectangle, by at least $1/k$; the assumption $\xi\le2/3$ ensures that Lemma~\ref{lem:epdelta} applies with this choice of $\ep$ and $\delta$.  The proof of Theorem~\ref{thm:planegraph} only requires the lattice spacing in the relevant bounded rectangle to be at most $\delta/100$ and the $\delta/5$-density condition there.  At scale $1/n$, this is guaranteed by
	\[
	n\ge 100/\delta=400k/\xi,
	\]
	which is exactly the lower bound on $\xi$ in the statement.
	
	Therefore Theorem~\ref{thm:planegraph} applies to this scaled copy of $G$.  The probability lower bound obtained in its proof depends on $\rho$, on $\delta$, and on the total length of the inner curves of $D$.  Here $\rho=1/8$, $\delta=\xi/(4k)$, and the total inner-curve length is $k-1$; the relevant curve separations are bounded below by $1/k$.  Hence the lower bound depends only on $k$ and $\xi$, not on $m/n$.
	
	It remains to check that $\ep$-compatibility implies $(k,\xi)$-approximate splittability.  Each vertical strip of $D$ contains exactly $mn/k$ grid vertices.  If a component is $\ep$-compatible with the $i$th strip, then any vertices it contains outside that strip lie in at most two vertical strips of Euclidean width $\ep$ and height $1$ in the scaled drawing.  \rev{Because the grid spacing is $1/n$, these two strips contain at most $(2\ep n+4)n$ vertices; the additive $4n$ term accounts for the boundary columns created by rounding to lattice columns.  With $\ep=\xi/(3k)$ and $\xi\ge 400k/n$, this is at most $\xi mn/k$ since $m\ge n$.}  The same estimate bounds the number of vertices of the $i$th strip that may be assigned to neighboring components.  Thus every component has between $(1-\xi)mn/k$ and $(1+\xi)mn/k$ vertices.  Since $mn=|V(G)|$, the spanning tree is $(k,\xi)$-approximately splittable.
	\end{proof}

	\subsection{Additive Approximate Balance}\label{subAdditive}
	
	We now turn to the proof of Theorem \ref{thmAdditiveError}. We begin by observing several useful facts implied by the definition of a uniform lattice sequence.
	
	\begin{observation}\label{obsAdditiveErrorRectangles}
	Let $(\{\Lambda_n\}, \rho, R, C_1, C_2)$ be a uniform lattice sequence. There exist positive constants $\cmin, \cmax$ such that, for all sufficiently large $w, h, n$, any $\frac{w}{n} \times \frac{h}{n}$ rectangle contains between $\cmin \cdot w \cdot h$ and $\cmax \cdot w \cdot h$ vertices in $\Lambda_n$.
	\end{observation}
	
	\begin{proof}
	Let $n$ be large enough so that the conditions of Definition~\ref{defStrongLatticeSequence} apply, and let $h, w \geq 4R$. \rev{In this proof we use the primal-vertex part of property~\eqref{itmStrongLatticeSequenceDense}.} 
	
	This rectangle contains at least $\lfloor w/(2R) \rfloor \times \lfloor h/(2R) \rfloor$ disjoint circles of radius $R/n$ placed in an evenly-spaced grid, and each must contain at least $C_1$ points. Note when $w \geq 4R$, then $\lfloor w/(2R) \rfloor \geq w/(2R) - 1 \geq w/(2R) - w/(4R) = w/(4R)$, and the same is true for $h$.  
	Therefore the number of points in this rectangle is least $C_1 \lfloor w/(2R) \rfloor \lfloor h/(2R) \rfloor \geq C_1 \left(\frac{w}{4R} \right) \left(\frac{h}{4R}\right)$.  Picking $\cmin = C_1 / 16 R^2$ ensures the desired lower bound is true. 
	
	This rectangle can be completely covered by the union of $\lceil w/R\rceil \times \lceil h/R \rceil$ balls of radius of $R/n$ placed in an evenly-spaced grid.  As each ball contains at most $C_2$ vertices, the  number of vertices in this box is at most $\lceil w/R\rceil \cdot \lceil h/R \rceil \cdot C_2 \leq \left(\frac{2w}{R} \right) \left( \frac{2h}{R}\right) C_2$.  Picking $\cmax = 4  C_2 / R^2$ ensures the desired upper bound is true. 
	\end{proof}
	
	Given $\cmin$ and $\cmax$, which depend only on the uniform lattice sequence, we define constants
	\begin{align*}
	c_1 &:= 1 - \left(\frac{1}{50} \cdot \frac{\cmin}{\cmax}\right),\\
	c_2 &:= \frac{1}{50} \cdot \frac{\cmin}{\cmax}.
	\end{align*}
	Term $c_1$ is meant to describe the rate at which a sequence of rectangles constructed in the proof are shrinking. Term $c_2$ is meant to describe the width of the interval through which we must leave each rectangle we consider. While $c_1 = 1-c_2$, this is simply a consequence of trying to choose constants that are convenient to work with; we will continue to use $c_1$ and $c_2$ in accordance with their conceptual meanings, as just described, rather than substituting one for another.  
	
	\begin{observation}\label{obsAdditiveErrorList}
	Let $(\{\Lambda_n\}, \rho, R, C_1, C_2)$ be a uniform lattice sequence, and let $\cmin, \cmax, c_1, c_2$ be as above. There exist positive constants $\rhat, R''\rev{,} \rho', \rho''$ such that the following hold for all sufficiently large~$n$. 
	\begin{enumerate}[({B}1)]
		\item\label{itmOAE1} Observation \ref{obsAdditiveErrorRectangles} holds for any $w, h \geq \frac{(1 - c_1)\rhat}{2}$.
		\item\label{itmOAE5} $\rhat \geq 50R$.
		\item\label{itmOAE4} $\rhat \geq \frac{c_1}{1 - c_1} R$.
		\item\label{itmOAE6} The distance between any two adjacent vertices in \neww{$\Lambda^*_n$} is at most $\frac{\rhat}{2n}\left(1 - c_1\right)$.
		\item\label{itmOAE2} A random walk in \neww{$\Lambda^*_n$} begun from \rev{a dual vertex} in a $\frac{2(1 - c_1)\rhat}{n} \times \frac{(5 - c_1)\rhat}{2n}$ rectangle that is horizontally-centered and at distance $\frac{2\rhat}{n}$ from the top will first leave the rectangle along the top with probability at least $\rho'$.
		\item\label{itmOAE3} For any $w \geq 2(1 - c_1) \rhat$ and $h \in [w - 2R, w]$, a random walk in \neww{$\Lambda^*_n$} begun from \rev{a dual vertex within $R/n$ of} the center of a $\frac{w}{n} \times \frac{h}{n}$ rectangle will first leave to a vertex above any given segment of the top of the rectangle of width $\frac{2c_2w}{n}$ with probability at least $\rho'$.
		\item\label{itmOAERDoublePrime} With probability at least $\rho''$, a random walk in $\Lambda^*_n$ begun from a \neww{dual vertex $v$} first exits \neww{any circle centered at $v$} of radius at least $\frac{R''}{n}$ to a point $v'$ such that the ray from $v$ to $v'$ passes through the arc of angle $\frac{\pi}{6}$ that is centered at the top of the circle.
	\end{enumerate}
	\end{observation}
	
	\begin{proof}
	Because Observation \ref{obsAdditiveErrorRectangles} holds whenever $w,h\ge 4R$, (B\ref{itmOAE1}) holds once $\rhat\ge 400R\cmax/\cmin$.  We enlarge $\rhat$ further, if necessary, so that
	\[
	\rhat\ge \max\left\{50R,\frac{c_1}{1-c_1}R,\frac{100R}{c_2^2}\right\}.
	\]
	This gives (B\ref{itmOAE5}) and (B\ref{itmOAE4}), and the last lower bound ensures that the circles used below in (B\ref{itmOAE3}) have radius at least $R/n$ even at the smallest allowed value of $w$.  Finally, (B\ref{itmOAE6}) follows from axiom~\eqref{itmStrongLatticeSequenceClose}, after increasing $\rhat$ if needed.
	
	We prove (B\ref{itmOAE2}) and (B\ref{itmOAE3}) by applying Lemma~\ref{lemCurveWalkSimple} to straight line segments inside the relevant rectangles.  For (B\ref{itmOAE2}), take the vertical segment from the starting dual vertex to a point just above the top side of the rectangle, and use a tube radius comparable to $(1-c_1)\rhat/n$.  This tube stays a positive fraction of the rectangle width away from the left, right, and bottom sides, and following it forces the walk to exit through the top.  The ratio of the segment length to the tube radius is bounded by a constant depending only on $c_1$, so Lemma~\ref{lemCurveWalkSimple} gives a positive lower bound.
	
	For (B\ref{itmOAE3}), fix the target segment on the top side and apply Lemma~\ref{lemCurveWalkSimple} to the straight segment from the starting vertex to the midpoint of the target segment, extended a distance $R/n$ past the top side.  Use tube radius $c_2w/(4n)$.  Since $w\ge 2(1-c_1)\rhat$ and $\rhat\ge 100R/c_2^2$, this radius is at least $R/n$, so property~\eqref{Strongcircleproperty} is available at the required scale.  The tube remains inside the rectangle until the top-side exit and forces the exit point to lie above the prescribed segment of width $2c_2w/n$.  Again the length-to-radius ratio is bounded by a constant depending only on $c_1,c_2$, so the probability is bounded below by a positive constant.  Let $\rho'$ be the smaller of the constants obtained in (B\ref{itmOAE2}) and (B\ref{itmOAE3}).
	
	For (B\ref{itmOAERDoublePrime}), apply Lemma~\ref{lemCurveWalkSimple} to the vertical segment from $v$ to the point directly above $v$ at distance $2r$, using tube radius $r/100$, where $r$ is the radius of the circle being exited.  If $R''$ is chosen sufficiently large compared to $R$, then $r/100\ge R/n$ for every $r\ge R''/n$.  A walk that follows this tube until it first exits the circle exits through the top sector of angular width at least $\pi/6$.  Lemma~\ref{lemCurveWalkSimple} therefore gives a positive lower bound $\rho''$, depending only on the uniform lattice-sequence constants.
	\end{proof}
	
	\new{We are now ready to prove Theorem \ref{thmAdditiveError}. To give an overview of the main idea behind the proof, let us first discuss why our previous arguments yield a multiplicative error and not an additive one. By property (\ref{itmStrongLatticeSequenceDense}) in the definition of a uniform lattice sequence, there is a correspondence between area in the plane and the number of vertices on either side of the random walk. We should thus be able to draw tubes that the random walks can stay in to guarantee approximately equal numbers of vertices on each side. Additively, the error can be bounded by the area of these tubes, which is an arbitrarily small constant times $n$. To reduce this error down to an absolute constant, we will keep track of the imbalance accumulated so far and adaptively change the trajectory of the path we hope the random walk will follow.
	
	However, the argument for showing that the walk will follow the path with constant probability involves a sequence of squares (or circles, as in Lemma~\ref{lemCurveWalkSimple}) each containing $\Theta(n)$ vertices that the walk is supposed to exit at particular points, so each of these steps introduces its own $\Theta(n)$ error. We must thus start shrinking the squares as we progress, taking smaller and smaller steps until the squares are of constant size. It turns out that we can get away with reducing the the size of the squares by a constant factor each time, so there are $O(\log(n))$ squares total. The probability bound is exponential in the number of squares, which is thus a polynomial function of $n$.}
	
	\begin{proof}[Proof of Theorem \ref{thmAdditiveError}]
	We show that there is a $1/\poly(n)$ probability that a random spanning tree of  $\Omega_{\Lambda_n}$ can be cut into pieces differing in size by an additive constant by examining the probability the tree resembles a particular plane partition consisting of the unit square divided into $k$ vertical pieces. For each $n$, we may use a slightly different vertical partition, chosen to ensure equal numbers of vertices in each region.
	
	Consider $\Omega_{\Lambda_n}$. We draw $k - 1$ vertical lines $L_1, L_2, \dots, L_{k - 1}$ to partition the unit square into $k$ vertical strips each containing exactly the same number of vertices of $\Omega_{\Lambda_n}$, up to an error of $\pm1$ (which is only necessary because the number of points in $\Omega_{\Lambda_n}$ may not be divisible by $k$). We can find the location of each $L_i$ by sweeping from left to right until we have the correct number of points. (If necessary, we can first perturb the embedding slightly so that no two vertices are vertically aligned.) Observe that there is an absolute bound on how close these lines can be to each other and to the left and right sides of the unit square for all sufficiently large $n$. This is because Observation \ref{obsAdditiveErrorRectangles} implies that if one strip has area $\frac{\cmax}{\cmin}$ times the area of another one, it definitely contains more vertices. So let $d_0 > 0$ be such that the horizontal distance between all vertical lines is at least $2d_0$. Let $\varepsilon := 4(1 - c_1)^2 d_0^2 < d_0$.
	
	We run the ordinary version of Wilson's algorithm (not the modified version described in Section \ref{sec:modified_wilson}), initiating the first $k - 1$ random walks $W_1, W_2, \dots, W_{k-1}$ from the bottom of the unit square, lower-bounding the probability each walk $W_i$ makes it to the top while roughly following $L_i$. Figure~\ref{figAdditiveBigPic} depicts a typical walk satisfying these properties, as well as several of the geometric objects described shortly in the proof to aid in the probability analysis.
	
	Each walk $W_i$ starts with a short segment $\beta_i$ from a dual vertex adjacent to the boundary cycle $C^*_n$ and proceeds inward, reaching a distance of at least $\frac{R}{n} + \frac{R''}{n}$ above the bottom of the unit square while staying within $\varepsilon$ of $L_i$, as depicted in the bottom-left of Figure \ref{figAdditiveBigPic}. From the definition of a uniform lattice sequence, it is straightforward to see that such a path must exist, and all such paths must have length bounded by a constant. Since uniform lattice sequences have at most a constant degree at every vertex, there is a constant probability that $W_i$ takes path $\beta_i$.
	
	From there, we next claim that there is a $\frac{1}{\poly(n)}$ probability that $W_i$ reaches a point within $\frac{\varepsilon}{2}$ of the point labeled $\gamma(0)$ in Figure \ref{figAdditiveBigPic}, which is the point on $L_i$ of distance $\varepsilon + \frac{R}{n}$ from the bottom of the unit square. We repeatedly apply (B\ref{itmOAERDoublePrime}) to a sequence of circles of exponentially increasing radius. Each circle is centered at the current position of the walk (the bottom three green points in Figure \ref{figAdditiveBigPic}) and extends downward to be tangent to the horizontal line which is a distance of $\frac{R}{n}$ to the bottom of the unit square, ensuring that the walk does not hit the boundary cycle $C^*_n$. At each step, the walk leaves the circle within the top arc of angle $\frac{\pi}{6}$ with probability at least $\rho''$, thus increasing its vertical distance from the $\frac{R}{n}$ line by a multiplicative factor of $1 + \cos(\pi/12)$ while remaining within the shaded triangular region, which is defined to have bottom angle $\pi/6$ and contain the initial vertex. After
	$$s(n) := \frac{\log(n\varepsilon/R'')}{\log(1 + \cos(\pi/12))}$$
	circles, the elevation will have increased from its initial value of $\frac{R''}{n}$ to $\varepsilon$, at which point it is clear from Figure \ref{figAdditiveBigPic} that the walk must have visited some point within the red circle around $\gamma(0)$ of radius $\frac{\varepsilon}{2}$. This happens with probability at least
	$$(\rho'')^{s(n)} = \Omega\left(\frac{1}{\poly(n)}\right).$$
	
	\begin{figure}\centering
		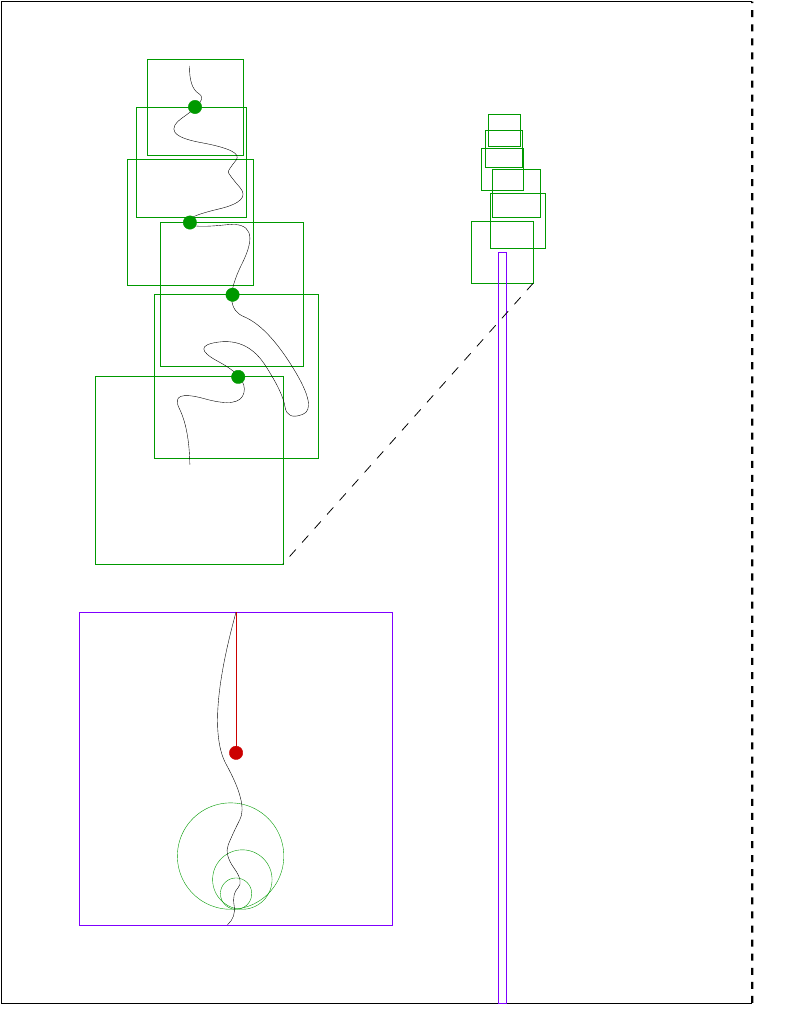
		\vspace{-.4cm}
		\caption{Illustration accompanying the proof of Theorem \ref{thmAdditiveError}, which gives an inverse polynomial lower bound on the probability of the first $k - 1$ walks of Wilson's algorithm follow paths like the windy thin black line on the right. In this diagram, we suppose $k = 2$ and there is roughly uniform density of dual vertices in the graph. Thus, there is only one such path, and it is supposed to split the unit square (which is the boundary of the figure, but also extending beyond the dashed line to the right) in two equal-area pieces. Parts of the path are magnified on the left. The dashed red curve is an erased loop, which we assume was traversed clockwise before continuing on the black path. The figure is drawn to-scale with the following parameters: $c_1 = \frac78$, $c_2 = \frac18$, $c_3 = 1.8$, $d_0 = \frac14$, $L = 6$, $\varepsilon = \frac{1}{256}$, $\frac{R}{n} = \frac{R''}{n} = \frac{1}{2048}$, $\frac{\rhat}{n} = \frac{1}{16}$. Aside from $c_1$ being too small and $c_2$ being too large, these are all valid parameter values.}
		\label{figAdditiveBigPic}
	\end{figure}
	
	As soon as this happens, we apply Lemma \ref{lemCurveWalkSimple} (which is valid as a uniform lattice sequence is indeed a lattice sequence) to the curve beginning at $\gamma(0)$ and proceeding upward to conclude that $W_i$ stays within a radius of $\varepsilon$ of $L_i$ until it is at a distance of $d_0$ from the top of the unit square with constant probability. Let us call the total probability of all of these events occurring $p_0(n)$, which we have shown is $\Omega\left(\frac{1}{\poly(n)}\right)$.
	
	At this point, $W_i$ has reached the vertex $v_0$ in Figure \ref{figAdditiveBigPic}; this is the first vertex encountered in $W_i$ that is of distance at most $d_0$ from the top. More generally, letting $d_j := (c_1)^j d_0$, for each $j = 1, 2, \dots$ let $v_j$ be the first vertex encountered in $W_i$ that is of distance at most $d_j$ from the top, assuming $W_i$ makes it that far. At each step, there is a non-self-intersecting path $P_{i, j}$ from the start of $W_i$ to $v_j$ that will define the ultimate partition of the dual tree assuming parts of it are not erased. Let $\imbalance(j)$ denote the imbalance of the partition of the vertices of $\Omega_{\Lambda_n}$ obtained by extending $P_{i, j}$ straight upward until it hi\neww{t}s the top of the unit square. Here imbalance is defined as the difference between the number of vertices on the left side and the target size of $\frac{i}{k}\abs{\Omega_{\Lambda_n}}$.
	
	Assuming $W_i$ makes it to each vertex $v_j$, we let $S_j$ be the rectangle centered at $v_j$ with width $w_j := 2(1 - c_1) d_j$ and height $h_j := 2(1 - c_1)d_j - 2e_j$, where $e_j$ is the difference between $d_j$ and the vertical distance between $v_j$ and the top of the unit square, which is at most $d_j$. For convenience, we define $v_j$, $w_j$, $h_j$, and $S_j$ for $j = -1, -2$ as well, in the obvious way. Note that $v_j$ and $S_j$ will still be contained within the unit square since we know $d_0 < \frac12$. We prove the following claim by induction on $j$:
	\begin{claim}\label{claImbalance}
		Let $0 \leq j \leq L$, where $L$ is such that $d_L = (c_1)^L d_0 \geq \frac{\rhat}{n}$. With probability at least $p_0(n) (\rho')^j$, all of the following events jointly occur:
		\begin{enumerate}[({C}1)]
			\item\label{itmCla1} The walk $W_i$ eventually reaches a first vertex $v_j$ that is distance at most $d_j$ from the top.
			\item\label{itmCla2} Each of $v_0, v_1, \dots, v_j$ are within
			$$\varepsilon_j' := \varepsilon + \left(\frac14 + c_2\right)(w_0 + w_1 + w_2 + \dots + w_{j - 1})$$
			of line $L_i$.
			\item\label{itmCla3} Each vertex encountered so far in $W_i$ has been within $\varepsilon'_L + w_0$ of $L_i$.
			
			\item\label{itmCla4}
			$$\abs{\imbalance(j)} \leq \cmax n^2\left( c_2 w_{j - 1} d_{j} + w_{j - 2}h_{j - 2} + w_{j - 1}\left(h_{j - 1} + \frac{R}{n}\right)\right).$$
			
			\item\label{itmCla5} For any vertex $v$ previously visited by $W_i$ within rectangle $S_j$, the sub-path of $P_{i, j}$ from $v$ to $v_j$ stays within $S_{j - 1}$ until reaching $v_j$.
		\end{enumerate}
	\end{claim}

	For the base case ($j = 0$), we have already noted that the walk makes it to $v_0$ while staying within the more stringent bound of $\varepsilon$ with probability at least $p_0(n)$. This satisfies (C\ref{itmCla1}), (C\ref{itmCla2}), and (C\ref{itmCla3}). When this happens, the maximum possible magnitude of imbalance is bounded by the initial imbalance of one plus the number of vertices in the $\varepsilon$-neighborhood of $L_i$, which is a rectangle of dimensions $2\ep \times (1-d_0 + e_0)$.  Assuming $n$ is sufficiently large so that Observation \ref{obsAdditiveErrorRectangles} applies, we may thus bound the absolute value of the imbalance by 
	\begin{align*} 1 + \cmax (2 \ep n) ( (1-d_0 + e_0) n) &= 1 + 2 \cmax n^2  \cdot 4(1 - c_1)^2 d_0^2  \cdot (1-d_0 + e_0) \\ &\leq 2 \cmax  n^2 w_0 h_0 \leq \cmax n^2 (w_{- 2} h_{- 2} + w_{-1} h_{-1}),\end{align*}
	which establishes (C\ref{itmCla4}). Finally, (C\ref{itmCla5}) holds by the proof of Lemma \ref{lemCurveWalkSimple}, that is, the way in which we've assured that our walk is staying within $\ep$ of line $L_i$.  This proof uses balls of small radius at most $\ep/2 = 2 (1-c_1)^2 d_0^2$ and bounds the probability the walk always makes progress toward a point further along $L_i$, that is, a point higher up along $L_i$. If such a path went below the bottom of $S_{-1}$, it would have had to backtrack through several circles of radius at most $\ep/2$ (which is much smaller than $h_{-1} = c_1^{-1} d_0$) in the wrong direction, and the fact that it does not do so is already captured by our probability $p_0(n)$. This concludes the base case $j = 0$. 
	
	For the inductive case, suppose the statement holds for $j - 1$. Suppose the random walk $W_i$ has just reached vertex $v_{j - 1}$ in a way that satisfies all five properties (C\ref{itmCla1}\neww{--}\ref{itmCla5}), which happens with probability at least $p_0(n)(\rho')^{j - 1}$. We have from (B\ref{itmOAE3}) (\rev{applied with $w=nw_{j-1}$ and $h=nh_{j-1}$}) that, with probability at least $\rho'$, $W_i$ will next leave $S_{j - 1}$ to a vertex within horizontal distance $\rev{c_2 w_{j-1}}$ of any given target $t$ within the top of $S_{j - 1}$. Since $p_0(n)(\rho')^{j - 1} \cdot \rho' = p_0(n)(\rho')^j$, it suffices to show that there is some target $t$ (that is, an $x$-coordinate) for which this event implies (C\ref{itmCla1}-\ref{itmCla5}) hold after step $j$.

	Consider how $\imbalance(j)$ changes as we vary $t$. Recall the imbalance is calculated by extending $P_{i,j}$ straight upward until it hits the top of the unit square.  We first focus on the contribution to the change in the imbalance from this extension of $P_{i,j}$ from $v_j$ to the top of the unit rectangle.     
	Specifically, if we change $t$ to be $\frac{1}{4}w_{j - 1}$ to the left or right, applying Observation \ref{obsAdditiveErrorRectangles} to the rectangle above $v_j$ to the top of the unit square, we know the imbalance will change by at least
	$$\cmin n^2 \cdot \frac14 w_{j - 1} \left(d_{j} - 2\frac{R}{n}\right). $$
	Here the $-2\frac{R}{n}$ comes from the fact that the height of the precise rectangle we choose is shorter by up to $\frac{R}{n}$ on the bottom because that is how far $v_j$ might be above the top of $S_{j - 1}$, and shorter by $\frac{R}{n}$ on the top because the vertices in $\Lambda_n$ within the top band of height $\frac{R}{n}$ might contain vertices that are outside of the boundary cycle $C^*$. Using (B\ref{itmOAE5}), we know that
	$$d_j \geq d_L \geq \frac{\rhat}{n} \geq 50 \frac{R}{n},$$
	so it follows that a lower bound on the imbalance change is given by
	$$\cmin n^2 \cdot \frac14 w_{j - 1} \left(d_{j} - 2\frac{R}{n}\right) \geq \frac14 \cdot \frac{48}{50} \cdot \cmin n^2 w_{j - 1}d_j = \frac{6}{25} \cmin n^2 w_{j - 1} d_j.$$
	On the other hand we are assuming the previous absolute imbalance is
	\begin{align*}
		\abs{\imbalance(j - 1)} &\leq \cmax n^2\left( c_2 w_{j - 2} d_{j - 1} + w_{j - 3}h_{j - 3} + w_{j - 2}\left(h_{j - 2} + \frac{R}{n}\right)\right)\\
		&\leq \cmax n^2\bigg( c_2 w_{j - 2} d_{j - 1} + w_{j - 3} \cdot 2 (1 - c_1) d_{j - 3} \push + w_{j - 2}\bigg(2 (1-c_1) d_{j-2} + \frac{R}{n}\bigg)\bigg) \\
		&= \cmax n^2 \bigg(\frac{1}{c_1^2}c_2 w_{j - 1} d_{j} + \frac{2(1-c_1)}{c_1^5}w_{j - 1}d_{j} \push + \frac{1}{c_1^2}w_{j - 1}\bigg(2 (1 - c_1) d_{j-1} + c_1\frac{R}{n}\bigg)\bigg)\\
		\intertext{ From how $c_1$ was chosen, we know, for example, that $1/c_1^2 \leq 1/c_1^5 \leq 2$:}
		&\leq \cmax n^2 \bigg(2c_2 w_{j - 1} d_{j} + 4 (1-c_1) w_{j - 1} d_j \push + 2 w_{j - 1}\bigg(2 (1-c_1) d_j + c_1\frac{R}{n}\bigg)\bigg) \\
		\intertext{Since (B\ref{itmOAE4}) implies $c_1 \frac{R}{n} \leq \neww{(1 - c_1)\frac{\rhat}{n}} \leq (1 - c_1)d_L \leq (1 - c_1) d_j$, we see this is }
		&\leq \cmax n^2 \bigg(2c_2 w_{j - 1} d_{j} + 4 (1 - c_1) w_{j - 1} d_j + 2w_{j - 1}\bigg(3(1 - c_1)d_j\bigg)\bigg) \\
		&= \cmax n^2 w_{j - 1} d_j \left(2c_2 + 10(1 - c_1)\right)\\
		&= \cmax n^2 w_{j - 1} d_j \left(\frac{\cmin}{25\cmax} + \frac{\cmin}{5\cmax}\right) \stext{from choice of $c_1,c_2$}\\
		&= \frac{6}{25} \cmin n^2 w_{j - 1} d_j.
	\end{align*}
	
	Thus, there must be some target $t$ in the middle half of the top boundary of $S_{j - 1}$ for which the imbalance contribution of vertices within $d_j - e_j$ of the top of the unit square exactly cancels out $\imbalance(j - 1)$ (up to a single vertex, perhaps). In other words, if $W_i$ were to proceed from $v_{j - 1}$ to a vertex $v_j$ with $x$-coordinate exactly at the target $t$, taking a path that went straight upward to the top of $S_{j - 1}$, then horizontally along the top of the square, then straight up to $v_j$, the partition would be balanced up to one vertex. However, the path it takes through $S_{j - 1}$ can introduce further imbalance of the following magnitudes:
	\begin{itemize}
		\item An imbalance of at most $\cmax n^2 c_2 w_{j - 1} d_{j}$ coming from the fact that $W_i$ might exit $S_{j - 1}$ a distance of up to $\pm c_2$ slightly to either side of the target $t$. (If there is a one-vertex imbalance from before, it can be accounted for in this term as well.)
		
		\item An imbalance of $\cmax n^2 w_{j - 1}(h_{j - 1} + e_j) \leq \neww{\cmax n^2 w_{j - 1}\left(h_{j - 1} + \frac{R}{n}\right)}$ coming from the fact that we can make an arbitrary partition of the vertices in $S_{j - 1}$, where the $e_j$ term accounts for including the vertices above $S_{j-1}$ and below $v_j$.
		
		\item An imbalance of $\cmax n^2 w_{j - 2}h_{j - 2}$ coming from loop erasures that affect the previous rectangle $S_{j - 2}$ as well. For example, this happens in Figure \ref{figAdditiveBigPic} when the red dashed loop is erased shortly after the walk visits $v_4$. Though we do not need to consider any rectangles further back in the sequence: By the inductive hypothesis of (C\ref{itmCla5}), such a loop cannot extend below the bottom of $S_{j - 2}$ (which, in Figure \ref{figAdditiveBigPic}, is the green rectangle centered at $v_3$).
	\end{itemize}
	Summing these up, we obtain the upper bound on $\abs{\imbalance(j)}$ from (C\ref{itmCla4}).
	
	The other four properties are easy to check. (C\ref{itmCla1}) is immediate because points above the top of $S_{j - 1}$ are within $d_j$ of the top of the unit square. (C\ref{itmCla2}) holds because we exit within a $c_2$ fraction of $w_{j - 1}$ from the target, which needs to be at most a $\frac14$ fraction away from the current position, so the total additional drift left or right is at most $(\frac14 + c_2)w_{j - 1}$. (C\ref{itmCla3}) is simply a crude upper-bound on how far to the side any walk could have traveled while staying within the left and right boundaries of the rectangles $S_j$, the biggest rectangle having width $w_0$. For (C\ref{itmCla5}), note that $S_j$ does not overlap with $S_{j - 2}$, so any such vertex $v$ must be in $S_{j - 1} \setminus S_{j - 2}$. Hence, $v$ must have been visited by $W_i$ after $v_{j - 1}$, so the path forward from $v$ must first exit $S_{j - 1}$ to $v_j$. Thus, by induction, Claim \ref{claImbalance} holds for all $0 \leq j \leq L$. 
	
	Returning to the proof of the theorem, we pick $L$ so that $\frac{\rhat}{n} \leq d_L \leq \frac{2\rhat}{n}$; let $c_3 \in [1, 2]$ be such that $d_L = (c_1)^L d_0 = \frac{c_3 \rhat}{n}$. Then with probability $p_0(n) (\rho')^L$, $W_i$ will satisfy (C\ref{itmCla1}-\ref{itmCla5}) with $j = L$. From vertex $v_L$, we know from (B\ref{itmOAE2}) that, with probability at least $\rho'$, $W_i$ will continue all the way to the top of the unit square, hitting $C^*$ before exiting the top of a final $\frac{2(1 - c_1)\rhat}{n} \times \frac{(5 - c_1)\rhat}{2n}$ rectangle, which is shown in purple at the very top of Figure \ref{figAdditiveBigPic}.\footnote{Note that this part of the figure is a bit misleading because it appears rather large in the figure, wider than the longer purple rectangle at the bottom. This is just due to the specific parameters we chose to make the figure readable. As $n$ goes to infinity, the top rectangle will actually shrink down to a point (which is necessary to get an additive constant imbalance!) while the bottom rectangle will remain relatively the same size and shape.} Observe that the imbalance before the final part of the walk starts is at most
	\begin{align*}
		|\imbalance(L)| &\leq \cmax n^2\left( c_2 w_{L - 1} d_{L} + w_{L - 2}h_{L - 2} + w_{L - 1}\left(h_{L - 1} + \frac{R}{n}\right)\right)\\
		&\leq \frac{1}{c_1^4}\cmax n^2\bigg(c_2 2(1 - c_1) d_L^2 + (2(1 - c_1))^2 d_L^2 \push + 2(1 - c_1) d_L^2 + 2(1 - c_1) d_L \frac{R}{n}\bigg)\\
		&\leq \frac{1}{c_1^4}\cmax n^2\bigg(c_2 2(1 - c_1) \frac{4 \rhat^2}{n^2} + (2(1 - c_1))^2 \frac{4 \rhat^2}{n^2} \push + 2(1 - c_1) \frac{4 \rhat^2}{n^2} + 2(1 - c_1) \frac{2 \rhat R}{n^2}\bigg).
	\end{align*}
	Since the $n^2$ terms cancel and $c_1$, $c_2$, $\rhat$, and $R$ are all constant, this is a constant. To bound the additional imbalance generated from this final part of the walk, observe that the final walk stays within a rectangle that does not overlap with $S_{L - 2}$. This is because the final walk starts at least a distance of $\frac{h_{L - 1}}{2}$ above the top of $S_{L - 2}$ and the distance from the start of the final walk to the bottom of the final rectangle is
	\begin{multline*}
		\frac{(5 - c_1)\rhat}{2n} - \frac{2\rhat}{n} = \frac{\rhat}{n}(1 - c_1) - \frac{\rhat}{2n} \left(1 - c_1\right) \\ \leq \frac{\rhat}{n}(1 - c_1) - e_{L - 1} \leq d_{L - 1}(1 - c_1) - e_{L - 1} = \frac{h_{L - 1}}{2},
	\end{multline*}
	where the first inequality follows from (B\ref{itmOAE6}). Thus, we may bound the imbalance by applying (B\ref{itmOAE1}) to just the final two rectangles. The imbalance is at most
	\begin{multline*}
		\cmax n^2 w_{L - 1}h_{L - 1} + \cmax \cdot (5 - c_1)(1 - c_1)\rhat^2 \\\leq \cmax n^2\left(\frac{2(1 - c_1)}{c_1}\cdot\frac{2\rhat}{n}\right)^2 + \cmax \cdot (5 - c_1) (1 - c_1)\rhat^2 ,
	\end{multline*}
	which is also a constant. Summing these imbalances, we know that the total imbalance is at most a constant, not depending on $n$. Also, applying (C\ref{itmCla2}) and (C\ref{itmCla3}), the maximum horizontal deviation of $W_i$ from $L_i$ is at most
	\begin{align*}
		\ep'_L + w_0 &\leq \ep + \left(\frac14 + c_2\right)\left(\sum_{j = 0}^\infty (c_1)^{j} w_0\right) + w_0\\
		&= 4(1 - c_1)^2 d_0^2 + \left(\frac14 + c_2\right)\left(\frac{w_0}{1 - c_1}\right) + w_0\\
		&= 4(1 - c_1)^2 d_0^2 + \left(\frac14 + c_2\right)(2d_0) + 2(1 - c_1) d_0\\
		& \leq d_0 \left( \frac{4}{50^2} + \frac{1}{2} + \frac{2}{50} + \frac{2}{50} \right)\\
		&\leq d_0,
	\end{align*}
	so by the way we chose $d_0$, no walk will hit the left or right sides of the unit square, and no two neighboring walks will collide with each other.
	
	We have shown that all $k - 1$ walks will make it to the top of the unit square and have constant imbalance with probability at least
	\begin{equation*}
		\left(p_0(n) (\rho')^{L + 1}\right)^{k - 1} = \left(p_0(n) \rho' \left(\frac{c_3 \rhat}{d_0 n}\right)^{(\log(\rho')/\log(c_1))}\right)^{k - 1} = \Omega\left(\neww{\frac{p_0(n)^{k - 1}}{n^{q}}}\right),
	\end{equation*}
	where
	$$q = \frac{(k - 1) \log(\rho')}{\log(c_1)}.$$
	Note that this is $\frac{1}{\poly(n)}$ since $p_0(n)$ is \neww{bounded below by an inverse polynomial in $n$} and $q$ is a constant that does not depend on $n$. When this happens, the primal tree must be splittable into $k$ pieces that are balanced up to an additive constant.
	\end{proof}
	
	\new{We remark that this proof gives a worse dependence on $k$ than in Theorem~\ref{thm:k-dist}: The probability of an approximately balanced split is at least
	$$\frac{1}{\beta^{k^3} n^{k^2}\poly(n)}$$
	for some constant $\beta$. The $n^{k^2}$ terms come from the increasing sequence of circles and decreasing sequence of squares. The proof shows that each segment of each random walk behaves as desired with probability at least $\frac{1}{\alpha n^k}$ for some constant $\alpha$, and there are $2(k - 1)$ such segments. The $\beta^{k^3}$ comes from the part in between, since $\frac{1}{\varepsilon} = \Theta(k^2)$ is the aspect ratio of the long skinny rectangles, and there are $k - 1$ of them.}

	\section{Experiments}\label{sec:experiments}
	
	\begin{figure}[ht]\centering
	\subfloat[$10 \times 10$ grid.]{\includegraphics[width=0.49\textwidth]{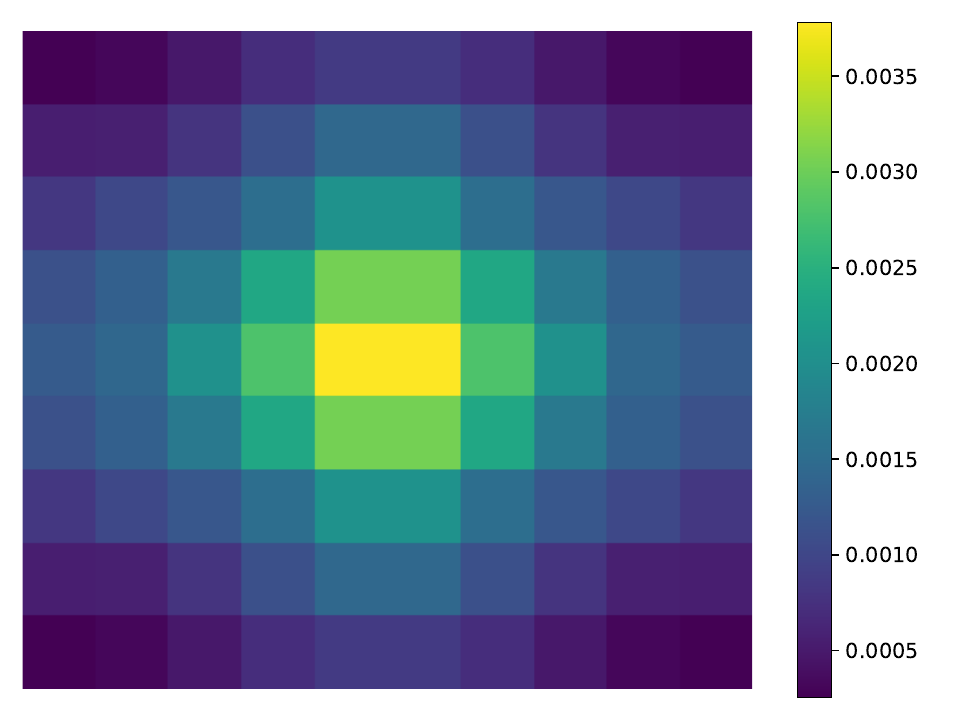}\label{figHeatmap10}}
	\hfill
	\subfloat[$50 \times 50$ grid.]{\includegraphics[width=0.502\textwidth]{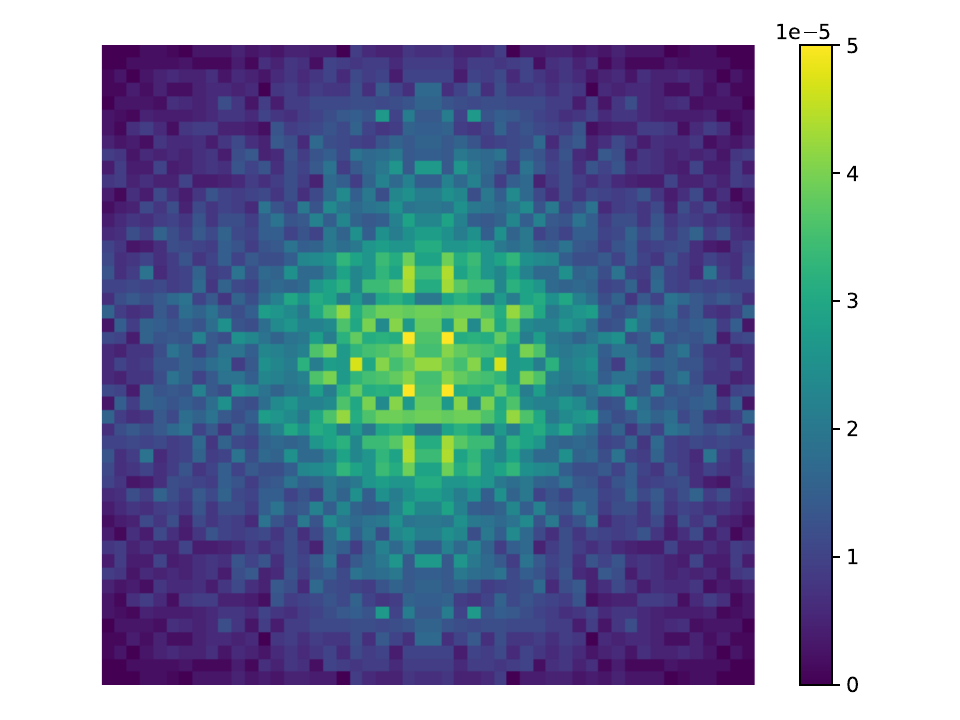}\label{figHeatmap50}}
	\caption{Heatmap of likelihoods that each vertical edge of the $10 \times 10$ and $50 \times 50$ grid graph is an exactly balanced split edge of a uniformly random spanning tree. Each pixel represents a vertical edge. Symmetry is enforced manually; for instance, the center two vertical edges of the $10 \times 10$ grid are exactly the same color not by coincidence, but because the experiment was only run for one of them, with both edges receiving the same color. 1,000,000 trials were run for each symmetry class of edges in each graph. In the $10 \times 10$ grid, the number of those trials yielding exact balance ranged from 255 (in the corners) to 3,781 (in the center). In the $50 \times 50$ grid, they ranged from 0 to 50.  In particular, this plot is still noisy at this number of trials.}
	\label{figHeatmapSmallMediumGrid}
	\end{figure}
	
	We visualize some of the results in this paper by running computational experiments with $k = 2$ on $10 \times 10$, $50 \times 50$, and $100 \times 100$ grid graphs. By running the first random walks of Wilson's algorithm on the dual grid, we empirically estimate the probability of various edges being contained in a random spanning tree and splitting it into two balanced pieces. The only concrete lower bounds we give on these probabilities are for central edges of the grid (in Lemma~\ref{lemCentralEdgeBound}). However, Figure \ref{figHeatmapSmallMediumGrid} suggests that, even though central edges have the highest probability, similar bounds, possibly worse by another factor of $n$, should apply for other edges as well.
	
	\begin{figure}[ht]\centering
	\includegraphics[scale=.128]{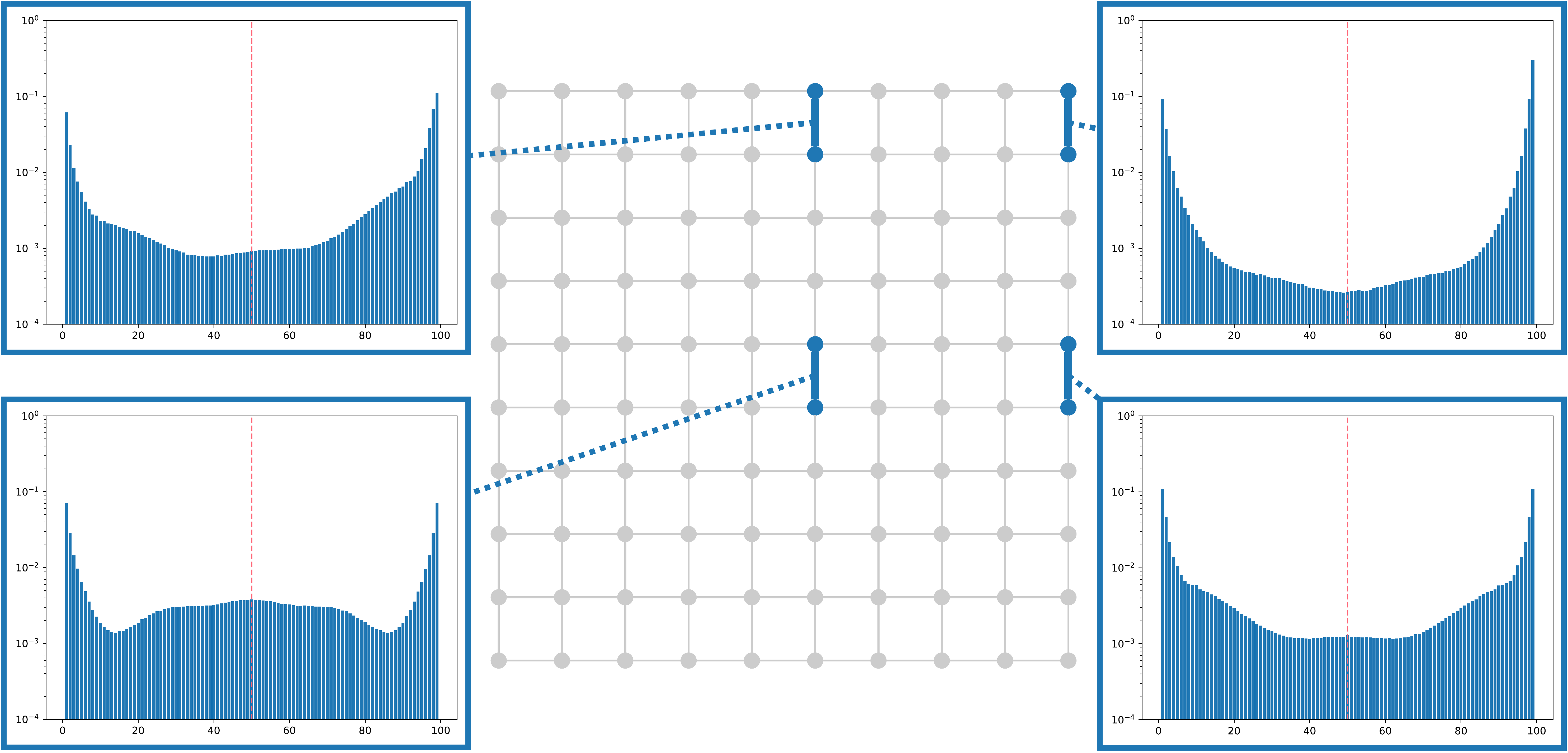}
	\caption{Distributions over component sizes after removing an edge from a random spanning tree for four specific edges in the $10 \times 10$ grid. The $x$-axis represents the size of the component below the edge, which can be any number between 1 and 99. Each bar represents a single value, with the one under the dashed red line representing exact balance. All plots are on the same logarithmic scale, with probability on the $y$-axis. Note that the probabilities sum to strictly less than one because sometimes the edge is not even included in the tree. The distribution is asymmetric for the two edges in the top row and symmetric for the edges in the middle row.}
	\label{figSplitDistributionsSmallGrid}
	\end{figure}
	
\begin{figure}[ht]\centering
	\subfloat[$50 \times 50$ grid, bin size 25.]{\includegraphics[width=0.495\textwidth]{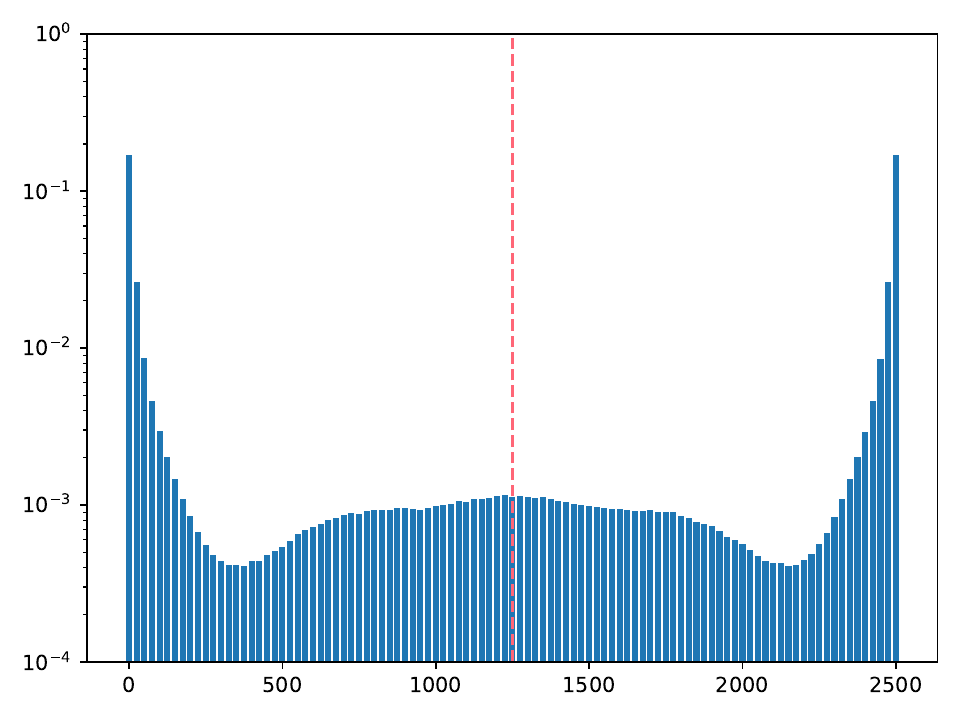}\label{figHistogram50}}
	\hfill
	\subfloat[$100 \times 100$ grid, bin size 101.]{\includegraphics[width=0.495\textwidth]{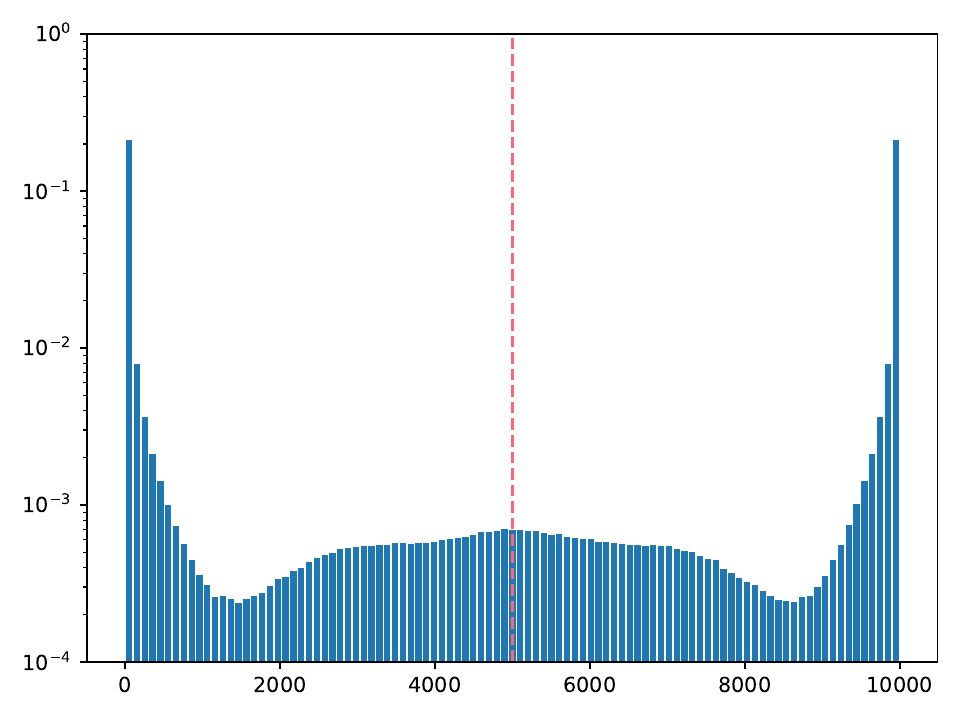}\label{figHistogram100}}
	\caption{Distribution over component sizes for the central edges in the $50 \times 50$ and $100 \times 100$ grids, similar to the bottom-left histogram in Figure \ref{figSplitDistributionsSmallGrid}. In the $50 \times 50$ grid, each bar (except the ones on the very end) represents 25 possible values, and in the $100 \times 100$ grid they represent 101 possible values. These plots are on the same logarithmic scales and have the same numbers of bars as in Figure \ref{figSplitDistributionsSmallGrid}.}
	\label{figHistogramMediumLargeGrid}
	\end{figure}

	When a given edge is in the tree but does not give a balanced split, how imbalanced is it? Figure \ref{figSplitDistributionsSmallGrid} shows the empirical distribution over component sizes for various edges in the $10 \times 10$ grid. As one can see, the most common component sizes are heavily imbalanced. Curiously though, at least for central edges on the $10 \times 10$ grid, being exactly balanced is more likely than being moderately balanced. 
	Figure \ref{figHistogramMediumLargeGrid} shows that this trend holds for $50 \times 50$ and $100 \times 100$ grids as well.
	
	Comparing these figures also illustrates the main point of Corollary \ref{cor:ep}. In each histogram, summing the area of a constant number of bars near the dashed line yields the probability that the parts are approximately balanced. As one can see, this probability does not appear to be vanishing as we increase the grid resolution; it should be lower-bounded by a constant.

	\section*{Acknowledgments}
	We thank Jacob Calvert for pointing out Lemma~\ref{lemSoS} could be proved more simply using the \neww{Cauchy--Schwarz} inequality. This material is based upon work supported by the National Science Foundation under Grant No. DMS-1928930 and by the Alfred P. Sloan Foundation under grant G-2021-16778, while the authors were in residence at the Simons Laufer Mathematical Sciences Institute (formerly MSRI) in Berkeley, California, during the Fall 2023 semester.  S. Cannon is supported in part by NSF CCF-2104795.  W. Pegden is supported in part by NSF DMS-2054503. J. Tucker-Foltz is supported in part by a Google PhD Fellowship.
	
	\FloatBarrier
	
	\appendix
	
	\section*{Appendix}
	
	This section includes proofs omitted from the main body of the paper.

	\bigskip
	
	To prove Proposition \ref{p.WilsonGrid}, we will use the following propositions:
	\begin{proposition}\label{p.commuteres}
		For vertices $u,v$ in a graph $G$ with $M$ edges, the commute time $\tau_u(v)+\tau_v(u)$ between $u$ and $v$ is equal to $2M\cR_{u,v}$, where $\cR_{u,v}$ denotes the effective resistance between $u$ and $v$.\qed
	\end{proposition}
	\noindent The above is proved, e.g., in \cite{electrical}. We will use the following bound on the effective resistance of the grid, which can be found, e.g., in Proposition 9.16 in \cite{markovmixing}:
	\begin{proposition}
		\noindent The effective resistance between opposite corners of a $k\times k$ grid graph is at most $2\log k$.\qed
	\end{proposition}
	\noindent Finally we recall the following monotonicity principles for effective resistance:
	\begin{proposition}\label{p.raleigh}
		If $G$ is a graph and $G'$ is obtained from $G$ by adding edges and/or gluing vertices, the effective resistance between any pair of vertices in $G'$ is at most the effective resistance between the corresponding pair of vertices in $G$.\qed
	\end{proposition}
	
	\begin{proof}[Proof of Proposition \ref{p.WilsonGrid}]
		Suppose the $G$ is an $m \times n$ grid graph, with $n\leq m$.
		By Propositions \ref{p.WilsonRunning} and \ref{p.commuteres}, it suffices that the effective resistance between any vertex $v$ in the dual $G^*$ and the root $r$ corresponding to the outer face of $G$ satisfies $\cR_{v,r}\leq 2\log n$.  We show this in the following way.  Let $\widetilde G$ be an $(m+1)\times (n+1)$ grid graph.  Observe that $G^*$ can be obtained from $\widetilde G$ by gluing all its boundary vertices (i.e., those incident with its outer face in the canonical drawing) to form a single vertex, which is then the root $r$ of $G^*$.  By Proposition \ref{p.raleigh}, for any vertex $v\in G^*$, an upper bound on the effective resistance between $v$ and $r$ in $G^*$ can be obtained by considering the effective resistance between $v$ and any boundary vertex in $\widetilde G$. To this end, simply consider a largest square subgrid $S$ of $\widetilde G$ which has $v$ as its corner vertex.  The opposite corner $u$ will necessarily be a boundary vertex of $\widetilde G$.  Even just in $S$, the effective resistance between $v$ and $u$ is at most $2\log n$, and thus by Proposition \ref{p.raleigh}, it is at most $2\log n$ in $\widetilde G$ as well, proving the proposition.
	\end{proof}
	
	\bigskip
	
	\begin{proof}[Proof of Theorem~\ref{thmAlgWilson}]
		We will refer to each time we begin again at Step 1 as a {\it round} of this algorithm. Note in Step~2, if there exist $k-1$ edges whose removal disconnects $T$ into $k$ components of equal size, this collection of edges is unique; there are never two distinct ways to divide a tree into $k$ components of equal size. 
		
		{\it Correctness}: Let $P$ be a particular balanced $k$-partition of $G$. Let $P_1$, $\ldots$, $P_k$ be the districts of $P$.
		In any given round of this algorithm, the probability we sample a spanning tree that can be divided to produce exactly $P$ is  
		\[
		\frac{\st(G / P) \prod_{i = 1}^k \st(P_i)}{ \st(G)}.
		\]
		We then accept this sample with probability $1/s = 1/\st(G / P)$. In all, the probability we sample partition $P$ is proportional to $\prod_{i = 1}^k \st(P_i)$, meaning we are sampling exactly from the spanning tree distribution.

		{\it Runtime}: In the interest of runtime, we can implement the first step by running Wilson's algorithm on the dual graph, with the vertex corresponding to the outer face of the grid as the root.  In this way, Wilson's algorithm takes $O(N\log N)$ steps to produce a uniformly random spanning tree of $G$.  The resulting uniformly random spanning tree of the dual graph can be converted to a uniformly random spanning tree of the grid graph in linear time, via the bijection in Lemma \ref{lemDuality}.
		
		Step 2 can be easily implemented using a greedy approach and depth-first search, which runs in time $O(N)$. Creating $G / P$ takes at most $O(N)$ time, and computing its spanning tree count is constant time whenever $k$ is a constant, as $G / P$ has only $k$ vertices. This means the runtime of each round of this algorithm is $O(N \log N)$ in expectation.

		The expected number of rounds until a random spanning tree of an $N$-vertex grid graph $G$ is found to be $k$-splittable is $O(N^{2k-2})$ by our Theorem~\ref{thm:k-split} when $k$ is constant.  Similarly as in the proof of Lemma \ref{lem:balancedprob}, a trivial upper bound on $s$ for any $G / P$ is $\binom{2N}{k-1} = O(N^{k-1})$, as there are at most $2N$ edges in $G$ and $k-1$ are in any spanning tree of $G / P$. This means the expected number of rounds we must do until a $k$-splittable partition is returned is $O(N^{3k-3})$ in expectation. 
		Altogether, this gives an expected \neww{runtime} of $O(N^{3k-2}\log N)$.  
	\end{proof}
	
	\bigskip
	
	\begin{proof}[Proof of Theorem~\ref{thmAlgUpDown}]
		We will refer to each time we begin again at Step 1 as a {\it round} of this algorithm.  The mixing time of the up-down walk on a graph with $N$ vertices is $O(N \log N)$, and each step can be implemented $O(\log N)$ amortized time \neww{(Theorem 3 in \cite{anari2021forestsampling}; see Section 4 of \cite{charikar2022complexity} for further discussion)}. It then takes $O(N)$ steps to check if the sampled $k$-forest is balanced. Thus each round of Step 1 takes total time $O(\kappa N\log^2 N)$. 
		
		In each such round we obtain a sample from the distribution on $k$-forests with total variation distance at most $e^{-\Omega(\kappa)}$.  By choosing $\kappa\geq \log^2(N)$, say, we have that the total variation distance is $O(\frac{1}{N^{\log N}})$.  Thus we will see a balanced $k$-forest with probability at least 
		\[
		\frac{1}{\beta^{k^2}n^{5k-5}m^{3k-3}}-O\big(\frac{1}{N^{\log N}}\big)=\Omega\bigg(\frac{1}{n^{5k-5}m^{3k-3}}\bigg)
		\]
		in each iteration of Step 1, by Theorem~\ref{thm:k-dist} and the definition of the total variation distance.
		
		Thus after $O(n^{5k-5}m^{3k-3})$ steps in expectation, we see a balanced $k$-forest in Step 2.  This is drawn from a distribution which had \neww{total variation (TV)} distance $e^{-\Omega(\kappa)}$ from uniform on all $k$-forests, conditioned then to have balanced component sizes.  Conditioning on this event inflates the TV distance by at most the \neww{reciprocal} of the probability of the event, so the final sample has TV distance
		\[
		\neww{O\big(n^{5k-5}m^{3k-3} e^{-\Omega(\kappa)}\big)=\frac{1}{N^{\log N-O(1)}}}
		\]
		from the uniform distribution on balanced $k$-component forests.
		
		Under this uniform distribution over forests, the probability a given partition $P$ is returned is proportional to the number of $k$-forests whose connected components are the districts of $P$, which is precisely the spanning tree distribution. 
	\end{proof}
	
	\bigskip
	
	\begin{proof}[Proof of Lemma \ref{lem:epdelta}]
		Choose $\ep>0$ small enough that distinct vertices of $D$ are at distance at least $3\ep$, and that every vertex of $D$ is at distance at least $3\ep$ from every curve of $D$ not incident to it.  For each curve $\gamma$, define $a_\gamma$ and $b_\gamma$ using this same value of $\ep$, as in the statement of the lemma.  Because distinct curves of $D$ meet only at endpoints, choosing $\ep$ sufficiently small also ensures that the compact images of the restricted curves $\gamma|_{[a_\gamma,b_\gamma]}$ are pairwise disjoint.  Let $s>0$ be the minimum distance between these finitely many compact sets.
		
		It remains to choose a uniform clearance inside each face.  Fix a face $\phi$ and let $K_\phi$ be the union of the curves of $D$ in the boundary of $\phi$.  Let
		\[
		L_{\phi,\ep}=\{x\in \phi: d(x,K_\phi)\ge \ep\}.
		\]
		For bounded faces this set is compact.  For the outer face, intersect it with a large closed disk containing the bounded region relevant to the theorem; points outside this disk can be joined through a still larger circle at a distance bounded below from the drawing.  It is therefore enough to work on compact sets.
		
		For any pair $x,y\in L_{\phi,\ep}$, the face $\phi$ is an open connected subset of the plane and hence is path connected, so there is a curve in $\phi$ joining $x$ to $y$.  The image of this curve is compact and disjoint from $K_\phi$, and therefore has some positive distance from $K_\phi$.  Moreover this positive clearance persists under sufficiently small perturbations of the endpoints, because the endpoints themselves lie at distance at least $\ep$ from $K_\phi$ and can be joined to nearby endpoints inside small balls disjoint from $K_\phi$.  These endpoint neighborhoods form an open cover of $L_{\phi,\ep}\times L_{\phi,\ep}$, so compactness gives a number $m_\phi>0$ such that any two points of $L_{\phi,\ep}$ can be joined in $\phi$ by a curve staying at distance at least $m_\phi$ from $K_\phi$.
		
		Now take
		\[
		0<\delta<\min\{\ep,\,s/3,\,m_\phi/3: \phi\text{ a face of }D\}.
		\]
		Then (A1) holds by the choice of $\ep$, (A2) holds because $3\delta<s$, and (A3) holds because $3\delta\le m_\phi$ in the common face containing the two points.
	\end{proof}

	\bibliographystyle{plain}
	\bibliography{treesplitting}
\end{document}